\documentclass{article}
\def \footer {DRAFT. Copyright is held by the author(s). May, 2017.}

\PassOptionsToPackage{numbers, sort&compress}{natbib}

\usepackage[final]{nips_2016}
\usepackage[utf8]{inputenc}
\usepackage[T1]{fontenc}
\usepackage[hidelinks]{hyperref}
\usepackage{url}
\usepackage{booktabs}
\usepackage{amsfonts}
\usepackage{nicefrac}
\usepackage{microtype}
\usepackage{graphicx}
\usepackage{todonotes}
\usepackage{subcaption}
\usepackage{amsmath}
\usepackage{amssymb}
\usepackage{multirow}
\usepackage{cleveref}
\usepackage{stfloats}

\providecommand{\norm}[1]{\lVert#1\rVert}
\newcommand\mybox[2][]{\tikz[overlay]\node[fill=blue!20,inner sep=2pt, anchor=text, rectangle, rounded corners=1mm,#1] {#2};\phantom{#2}}

\newcommand*{\uset}[3][0pt]{%
  \begingroup
    \renewcommand*{\arraystretch}{0}%
    \:\begin{array}[t]{@{}c@{}}%
      #2\\[{#1}]%
      \scriptstyle #3%
    \end{array}\:%
  \endgroup
}

\hypersetup{%
 colorlinks=true,
 linkcolor=blue,
 citecolor=blue,
 urlcolor=blue
}

\title{Neural Models for Information Retrieval}

\author{
  Bhaskar Mitra \\
  Microsoft, UCL\thanks{The author is a part-time PhD student at University College London.} \\
  Cambridge, UK \\
  \texttt{bmitra@microsoft.com} \\
  \And
  Nick Craswell \\
  Microsoft \\
  Bellevue, USA \\
  \texttt{nickcr@microsoft.com} \\
}

\begin{document}

\maketitle

\begin{abstract}

Neural ranking models for information retrieval (IR) use shallow or deep neural networks to rank search results in response to a query. Traditional learning to rank models employ machine learning techniques over hand-crafted IR features. By contrast, neural models learn representations of language from raw text that can bridge the gap between query and document vocabulary. Unlike classical IR models, these new machine learning based approaches are data-hungry, requiring large scale training data before they can be deployed. This tutorial introduces basic concepts and intuitions behind neural IR models, and places them in the context of traditional retrieval models. We begin by introducing fundamental concepts of IR and different neural and non-neural approaches to learning vector representations of text. We then review shallow neural IR methods that employ pre-trained neural term embeddings without learning the IR task end-to-end. We introduce deep neural networks next, discussing popular deep architectures. Finally, we review the current DNN models for information retrieval. We conclude with a discussion on potential future directions for neural IR.

\end{abstract}


\section{Introduction}
\label{sec:intro}

Since the turn of the decade, there have been dramatic improvements in performance in computer vision, speech recognition, and machine translation tasks, witnessed in research and in real-world applications \cite{lecun2015deep}. These breakthroughs were largely fuelled by recent advances in neural network models, usually with multiple hidden layers, known as deep architectures \cite{krizhevsky2012imagenet, lecun2015deep, hinton2012deep, bahdanau2014neural, deng2014deep}. Exciting novel applications, such as conversational agents \cite{vinyals2015neural, sordoni2015neural}, have also emerged, as well as game-playing agents with human-level performance \cite{silver2016mastering, mnih2015human}. Work has now begun in the information retrieval (IR) community to apply these neural methods, leading to the possibility of advancing the state of the art or even achieving breakthrough performance as in these other fields.

Retrieval of information can take many forms. Users can express their information need in the form of a text query---by typing on a keyboard, by selecting a query suggestion, or by voice recognition---or the query can be in the form of an image, or in some cases the need can even be implicit. Retrieval can involve ranking existing pieces of content, such as documents or short-text answers, or composing new responses incorporating retrieved information. Both the information need and the retrieved results may use the same modality (e.g., retrieving text documents in response to keyword queries), or different ones (e.g., image search using text queries). Retrieval systems may consider user history, physical location, temporal changes in information, or other context when ranking results. They may also help users formulate their intent (e.g., via query auto-completion or query suggestion) and/or extract succinct summaries of results for easier inspection.

\emph{Neural IR} refers to the application of shallow or deep neural networks to these retrieval tasks. This tutorial serves as an introduction to neural methods for ranking documents in response to a query, an important IR task. A search query may typically contain a few terms, while the document length, depending on the scenario, may range from a few terms to hundreds of sentences or more. 
Neural models for IR use vector representations of text, and usually contain a large number of parameters that needs to be tuned. ML models with large set of parameters typically require a large quantity of training data \cite{taylor2006optimisation}. Unlike traditional \emph{learning to rank} (L2R) approaches that train ML models over a set of hand-crafted features, neural models for IR typically accept the raw text of a query and document as input. Learning suitable representations of text also demands large-scale datasets for training \cite{mitra2016learning}. Therefore, unlike classical IR models, these neural approaches tend to be data-hungry, with performance that improves with more training data.

Text representations can be learnt in an unsupervised or supervised fashion. The supervised approach uses IR data such as labeled query-document pairs, to learn a representation that is optimized end-to-end for the task at hand. If sufficient IR labels are not available, the unsupervised approach learns a representation using just the queries and/or documents. In the latter case, different unsupervised learning setups may lead to different vector representations, that differ in the \emph{notion of similarity} that they capture between represented items. When applying such representations, the choice of unsupervised learning setup should be carefully considered, to yield a notion of text similarity that is suitable for the target task. Traditional IR models such as \emph{Latent Semantic Analysis} (LSA) \cite{deerwester1990indexing} learn dense vector representations of terms and documents. Neural representation learning models share some commonalities with these traditional approaches. Much of our understanding of these traditional approaches from decades of research can be extended to these modern representation learning models.

In other fields, advances in neural networks have been fuelled by specific datasets and application needs. For example, the datasets and successful architectures are quite different in visual object recognition, speech recognition, and game playing agents. While IR shares some common attributes with the field of natural language processing, it also comes with its own set of unique challenges. IR systems must deal with short queries that may contain previously unseen vocabulary, to match against documents that vary in length, to find relevant documents that may also contain large sections of irrelevant text. IR systems should learn patterns in query and document text that indicate relevance, even if query and document use different vocabulary, and even if the patterns are task-specific or context-specific.

\begin{figure}
\center
\includegraphics[width=0.6\linewidth]{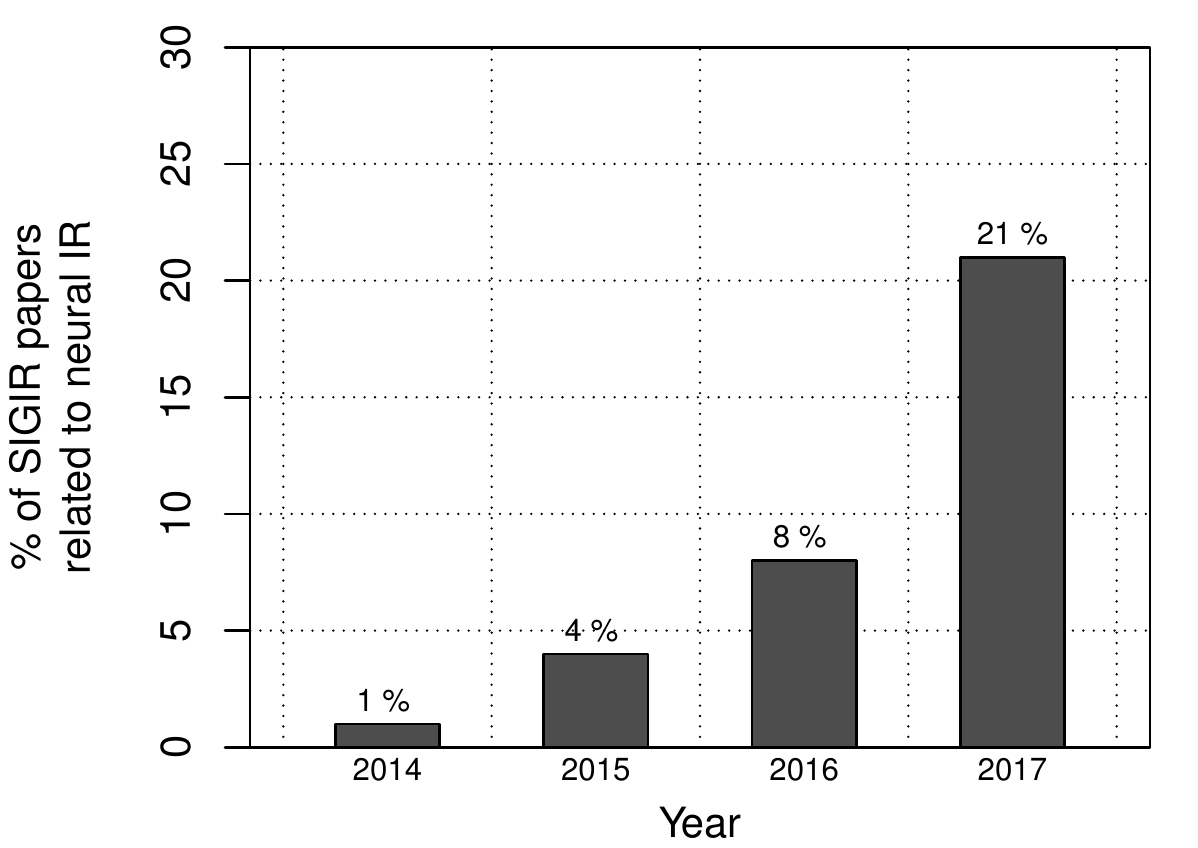}
\caption{The percentage of neural IR papers at the ACM SIGIR conference---as determined by a manual inspection of the paper titles---shows a clear trend in the growing popularity of the field.}
\label{fig:sigirtrend}
\end{figure}

The goal of this tutorial is to introduce the fundamentals of neural IR, in context of traditional IR research, with visual examples to illustrate key concepts and a consistent mathematical notation for describing key models. Section~\ref{sec:ir} presents a survey of IR tasks, challenges, metrics and non-neural models. Section~\ref{sec:anatomy} provides a brief overview of neural IR models and a taxonomy for different neural approaches to IR. Section~\ref{sec:wordrep} introduces neural and non-neural methods for learning term embeddings, without the use of supervision from IR labels, and with a focus on the notion of similarity. Section~\ref{sec:embir} surveys some specific approaches for incorporating such embeddings in IR. Section~\ref{sec:dnn} introduces the fundamentals of deep models that are used in IR so far, including popular architectures and toolkits. Section~\ref{sec:dnnir} surveys some specific approaches for incorporating deep neural networks in IR. Section~\ref{sec:conclusion} is our discussion, including future work, and conclusion.

\paragraph*{Motivation for this tutorial}

Neural IR is an emerging field. Research publication in the area has been increasing (Figure \ref{fig:sigirtrend}), along with relevant workshops \cite{craswellneu, craswellreport, craswell2017neu}, tutorials \cite{li2016deep, mitra2017neural, kenter2017neural}, and plenary talks \cite{manning2016understanding, craswell2017neural}. Because this growth in interest is fairly recent, some researchers with IR expertise may be unfamiliar with neural models, and other researchers who have already worked with neural models may be unfamiliar with IR. The purpose of this tutorial is to bridge the gap, by describing the relevant IR concepts and neural methods in the current literature.

\section{Fundamentals of text retrieval}
\label{sec:ir}

We focus on text retrieval in IR, where the user enters a text query and the system returns a ranked list of search results. Search results may be passages of text or full text documents. The system's goal is to rank the user's preferred search results at the top. This problem is a central one in the IR literature, with well understood challenges and solutions. This section provides an overview of those, such that we can refer to them in subsequent sections.

\subsection{IR tasks}

Text retrieval methods for full text documents and for short text passages have application in ad hoc retrieval systems and question answering systems respectively.

\paragraph*{Ad-hoc retrieval} Ranked document retrieval is a classic problem in information retrieval, as in the main task of the Text Retrieval Conference \cite{voorhees2005trec}, and performed by popular search engines such as Google, Bing, Baidu, or Yandex. TREC tasks may offer a choice of query length, ranging from a few words to a few sentences, whereas search engine queries tend to be at the shorter end of the range. In an operational search engine, the retrieval system uses specialized index structures to search potentially billions of documents. The results ranking is presented in a search engine results page (SERP), with each result appearing as a summary and a hyperlink. The engine can instrument the SERP, gathering implicit feedback on the quality of search results such as click decisions and dwell times.

A ranking model can take a variety of input features. Some ranking features may depend on the document alone, such as how popular the document is with users, how many incoming links it has, or to what extent document seems problematic according to a Web spam classifier. Other features depend on how the query matches the text content of the document. Still more features match the query against document metadata, such as referred text of incoming hyperlink anchors, or the text of queries from previous users that led to clicks on this document. Because anchors and click queries are a succinct description of the document, they can be a useful source of ranking evidence, but they are not always available. A newly created document would not have much link or click text. Also, not every document is popular enough to have past links and clicks, but it still may be the best search result for a user's rare or tail query. In such cases, when text metadata is unavailable, it is crucial to estimate the document's relevance primarily based on its text content.

In the text retrieval community, retrieving documents for short-text queries by considering the long body text of the document is an important challenge. The \emph{ad-hoc} and \emph{Web} tracks\footnote{\url{http://www10.wwwconference.org/cdrom/papers/317/node2.html}} at the popular Text REtrieval Conference (TREC) \cite{voorhees2000overview} focus specifically on this task. The TREC participants are provided a set of, say fifty, search queries and a document collection containing 500-700K newswire and other documents. Top ranked documents retrieved for each query from the collection by different competing retrieval systems are assessed by human annotators based on their relevance to the query. Given a query, the goal of the IR model is to rank documents with better assessor ratings higher than the rest of the documents in the collection. In Section \ref{sec:ir-metrics}, we describe popular IR metrics for quantifying model performance given the ranked documents retrieved by the model and the corresponding assessor judgments for a given query.

\paragraph*{Question-answering} Question-answering tasks may range from choosing between multiple choices (typically entities or binary true-or-false decisions) \cite{richardson2013mctest, hermann2015teaching, hill2015goldilocks, weston2015towards} to ranking spans of text or passages \cite{voorhees2000building, yang2015wikiqa, rajpurkar2016squad, agichtein2015overview, ferrucci2010building}, and may even include synthesizing textual responses by gathering evidence from one or more sources \cite{nguyen2016ms, mitraproposal}. TREC question-answering experiments \cite{voorhees2000building} has participating IR systems retrieve spans of text, rather than documents, in response to questions. IBM's DeepQA \cite{ferrucci2010building} system---behind the Watson project that famously demonstrated human-level performance on the American TV quiz show, "Jeopardy!"---also has a primary search phase, whose goal is to find as many potentially answer-bearing passages of text as possible. With respect to the question-answering task, the scope of this tutorial is limited to ranking answer containing passages in response to natural language questions or short query texts.

Retrieving short spans of text pose different challenges than ranking documents. Unlike the long body text of documents, single sentences or short passages tend to be on point with respect to a single topic. However, answers often tend to use different vocabulary than the one used to frame the question. For example, the span of text that contains the answer to the question "what year was Martin Luther King Jr. born?" may not contain the term "year". However, the phrase "what year" implies that the correct answer text should contain a year---such as `1929' in this case. Therefore, IR systems that focus on the question-answering task need to model the patterns expected in the answer passage based on the intent of the question.

\subsection{Desiderata of IR models}
\label{sec:ir-desiderata}
Before we describe any specific IR model, it is important for us to discuss the attributes that we desire from a good retrieval system. For any IR system, the relevance of the retrieved items to the input query is of foremost importance. But relevance measurements can be nuanced by the properties of \emph{robustness}, \emph{sensitivity} and \emph{efficiency} that we expect the system to demonstrate. These attributes not only guide our model designs but also serve as yard sticks for comparing the different neural and non-neural approaches.

\paragraph*{Semantic understanding}
Most traditional approaches for ad-hoc retrieval count repititions of the query terms in the document text. \emph{Exact term matching} between the query and the document text, while simple, serves as a foundation for many IR systems. Different weighting and normalization schemes over these counts leads to a variety of TF-IDF models, such as BM25 \cite{robertson2009probabilistic}.

However, by only inspecting the query terms the IR model ignores all the evidence of \emph{aboutness} from the rest of the document. So, when ranking for the query ``Australia'', only the occurrences of ``Australia'' in the document are considered, although the frequency of other words like ``Sydeny'' or ``kangaroo'' may be highly informative. In the case of the query ``what channel are the seahawks on today'', the query term ``channel'' implies that the IR model should pay attention to occurrences of ``ESPN'' or ``Sky Sports'' in the document text---none of which appears in the query itself. 

Semantic understanding, however, goes beyond mapping query terms to document terms. A good IR model may consider the terms ``hot'' and ``warm'' related, as well as the terms ``dog'' and ``puppy''---but must also distinguish that a user who submits the query ``hot dog'' is not looking for a "warm puppy" \cite{levy2011plex}. At the more ambitious end of the spectrum, semantic understanding would involve logical reasons by the IR system---so for the query ``concerts during SIGIR'' it associates a specific edition of the conference (the upcoming one) and considers both its location and dates when recommending concerts nearby during the correct week.

These examples motivate that IR models should have some latent representations of intent as expressed by the query and of the different topics in the document text---so that \emph{inexact matching} can be performed that goes beyond lexical term counting.

\paragraph*{Robustness to rare inputs}

\begin{figure*}
\center
\begin{subfigure}{0.48\textwidth}
    \includegraphics[width=\textwidth]{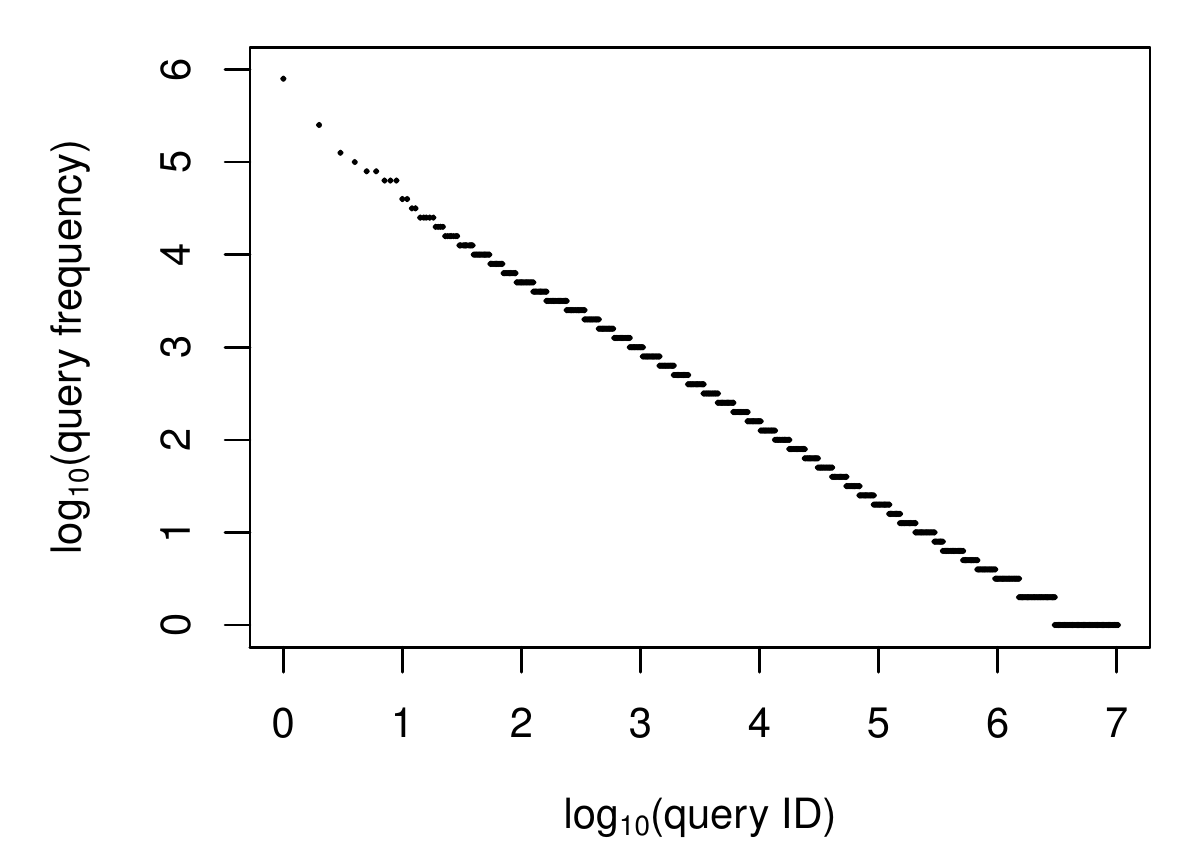}
    \caption{Distribution of query impressions}
    \label{fig:aol-querydistrib}
\end{subfigure}
\begin{subfigure}{0.48\textwidth}
    \includegraphics[width=\textwidth]{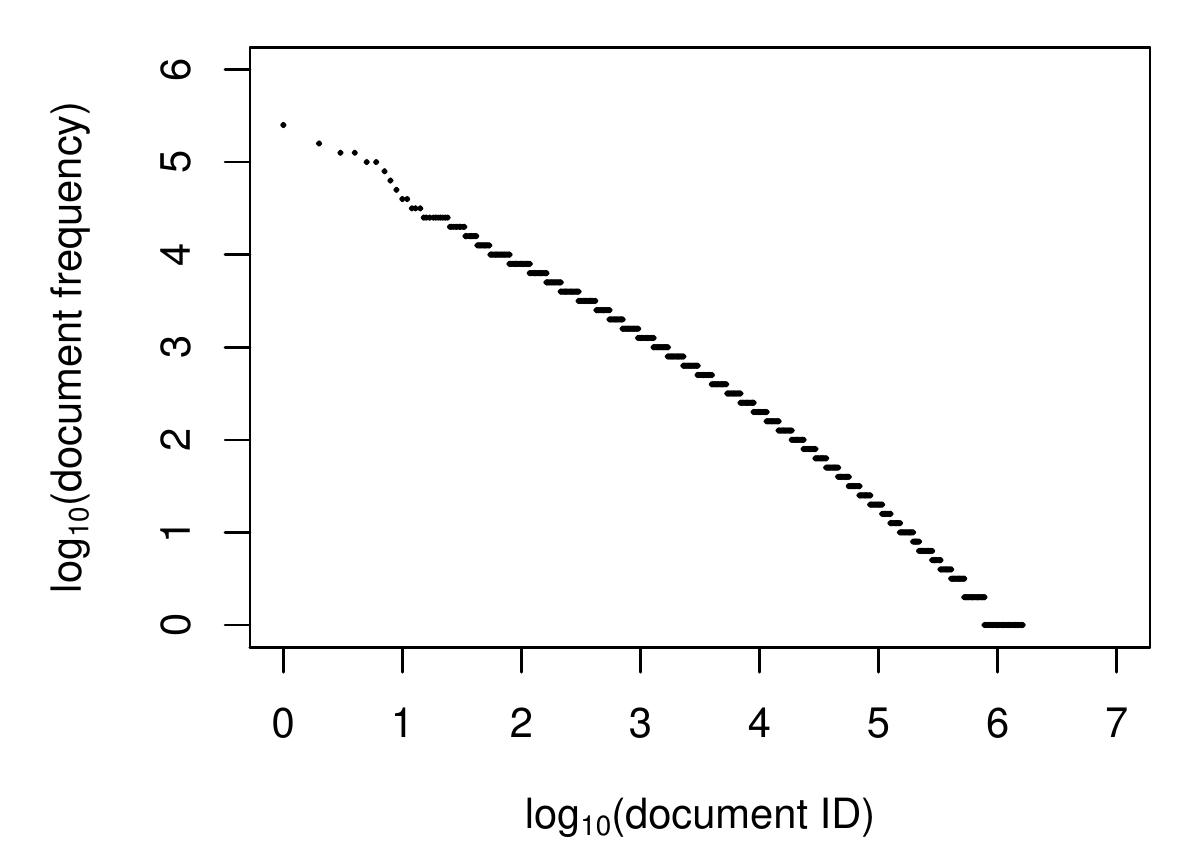}
    \caption{Distribution of document clicks}
    \label{fig:aol-docdistrib}
\end{subfigure}
\caption{A Log-Log plot of frequency versus rank for query impressions and document clicks in the AOL query logs \cite{Pass:2006}. The plots highlight that these quantities follow a Zipfian distribution.}
\label{fig:querydistrib}
\end{figure*}

Query frequencies in most IR setups follow a Zipfian distribution \cite{xie2002locality} (see Figure \ref{fig:querydistrib}). In the publicly available AOL query logs \cite{Pass:2006}, for example, more than 70\% of the distinct queries are seen only once in the period of three months from which the queries are sampled. In the same dataset, more than 50\% of the distinct documents are clicked only once. A good IR method must be able to retrieve these infrequently searched-for documents, and perform reasonably well on queries containing terms that appear extremely rarely, if ever, in its historical logs.

Many IR models that learn latent representations of text from data often naively assume a fixed size vocabulary. These models perform poorly when the query consists of terms rarely (or never) seen in the training data. Even if the model does not assume a fixed vocabulary, the quality of the latent representations may depend heavily on how frequently the terms under consideration appear in the training dataset. \emph{Exact matching} models, like BM25 \cite{robertson2009probabilistic}, on the other hand can precisely retrieve documents containing rare terms.

Semantic understanding in an IR model cannot come at the cost of poor retrieval performance on queries containing rare terms. When dealing with a query such as ``pekarovic land company'' the IR model will benefit from considering exact matches of the rare term ``pekarovic''. In practice an IR model may need to effectively trade-off exact and inexact matching for a query term. However, the decision of when to perform exact matching can itself be informed by semantic understanding of the context in which the terms appear in addition to the terms themselves.

\paragraph*{Robustness to corpus variance} An interesting consideration for IR models is how well they perform on corpuses whose distributions are different from the data that the model was trained on. Models like BM25 \cite{robertson2009probabilistic} have very few parameters and often demonstrate reasonable performance ``out of the box'' on new corpuses with little or no additional tuning of parameters. Deep learning models containing millions (or even billions) of parameters, on the other hand, are known to be more sensitive to distributional differences between training and evaluation data, and has been shown to be especially vulnerable to adversarial inputs \cite{szegedy2013intriguing}. 

Some of the variances in performance of deep models on new corpuses is offset by better retrieval on the test corpus that is distributionally closer to the training data, where the model may have picked up crucial corpus specific patterns. For example, it maybe understandable if a model that learns term representations based on the text of Shakespeare's Hamlet is effective at retrieving passages relevant to a search query from The Bard's other works, but performs poorly when the retrieval task involves a corpus of song lyrics by Jay-Z. However, the poor performances on new corpus can also be indicative that the model is overfitting, or suffering from the Clever Hans\footnote{\url{https://en.wikipedia.org/wiki/Clever_Hans}} effect \cite{sturm2014simple}. For example, an IR model trained on recent news corpus may learn to associate ``Theresa May'' with the query ``uk prime minister'' and as a consequence may perform poorly on older TREC datasets where the connection to ``John Major'' may be more appropriate.

ML models that are hyper-sensitive to corpus distributions may be vulnerable when faced with unexpected changes in distributions or ``black swans''\footnote{\url{https://en.wikipedia.org/wiki/Black_swan_theory}} in the test data. This can be particularly problematic when the test distributions naturally evolve over time due to underlying changes in the user population or behavior. The models, in these cases, may need to be re-trained periodically, or designed to be invariant to such changes.

\begin{figure}
\center
\includegraphics[width=0.95\linewidth]{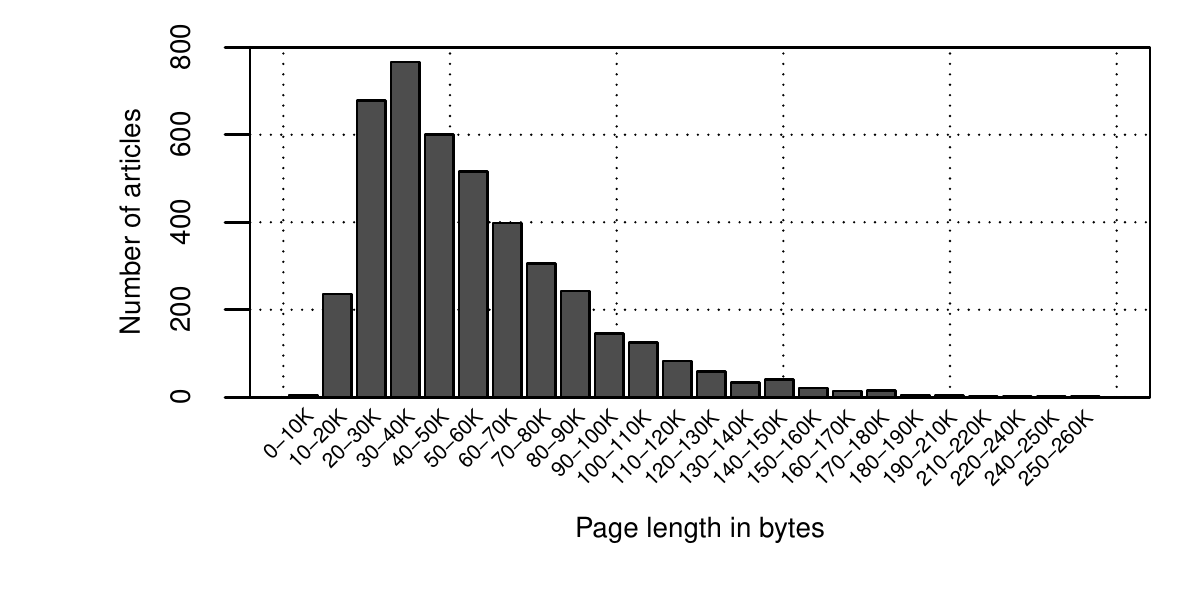}
\caption{Distribution of Wikipedia featured articles by document length (in bytes) as of June 30, 2014. Source: \url{https://en.wikipedia.org/wiki/Wikipedia:Featured_articles/By_length}.}
\label{fig:wikidoclength}
\end{figure}

\paragraph*{Robustness to variable length inputs}

A typical text collection contains documents of varied lengths (see Figure \ref{fig:wikidoclength}). For a given query, a good IR system must be able to deal with documents of different lengths without over-retrieving either long or short documents. Relevant documents may contain irrelevant sections, and the relevant content may either be localized in a single section of the document, or spread over different sections. Document length normalization is well-studied in the context of IR models (e.g., pivoted length normalization \cite{singhal1996pivoted}), and this existing research should inform the design of any new IR models.

\paragraph*{Robustness to errors in input}
No IR system should assume error-free inputs---neither when considering the user query nor when inspecting the documents in the text collection. While traditional IR models have typically involved specific components for error correction---such as automatic spell corrections over queries---new IR models may adopt different strategies towards dealing with such errors by operating at the character-level and/or by learning better representations from noisy texts.

\paragraph*{Sensitivity to context}
Retrieval in the wild can leverage many implicit and explicit context information.\footnote{As an extreme example, in the \emph{proactive retrieval} scenario the retrieval can be triggered based solely on implicit context without any explicit query submission from the user.} The query ``weather'' can refer to the weather in Seattle or in London depending on where the user is located. An IR model may retrieve different results for the query ``decorations'' depending on the time of the year. The query ``giants match highlights'' can be better disambiguated if the IR system knows whether the user is a fan of baseball or American football, whether she is located on the East or the West coast of USA, or if the model has knowledge of recent sport fixtures. In a conversational IR system, the correct response to the question "When did she become the prime minister?" would depend on disambiguating the correct entity based on the context of references made in the previous turns of the conversation. Relevance, therefore, in many applications is situated in the user and task context, and is an important consideration in the design of IR systems.

\paragraph*{Efficiency}
Efficiency of retrieval is one of the salient points of any retrieval system. A typical commercial Web search engine may deal with tens of thousands of queries per second\footnote{\url{http://www.internetlivestats.com/one-second/\#google-band}}---retrieving results for each query from an index containing billions of documents. Search engines typically involve large multi-tier architectures and the retrieval process generally consists of multiple stages of pruning the candidate set of documents \cite{matveeva2006high}. The IR model at the bottom of this \emph{telescoping} setup may need to sift through billions of documents---while the model at the top may only need to re-rank between tens of promising documents. The retrieval approaches that are suitable at one level of the stack may be highly impractical at a different step---models at the bottom need to be \emph{fast} but mostly focus on eliminating irrelevant or junk results, while models at the top tend to develop more sophisticated notions of \emph{relevance}, and focus on distinguishing between documents that are much closer on the relevance scale. So far, much of the focus on neural IR approaches have been limited to re-ranking top-$n$ documents.

\bigskip
While this list of desired attributes of an IR model is in no way complete, it serves as a reference for comparing many of the neural and non-neural approaches described in the rest of this tutorial.

\subsection{Notation}
\label{sec:ir-notations}

\begin{table}
\caption{Notation used in this tutorial.}
\label{tbl:ir-notations}
\begin{center}
\resizebox{.99\textwidth}{!}{
\begin{tabular}{lcl}
\midrule
Meaning & & Notation \\
\midrule
Single query & & $q$ \\
Single document & & $d$ \\
Set of queries & & $Q$ \\
Collection of documents & & $D$ \\
Term in query $q$ & & $t_q$ \\
Term in document $d$ & & $t_d$ \\
Full vocabulary of all terms & & $T$ \\
Set of ranked results retrieved for query $q$ & & $R_q$ \\
Result tuple (document $d$ at rank $i$) & & $\langle i,d \rangle$, where $\langle i,d \rangle \in R_q$ \\
Ground truth relevance label of document $d$ for query $q$ & & $rel_q(d)$ \\
$d_i$ is more relevant than $d_j$ for query $q$ & & $rel_q(d_i) > rel_q(d_j)$, or succinctly $d_i \uset{\succ}{q} d_j$ \\
Frequency of term $t$ in document $d$ & & $tf(t,d)$ \\
Number of documents in $D$ that contains term $t$ & & $df(t)$ \\
Vector representation of text $z$ & & $\vec{v}_z$ \\
Probability function for an event $\mathcal{E}$ & & $p(\mathcal{E})$ \\
\bottomrule
\end{tabular}
}
\end{center}
\end{table}

We adopt some common notation for this tutorial shown in Table \ref{tbl:ir-notations}. We use lower-case to denote vectors (e.g., $\vec{x}$) and upper-case for tensors of higher dimensions (e.g., $X$). The ground truth $rel_q(d)$ in Table \ref{tbl:ir-notations} may be based on either manual relevance annotations or be implicitly derived from user behaviour on SERP (e.g., from clicks).

\subsection{Metrics}
\label{sec:ir-metrics}
A large number of IR studies \cite{guan2007eye, joachims2007evaluating, granka2004eye, joachims2005accurately, mitra2014user, hofmann2014eye, lagun2014towards, diaz2013robust} have demonstrated that users of retrieval systems tend to pay attention mostly to top-ranked results. IR metrics, therefore, focus on rank-based comparisons of the retrieved result set $R$ to an ideal ranking of documents, as determined by manual judgments or implicit feedback from user behaviour data. These metrics are typically computed at a rank position, say $k$, and then averaged over all queries in the test set. Unless otherwise specified, $R$ refers to the top-$k$ results retrieved by the model. Next, we describe a few popular metrics used in IR evaluations.

\paragraph*{Precision and recall}
Precision and recall both compute the fraction of relevant documents retrieved for a query $q$, but with respect to the total number of documents in the retrieved set $R_q$ and the total number of relevant documents in the collection $D$, respectively. Both metrics assume that the relevance labels are binary.

\begin{align}
\label{eqn:ir-precrecall}
Precision_q &= \frac{\sum_{\langle i,d \rangle \in R_q}{rel_q(d)}}{|R_q|} \\
Recall_q &= \frac{\sum_{\langle i,d \rangle \in R_q}{rel_q(d)}}{\sum_{d \in D}{rel_q(d)}}
\end{align}

\paragraph*{Mean reciprocal rank (MRR)}

Mean reciprocal rank \cite{craswell2009mean} is also computed over binary relevance judgments. It is given as the reciprocal rank of the first relevant document averaged over all queries.

\begin{align}
\label{eqn:ir-mrr}
RR_q &= \max_{\langle i,d \rangle \in R_q}{\frac{rel_q(d)}{i}}
\end{align}

\paragraph*{Mean average precision (MAP)}
The average precision \cite{zhu2004recall} for a ranked list of documents $R$ is given by,

\begin{align}
\label{eqn:ir-avgprec}
AveP_q &= \frac{\sum_{\langle i,d \rangle \in R_q}{Precision_{q,i} \times rel_q(d)}}{\sum_{d \in D}{rel_q(d)}}
\end{align}

where, $Precision_{q,i}$ is the precision computed at rank $i$ for the query $q$. The average precision metric is generally used when relevance judgments are binary, although variants using graded judgments have also been proposed \cite{robertson2010extending}. The mean of the average precision over all queries gives the MAP score for the whole set.

\paragraph*{Normalized discounted cumulative gain ($NDCG$)}
There are few different variants of the discounted cumulative gain ($DCG_q$) metric \cite{jarvelin2002cumulated} which can be used when graded relevance judgments are available for a query $q$---say, on a five-point scale between zero to four. A popular incarnation of this metric is as follows.

\begin{align}
\label{eqn:ir-dcg}
\mathrm DCG_q &= \sum _{\langle i,d \rangle \in R_q} \frac {2^{rel_q(d)}-1}{\log _{2}(i+1)}
\end{align}

The ideal DCG ($IDCG_q$) is computed the same way but by assuming an ideal rank order for the documents up to rank $k$. The normalized DCG ($NDCG_q$) is then given by,

\begin{align}
\label{eqn:ir-ndcg}
\mathrm  NDCG_q &= \frac{DCG_q}{IDCG_q}
\end{align}

\subsection{Traditional IR models}
\label{sec:ir-models}
In this section, we introduce a few of the traditionally popular IR approaches. The decades of insights from these IR models not only inform the design of our new neural based approaches, but these models also serve as important baselines for comparison. They also highlight the various desiderata that we expect the neural IR models to incorporate.

\paragraph*{TF-IDF}
There is a broad family of statistical functions in IR that consider the number of occurrences of each query term in the document (term-frequency) and the corresponding inverse document frequency of the same terms in the full collection (as an indicator of the informativeness of the term). One theoretical basis for such formulations is the probabilistic model of IR that yielded the popular BM25 \cite{robertson2009probabilistic} ranking function.

\begin{align}
\label{eqn:ir-bm25}
\displaystyle BM25(q,d)=\sum_{t_q \in q}{idf(t_q)\cdot {\frac {tf(t_q,d) \cdot (k_1+1)}{tf(t_q,d)+k_1 \cdot \left(1-b+b \cdot {\frac{|d|}{avgdl}}\right)}}}
\end{align}

where, $avgdl$ is the average length of documents in the collection $D$, and $k_{1}$ and $b$ are parameters that are usually tuned on a validation dataset. In practice, $k_1$ is sometimes set to some default value in the range $[1.2,2.0]$ and $b$ as $0.75$. The $idf(t)$ is popularly computed as,

\begin{align}
\label{eqn:ir-idf}
idf(t) = \log \frac{|D| - df(t) + 0.5}{df(t) + 0.5}
\end{align}

BM25 aggregates the contributions from individual terms but ignores any phrasal or proximity signals between the occurrences of the different query terms in the document. A variant of BM25 \cite{zaragoza2004microsoft} also considers documents as composed of several fields (such as, title, body, and anchor texts).

\paragraph*{Language modelling (LM)}
In the language modelling based approach \cite{ponte1998language, hiemstra2001using, zhai2001study}, documents are ranked by the posterior probability $p(d|q)$.

\begin{align}
p(d|q) = \frac{p(q|d).p(d)}{\sum_{\bar{d} \in D}{p(q|\bar{d}).p(\bar{d})}} &\propto p(q|d).p(d) \\
          &= p(q|d) \qquad\text{, assuming p(d) is uniform} \\
\label{eqn:ir-lm1}
          &= \prod_{t_q \in q}{p(t_q|d)} \\
          &= \prod_{t_q \in q}{\bigg(\lambda \hat{p}(t_q|d) + (1 - \lambda) \hat{p}(t_q|D)\bigg)} \\
          &= \prod_{t_q \in q}{\bigg(\lambda \frac{tf(t_q,d)}{|d|} + (1 - \lambda) \frac{\sum_{\bar{d} \in D}{tf(t_q,\bar{d})}}{\sum_{\bar{d} \in D}{|\bar{d}|}}\bigg)}
\label{eqn:ir-lm}
\end{align}

where, $\hat{p}(\mathcal{E})$ is the maximum likelihood estimate (MLE) of the probability of event $\mathcal{E}$. $p(q|d)$ indicates the probability of generating query $q$ by randomly sampling terms from document $d$. For smoothing, terms are sampled from both the document $d$ and the full collection $D$---the two events are treated as mutually exclusive, and their probability is given by $\lambda$ and $(1 - \lambda)$, respectively.

\bigskip
Both TF-IDF and language modelling based approaches estimate document relevance based on the count of only the query terms in the document. The position of these occurrences and the relationship with other terms in the document are ignored.

\paragraph*{Translation models}

\citet{berger1999information} proposed an alternative method to estimate $p(t_q|d)$ in the language modelling based IR approach (Equation \ref{eqn:ir-lm1}), by assuming that the query $q$ is being generated via a "translation" process from the document $d$.

\begin{align}
p(t_q|d) &= \sum_{t_d \in d}{p(t_q|t_d)\cdot p(t_d|d)}
\label{eqn:ir-tm}
\end{align}

The $p(t_q|t_d)$ component allows the model to garner evidence of relevance from non-query terms in the document. \citet{berger1999information} propose to estimate $p(t_q|t_d)$ from query-document paired data similar to popular techniques in statistical machine translation \cite{brown1990statistical, brown1993mathematics}---but other approaches for estimation have also been explored \cite{zuccon2015integrating}.

\paragraph*{Dependence model}

None of the three IR models described so far consider proximity between query terms. To address this, \citet{metzler2005markov} proposed a linear model over proximity-based features.

\begin{align}
\begin{split}
DM(q, d) &= (1 - \lambda_{ow} - \lambda_{uw})\sum_{t_q \in q}{log\Bigg((1-\alpha_d)\frac{tf(t_q, d)}{|d|}+\alpha_d\frac{\sum_{\bar{d} \in D}{tf(t_q, \bar{d})}}{\sum_{\bar{d} \in D}{|\bar{d}|}}\Bigg)} \\ &+ \lambda_{ow}\sum_{c_q \in ow(q)}{log\Bigg((1-\alpha_d)\frac{tf_{\#1}(c_q, d)}{|d|}+\alpha_d\frac{\sum_{\bar{d} \in D}{tf_{\#1}(c_q, \bar{d})}}{\sum_{\bar{d} \in D}{|\bar{d}|}}\Bigg)} \\ &+ \lambda_{uw}\sum_{c_q \in uw(q)}{log\Bigg((1-\alpha_d)\frac{tf_{\#uwN}(c_q, d)}{|d|}+\alpha_d\frac{\sum_{\bar{d} \in D}{tf_{\#uwN}(c_q, \bar{d})}}{\sum_{\bar{d} \in D}{|\bar{d}|}}\Bigg)}
\label{eqn:ir-tm}
\end{split}
\end{align}

where, $ow(q)$ and $uw(q)$ are the set of all contiguous $n$-grams (or phrases) and the set of all bags of terms that can be generated from query $q$. $tf_{\#1}$ and $tf_{\#uwN}$ are the ordered-window and unordered-window operators from Indri \cite{strohman2005indri}. Finally, $\lambda_{ow}$ and $\lambda_{uw}$ are the tunable parameters of the model.

\paragraph*{Pseudo relevance feedback (PRF)}
PRF-based methods, such as Relevance Models (RM) \cite{lavrenko2008generative, lavrenko2001relevance}, typically demonstrate strong performance at the cost of executing an additional round of retrieval. The set of ranked documents $R_1$ from the first round of retrieval is used to select expansion terms to augment the query for the second round of retrieval. The ranked set $R_2$ from the second round are presented to the user.

The underlying approach to scoring a document in RM is by computing the KL divergence \cite{lafferty2001document} between the query language model $\theta_q$ and the document language model $\theta_d$.

\begin{align}
\label{eqn:ir-kld}
score(q, d) &= - \sum_{t \in T} p(t|\theta_q) log \frac{p(t|\theta_q)}{p(t|\theta_d)}
\end{align}

Without PRF,

\begin{align}
\label{eqn:ir-qlm}
p(t|\theta_q) &= \frac{tf(t,q)}{|q|}
\end{align}

But under the popular RM3 \cite{abdul2004umass} formulation the new query language model $\bar{\theta_q}$ is estimated by,

\begin{align}
\label{eqn:ir-rm3}
p(t|\bar{\theta_q}) &= \alpha \frac{tf(t,q)}{|q|} + (1 - \alpha) \sum_{d \in R_1}{p(t|\theta_d)p(d) \prod_{\bar{t} \in q}{p(\bar{t}|\theta_d)}}
\end{align}

By expanding the query using the results from the first round of retrieval PRF based approaches tend to be more robust to the vocabulary mismatch problem plaguing many other traditional IR models.

\subsection{Learning to rank (L2R)}
\label{sec:ir-l2r}
In learning to rank, a query-document pair is represented by a vector of numerical features $\vec{x} \in \mathbb{R}^n$, and a model $f: \vec{x} \rightarrow \mathbb{R}$ is trained that maps the feature vector to a real-valued score. The training dataset for the model consists of a set of queries and a set of documents per query. Depending on the flavour of L2R, in addition to the feature vector, each query-document pair in the training data is augmented with some relevance information. \citet{Liu:2009} categorized the different L2R approaches based on their training objectives.

\begin{itemize}
  \item In the \emph{pointwise approach}, the relevance information $rel_q(d)$ is in the form of a numerical value associated with every query-document pair with feature vector $\vec{x}_{q,d}$. The numerical relevance label can be derived from binary or graded relevance judgments or from implicit user feedback, such as clickthrough information. A regression model is typically trained on the data to predict the numerical value $rel_q(d)$ given $\vec{x}_{q,d}$.
  \item In the \emph{pairwise approach}, the relevance information is in the form of preferences between pairs of documents with respect to individual queries (e.g., $d_i \uset{\succ}{q} d_j$). The ranking problem in this case reduces to binary classification for predicting the more relevant document.
  \item Finally, the \emph{listwise approach} involves directly optimizing for a rank-based metric---which is difficult because these metrics are often not continuous (and hence not differentiable) with respect to the model parameters.
\end{itemize}

The input features for L2R models typically belong to one of three categories.

\begin{itemize}
  \item \emph{Query-independent} or \emph{static} features (e.g., PageRank or spam score of the document)
  \item \emph{Query-dependent} or \emph{dynamic} features (e.g., BM25)
  \item \emph{Query-level} features (e.g., number of words in query)
\end{itemize}

Many machine learning models---including support vector machines, neural networks, and boosted decision trees---have been employed over the years for the learning to rank task, and a correspondingly large number of different loss functions have been explored. Next, we briefly describe RankNet \cite{burges2005learning} that has been a popular choice for training neural L2R models and was also---for many years---an industry favourite, such as at the commercial Web search engine Bing.\footnote{\url{https://www.microsoft.com/en-us/research/blog/ranknet-a-ranking-retrospective/}}

\paragraph*{RankNet}
RankNet \cite{burges2005learning} is \emph{pairwise} loss function. For a given query $q$, a pair of documents $\langle d_i,d_j \rangle$, with different relevance labels, such that $d_i \uset{\succ}{q} d_j$, and feature vectors $\langle \vec{x}_i,\vec{x}_j \rangle$, is chosen. The model $f: \mathbb{R}^n \rightarrow \mathbb{R}$, typically a neural network but can also be any other machine learning model whose output is differentiable with respect to its parameters, computes the scores $s_i = f(\vec{x}_i)$ and $s_j = f(\vec{x}_j)$, such that ideally $s_i > s_j$. Given the output scores $\langle s_i,s_j \rangle$ from the model corresponding to the two documents, the probability that $d_i$ would be ranked higher than $d_j$ is given by,

\begin{align}
\label{eqn:ir-ranknet-prob}
p_{ij} \equiv p(d_i \uset{\succ}{q} d_j) \equiv \frac{1}{1 + e^{-\sigma (s_i-s_j)}}
\end{align}

where, $\sigma$ determines the shape of the sigmoid. Let $S_{ij} \in \{-1, 0, +1\}$ be the true preference label between $d_i$ and $d_j$ for the training sample--- denoting $d_i$ is more, equal, or less relevant than $d_j$, respectively. Then the desired probability of ranking $d_i$ over $d_j$ is given by $\bar{p}_{ij} = \frac{1}{2}(1+S_{ij})$. The cross-entropy loss $\mathcal{L}$ between the desired probability $\bar{p}_{ij}$ and the predicted probability $p_{ij}$ is given by,

\begin{align}
\label{eqn:ir-ranknet-loss}
\mathcal{L} &= -\bar{p}_{ij} log(p_{ij}) - (1 - \bar{p}_{ij})log(1 - p_{ij}) \\
   &= \frac{1}{2}(1 - S_{ij})\sigma(s_i - s_j)+log(1 + e^{-\sigma(s_i - s_j)}) \\
   &= log(1 + e^{-\sigma(s_i - s_j)}) \quad \text{if, documents are ordered such that } d_i \uset{\succ}{q} d_j (S_{ij} = 1)
\end{align}

Note that $\mathcal{L}$ is differentiable with respect to the model output $s_i$ and hence the model can be trained using gradient descent. We direct the interested reader to \cite{burges2010ranknet} for more detailed derivations for computing the gradients for RankNet and for the evolution to the \emph{listwise} models LambdaRank \cite{burges2006learning} and LambdaMART \cite{wu2010adapting}.


\section{Anatomy of a neural IR model}
\label{sec:anatomy}

\begin{figure*}
\center
\includegraphics[width=0.6\textwidth]{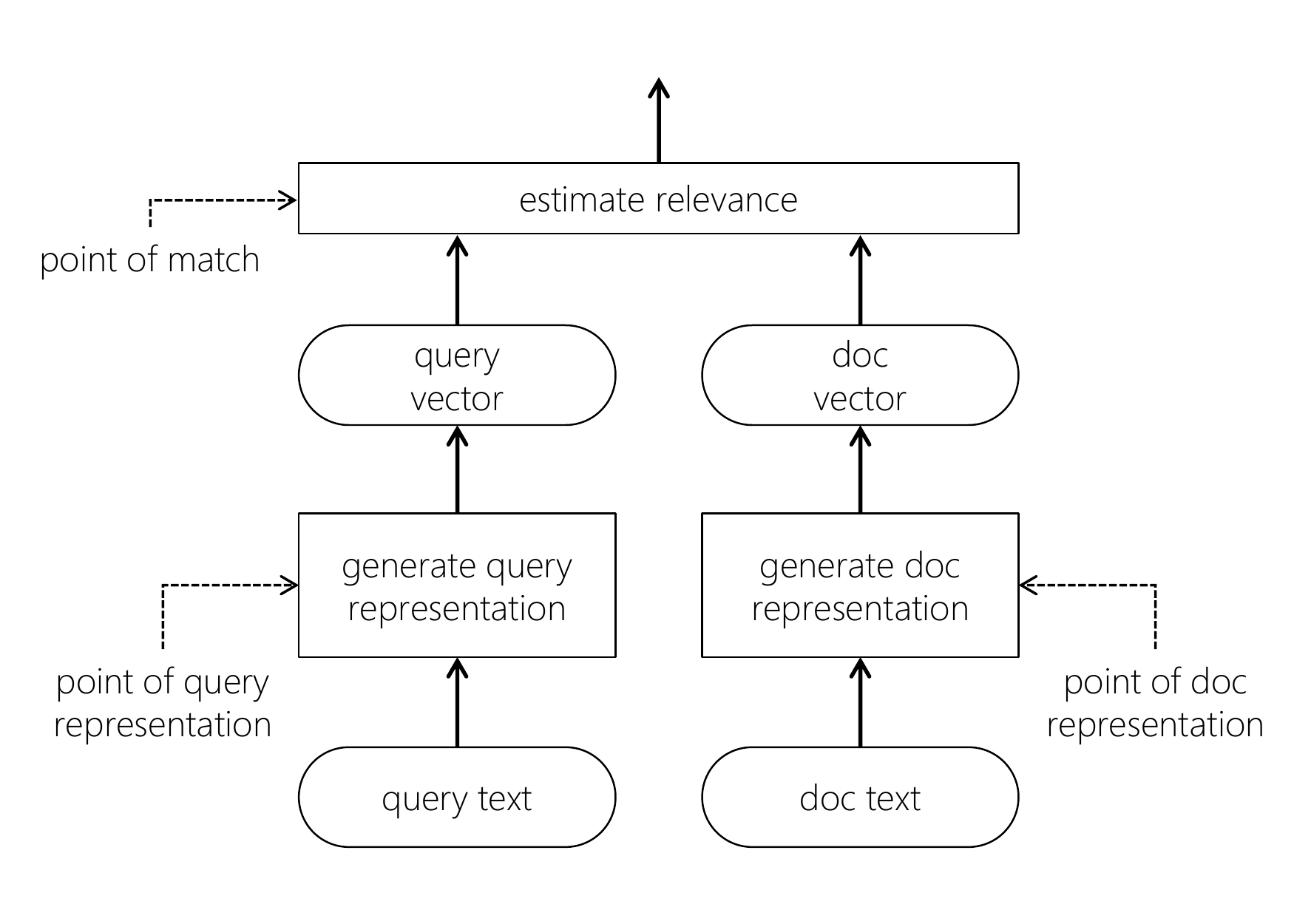}
\caption{Document ranking typically involves a query and a document representation steps, followed by a matching stage. Neural models can be useful either for generating good representations or in estimating relevance, or both.}
\label{fig:anatomy1}
\end{figure*}

\begin{figure*}
\center
\begin{subfigure}{0.49\textwidth}
    \includegraphics[width=\textwidth]{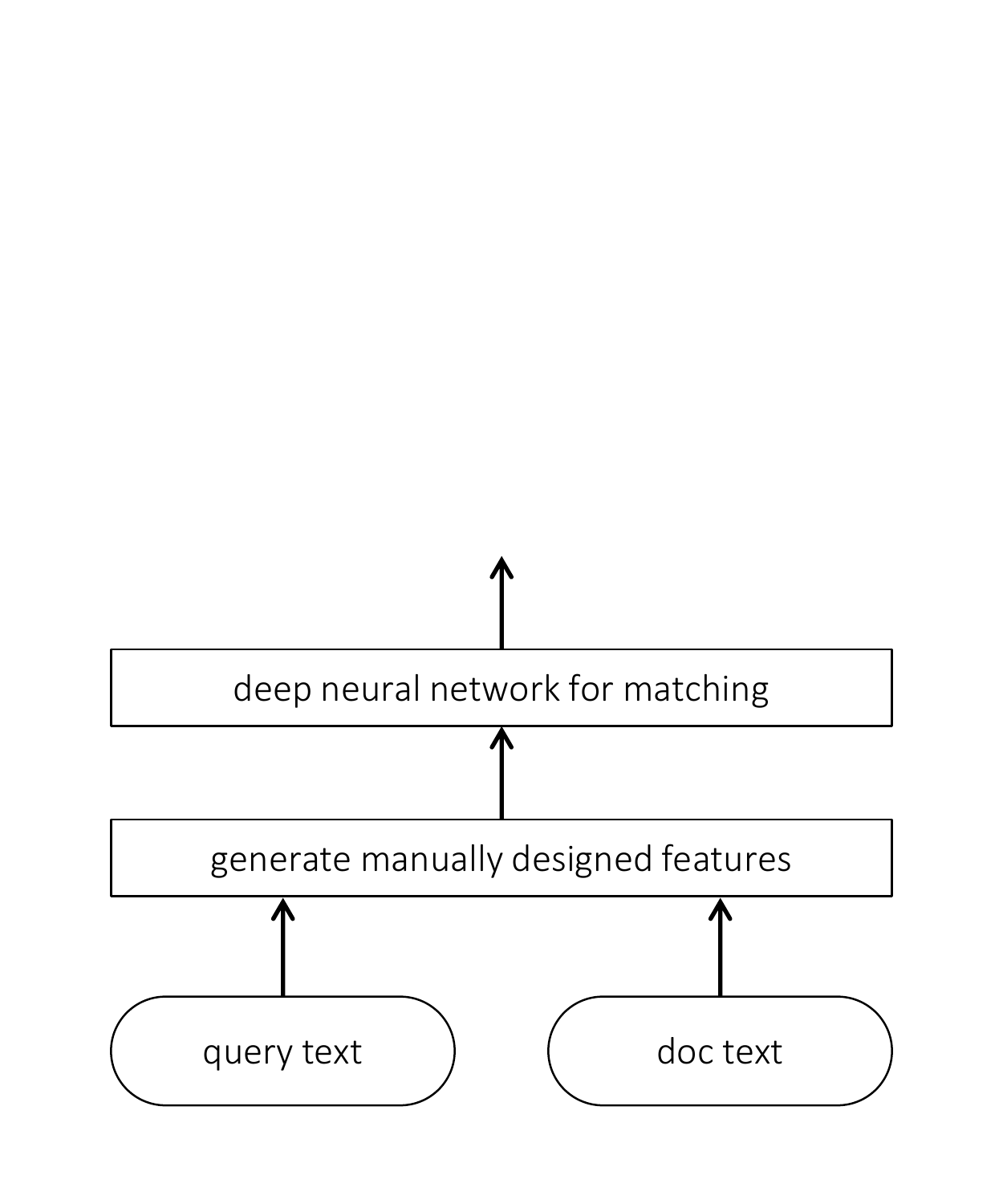}
    \caption{Learning to rank using manually designed features (e.g., \citet{Liu:2009})}
    \label{fig:anatomy2}
\end{subfigure}
\hfill
\begin{subfigure}{0.49\textwidth}
    \includegraphics[width=\textwidth]{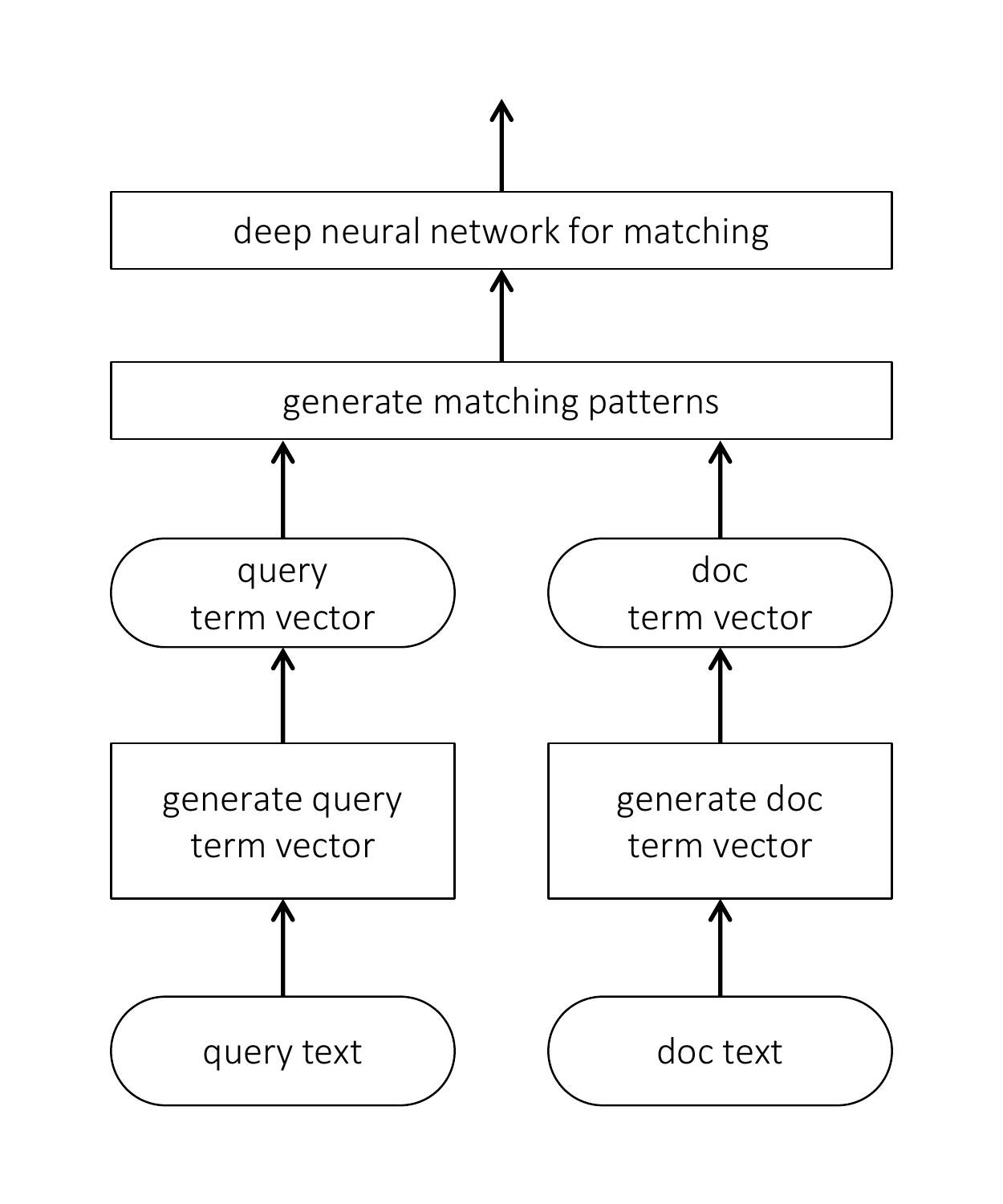}
    \caption{Estimating relevance from patterns of exact matches (e.g., \cite{guo2016deep, mitra2016learning})}
    \label{fig:anatomy3}
\end{subfigure}
\hfill
\begin{subfigure}{0.49\textwidth}
    \includegraphics[width=\textwidth]{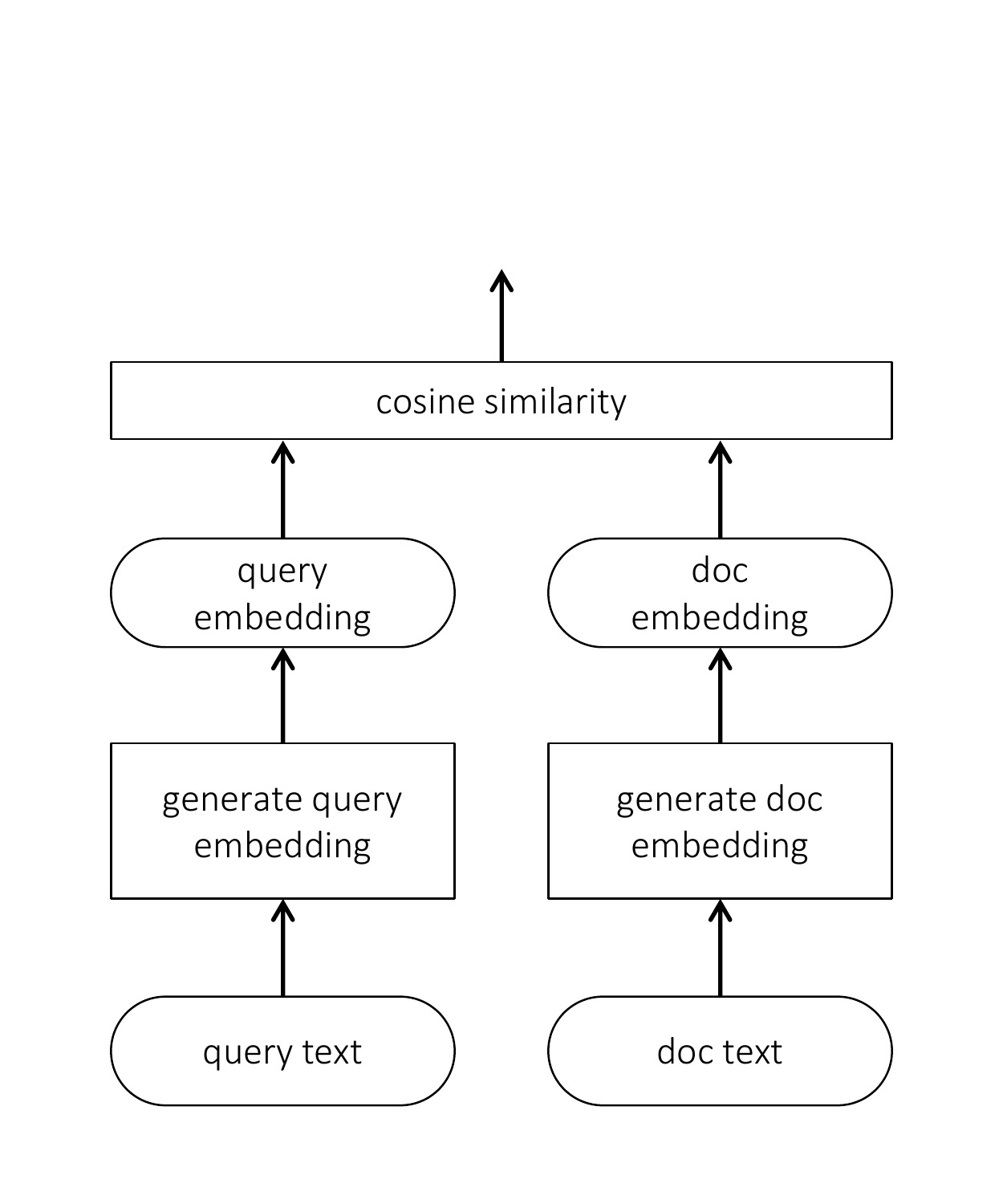}
    \caption{Learning query and document representations for matching (e.g., \cite{huang2013learning, mitra2016desm})}
    \label{fig:anatomy4}
\end{subfigure}
\hfill
\begin{subfigure}{0.49\textwidth}
    \includegraphics[width=\textwidth]{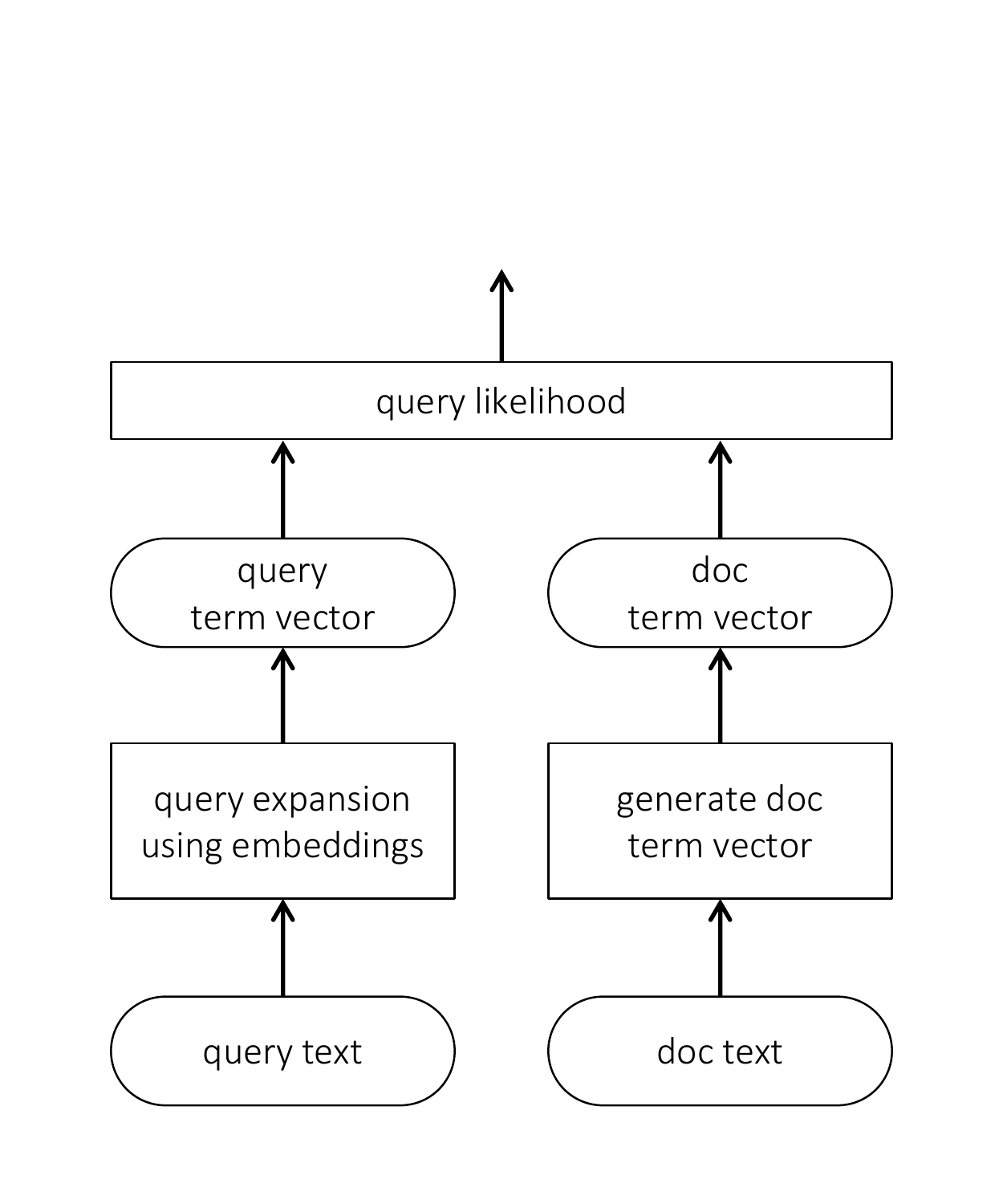}
    \caption{Query expansion using neural embeddings (e.g., \cite{roy2016using, diaz2016query})}
    \label{fig:anatomy5}
\end{subfigure}
\caption{Examples of different neural approaches to IR. In (a) and (b) the neural network is only used at the point of matching, whereas in (c) the focus is on learning effective representations of text using neural methods. Neural models can also be used to expand or augment the query before applying traditional IR techniques, as shown in (d).}
\label{fig:anatomy}
\end{figure*}

At a high level, document ranking comprises of performing three primary steps---generate a representation of the query that specifies the information need, generate a representation of the document that captures the distribution over the information contained, and match the query and the document representations to estimate their mutual relevance. All existing neural approaches to IR can be broadly categorized based on whether they influence the query representation, the document representation, or in estimating relevance. A neural approach may impact one or more of these stages shown in Figure \ref{fig:anatomy1}.

Neural networks are popular as learning to rank models discussed in Section \ref{sec:ir-l2r}. In these models, a joint representation of the query and the document is generated using manually designed features and the neural network is used only at the \emph{point of match} to estimate relevance, as shown in Figure \ref{fig:anatomy2}. In Section \ref{sec:dnnir-lexical}, we will discuss deep neural network models, such as \cite{guo2016deep, mitra2016learning}, that estimate relevance based on patterns of exact query term matches in the document. Unlike traditional learning to rank models, however, these architectures (shown in Figure \ref{fig:anatomy3}) depend less on manual feature engineering and more on automatically detecting regularities in good matching patterns.

In contrast, many (shallow and deep) neural IR models depend on learning good low-dimensional vector representations---or \emph{embeddings}---of query and document text, and using them within traditional IR models or in conjunction with simple similarity metrics (e.g., cosine similarity). These models shown in Figure \ref{fig:anatomy4} may learn the embeddings by optimizing directly for the IR task (e.g., \cite{huang2013learning}), or separately in an unsupervised fashion (e.g., \cite{mitra2016desm}). Finally, Figure \ref{fig:anatomy5} shows IR approaches where the neural models are used for query expansion \cite{diaz2016query, roy2016using}.

While the taxonomy of neural approaches described in this section is rather simple, it does provide an intuitive framework for comparing the different neural approaches in IR, and highlights the similarities and distinctions between these different techniques.

\section{Term representations}
\label{sec:wordrep}

\subsection{A tale of two representations}
\label{sec:wordrep-vsm}

\begin{figure*}
\center
\begin{subfigure}{0.49\textwidth}
    \includegraphics[width=\textwidth]{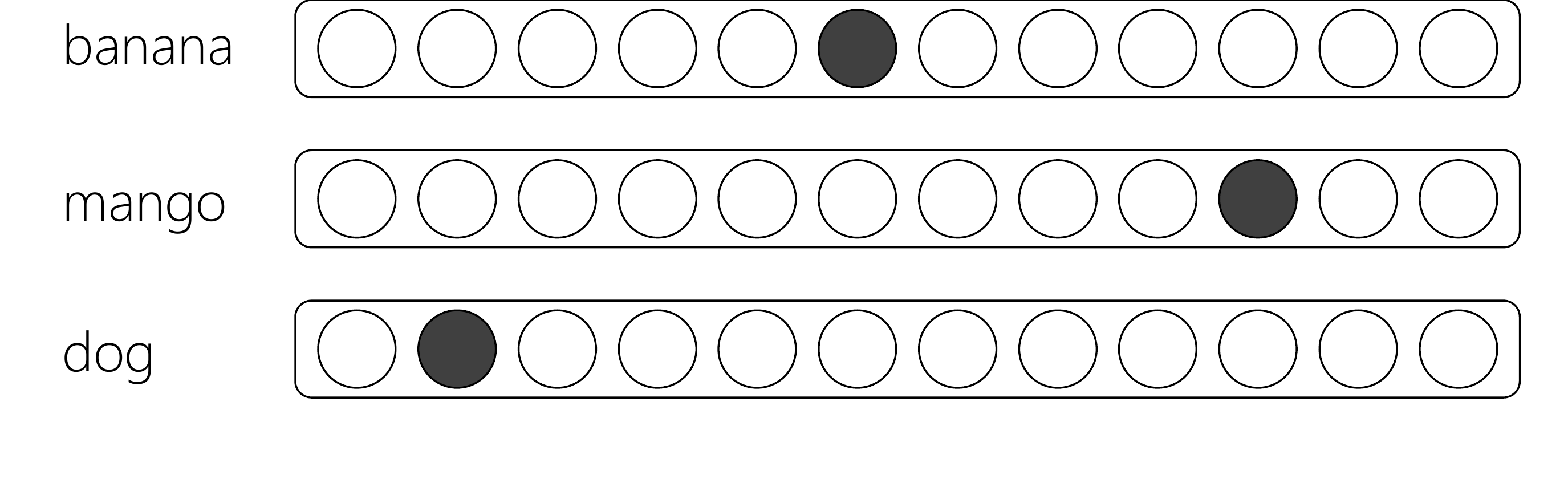}
    \caption{Local representation}
    \label{fig:localrepresentations}
\end{subfigure}
\begin{subfigure}{0.49\textwidth}
    \includegraphics[width=\textwidth]{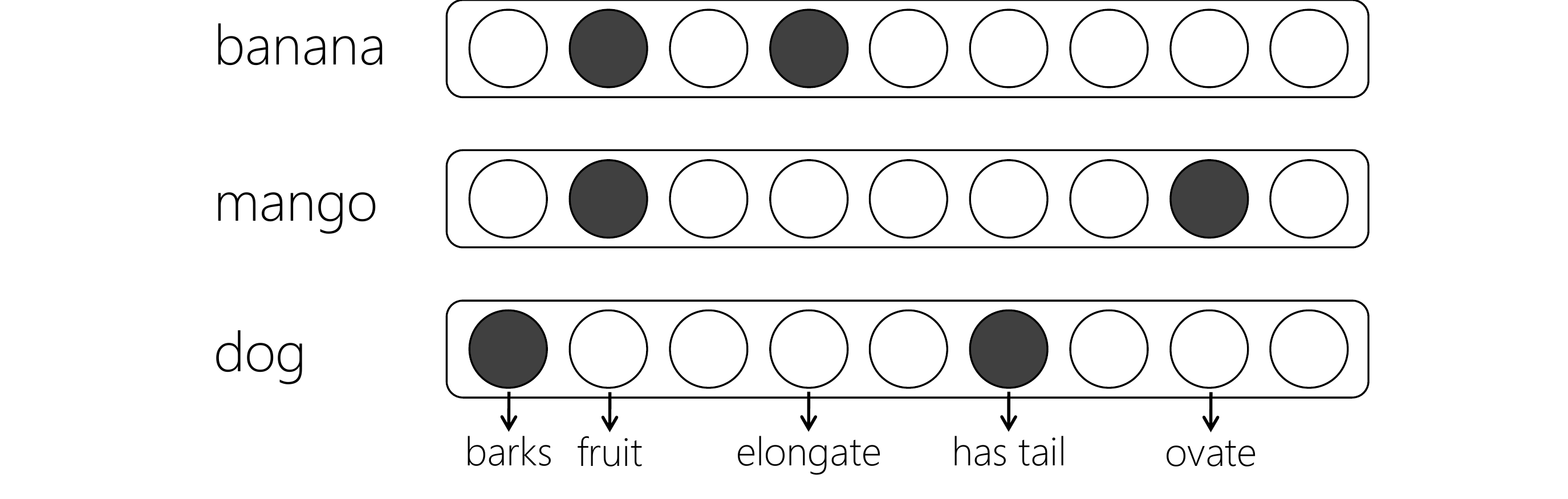}
    \caption{Distributed representation}
    \label{fig:distribrepresentations}
\end{subfigure}
\caption{Under local representations the terms ``banana'', ``mango'', and ``dog'' are distinct items. But distributed vector representations may recognize that ``banana'' and ``mango'' are both fruits, but ``dog'' is different.}
\end{figure*}

Vector representations are fundamental to both information retrieval and machine learning. In IR, terms are typically the smallest unit of representation for indexing and retrieval. Therefore, many IR models---both neural and non-neural---focus on learning good vector representations of terms.

Different vector representations exhibit different levels of generalization---some consider every term as distinct entities while others learn to identify common attributes. Different representation schemes derive different notions of similarity between terms from the definition of the corresponding vector spaces. Some representations operate over fixed-size vocabularies, while the design of others obviate such constraints. They also differ on the properties of compositionality that defines how representations for larger units of information, such as passages and documents, can be derived from individual term vectors. These are some of the important considerations for choosing a term representation suitable for a specific task.

\paragraph*{Local representations} Under local (or \emph{one-hot}) representations, every term in a fixed size vocabulary $T$ is represented by a binary vector $\vec{v} \in \{0, 1\}^{|T|}$,  where only one of the values in the vector is one and all the others are set to zero. Each position in the vector $\vec{v}$ corresponds to a term. The term ``banana'', under this representation, is given by a vector that has the value one in the position corresponding to ``banana'' and zero everywhere else. Similarly, the terms ``mango'' and ``dog'' are represented by setting different positions in the vector to one.

Figure \ref{fig:localrepresentations} highlights that under this scheme each term is a unique entity, and ``banana'' is as distinct from ``dog'' as it is from ``mango''. Terms outside of the vocabulary either have no representation, or are denoted by a special ``UNK'' symbol, under this scheme.

\paragraph*{Distributed representations} Under distributed representations every term is represented by a vector $\vec{v} \in \mathbb{R}^{|k|}$. $\vec{v}$ can be a sparse or a dense vector---a vector of hand-crafted features or a learnt representation in which the individual dimensions are not interpretable in isolation. The key underlying hypothesis for any distributed representation scheme, however, is that by representing a term by its attributes allows for defining some notion of similarity between the different terms based on the chosen properties. For example, in Figure \ref{fig:distribrepresentations} ``banana'' is more similar to ``mango'' than ``dog'' because they are both fruits, but yet different because of other properties that are not shared between the two, such as shape.

\begin{figure*}
\center
\begin{subfigure}{0.49\textwidth}
    \includegraphics[width=\textwidth]{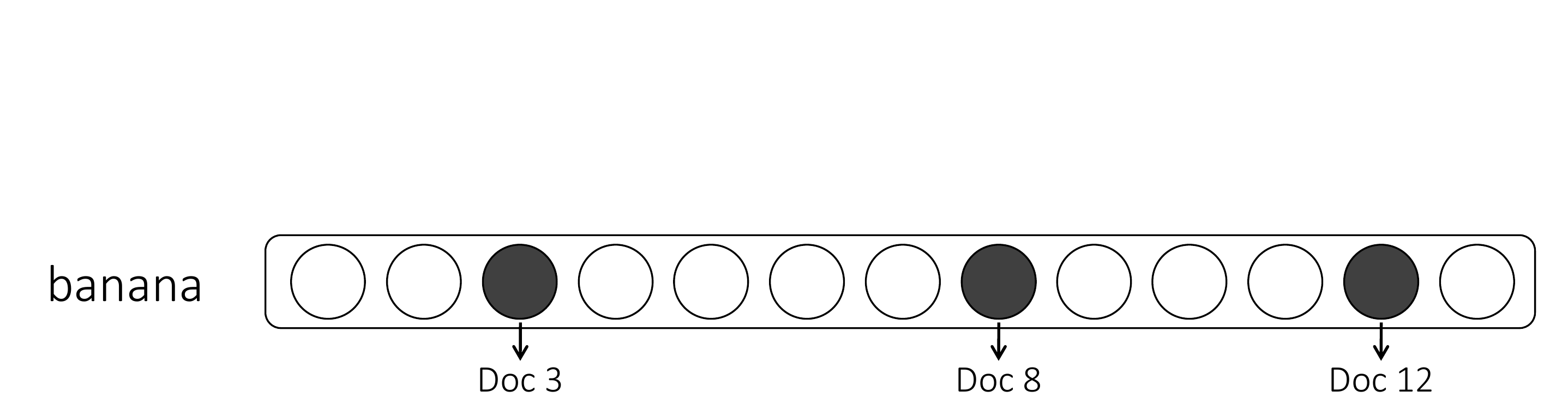}
    \caption{In-document features}
    \label{fig:banana1}
\end{subfigure}
\begin{subfigure}{0.39\textwidth}
    \includegraphics[width=\textwidth]{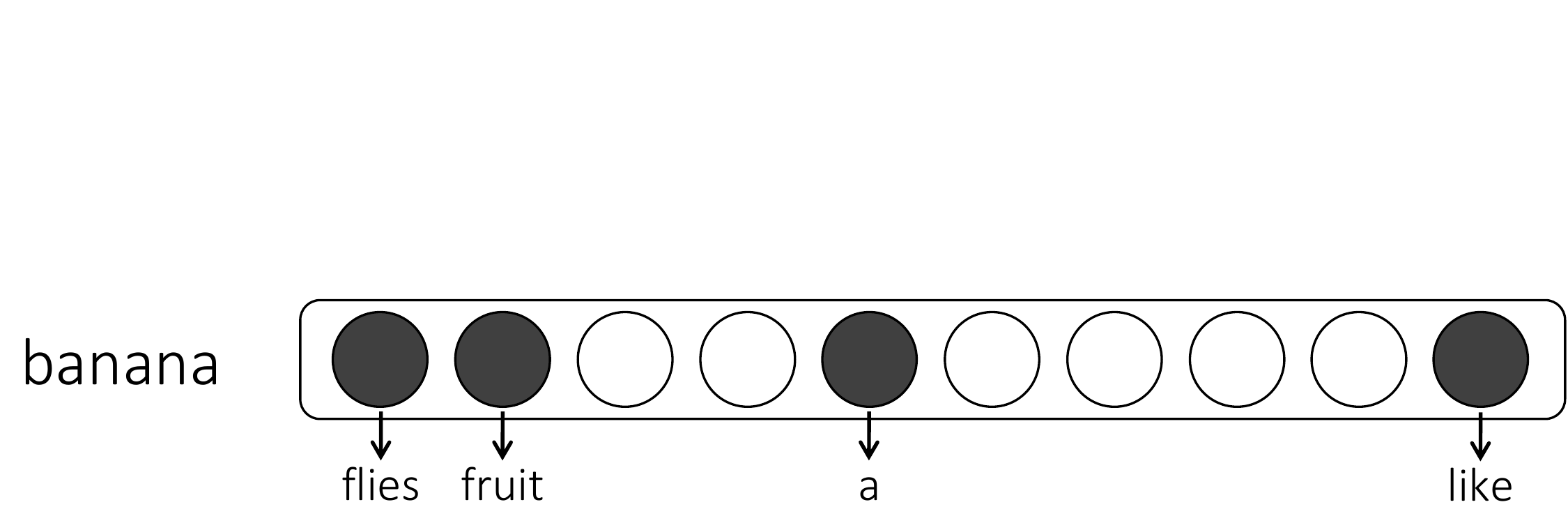}
    \caption{Neighbouring-word features}
    \label{fig:banana2}
\end{subfigure}
\begin{subfigure}{0.60\textwidth}
    \includegraphics[width=\textwidth]{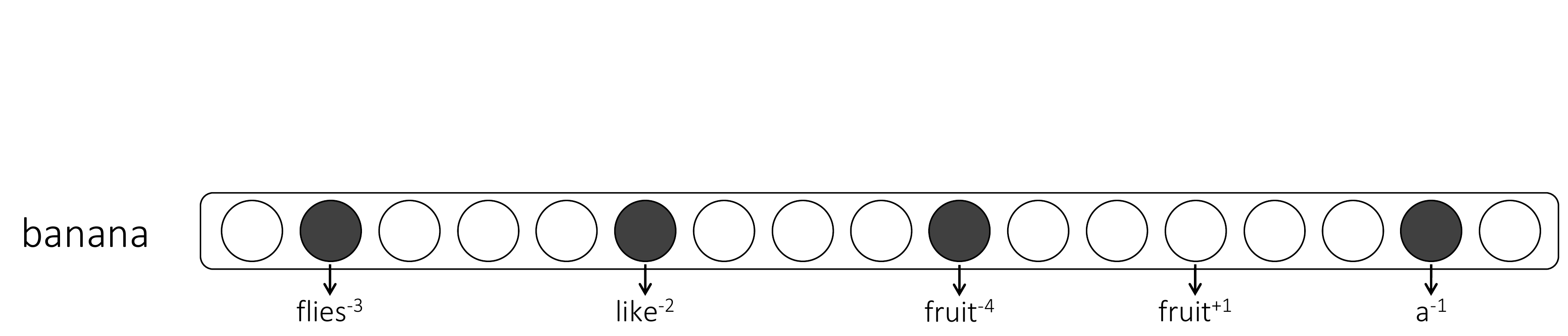}
    \caption{Neighbouring-word w/ distance features}
    \label{fig:banana3}
\end{subfigure}
\begin{subfigure}{0.60\textwidth}
    \includegraphics[width=\textwidth]{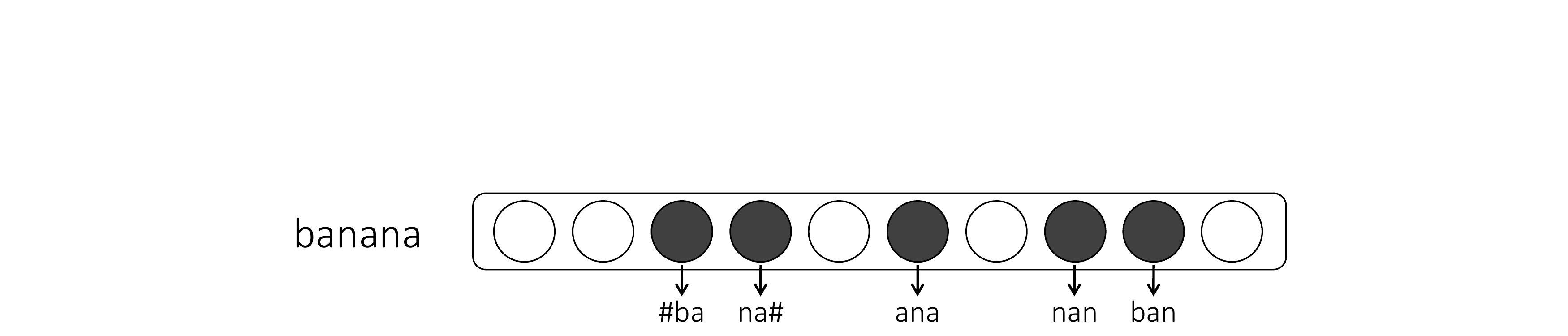}
    \caption{Character-trigraph features}
    \label{fig:banana4}
\end{subfigure}
\caption{Examples of different feature-based distributed representations of the term ``banana''. The representations in (a), (b), and (c) are based on external contexts in which the term frequently occurs, while (d) is based on properties intrinsic to the term. The representation scheme in (a) depends on the documents containing the term, while the scheme shown in (b) and (c) depends on other terms that appears in its neighbourhood. The scheme (b) ignores inter-term distances. Therefore, in the sentence ``Time flies like an arrow; fruit flies like a banana'', the feature ``$\text{fruit}$'' describes both the terms ``banana'' and ``arrow''. However, in the representation scheme of (c) the feature ``$\text{fruit}^{-4}$'' is positive for ``banana'', and the feature ``$\text{fruit}^{+1}$'' for ``arrow''.}
\label{fig:banana}
\end{figure*}

A key consideration in any feature based distributed representation is the choice of the features themselves. A popular approach involves representing terms by features that capture their distributional properties. This is motivated by the \emph{distributional hypothesis} \cite{harris1954distributional} that states that terms that are used (or occur) in similar context tend to be semantically similar. \citet{firth1957synopsis} famously purported this idea of \emph{distributional semantics}\footnote{Readers should take note that while many \emph{distributed} representations take advantage of \emph{distributional} properties, the two concepts are not synonymous. A term can have a distributed representation based on non-distributional features---e.g., parts of speech classification and character trigraphs in the term.} by stating ``\emph{a word is characterized by the company it keeps}''. However, both \emph{distribution} and \emph{semantics} by themselves are not well-defined and under different context may mean very different things. Figure \ref{fig:banana} shows three different sparse vector representations of the term ``banana'' corresponding to different distributional feature spaces---documents containing the term (e.g., LSA \cite{deerwester1990indexing}), neighbouring words in a window (e.g., HAL \cite{lund1996producing}, COALS \cite{rohde2006improved}, and \cite{bullinaria2007extracting}), and neighbouring words with distance (e.g., \cite{levy2014linguistic}). Finally, Figure \ref{fig:banana4} shows a vector representation of ``banana'' based on the character trigraphs in the term itself---instead of external contexts in which the term occurs. In Section \ref{sec:wordrep-sim} we will discuss how choosing different distributional features for term representation leads to different nuanced notions of semantic similarity between them.

When the vectors are high-dimensional, sparse, and based on distributional feature they are referred to as \emph{explicit} vector representations \cite{levy2014linguistic}. On the other hand, when the vectors are dense, small ($k \ll |T|$), and learnt from data then they are commonly referred to as \emph{embeddings}. For both explicit and embedding based representations several distance metrics can be used to define similarity between terms, although cosine similarity is commonly used.

\begin{align}
sim(\vec{v}_i, \vec{v}_j) = cos(\vec{v}_i, \vec{v}_j) = \frac{\vec{v}_i^{\;\intercal} \vec{v}_j}{\norm{\vec{v}_i}\norm{\vec{v}_j}}
\end{align}

Most embeddings are learnt from explicit vector space representations, and hence the discussions in \ref{sec:wordrep-sim} about different notions of similarity are also relevant to the embedding models. In Section \ref{sec:wordrep-explicit} and \ref{sec:wordrep-embeddings} we briefly discuss explicit and embedding based representations.

\begin{figure*}
\center
\includegraphics[width=0.55\textwidth]{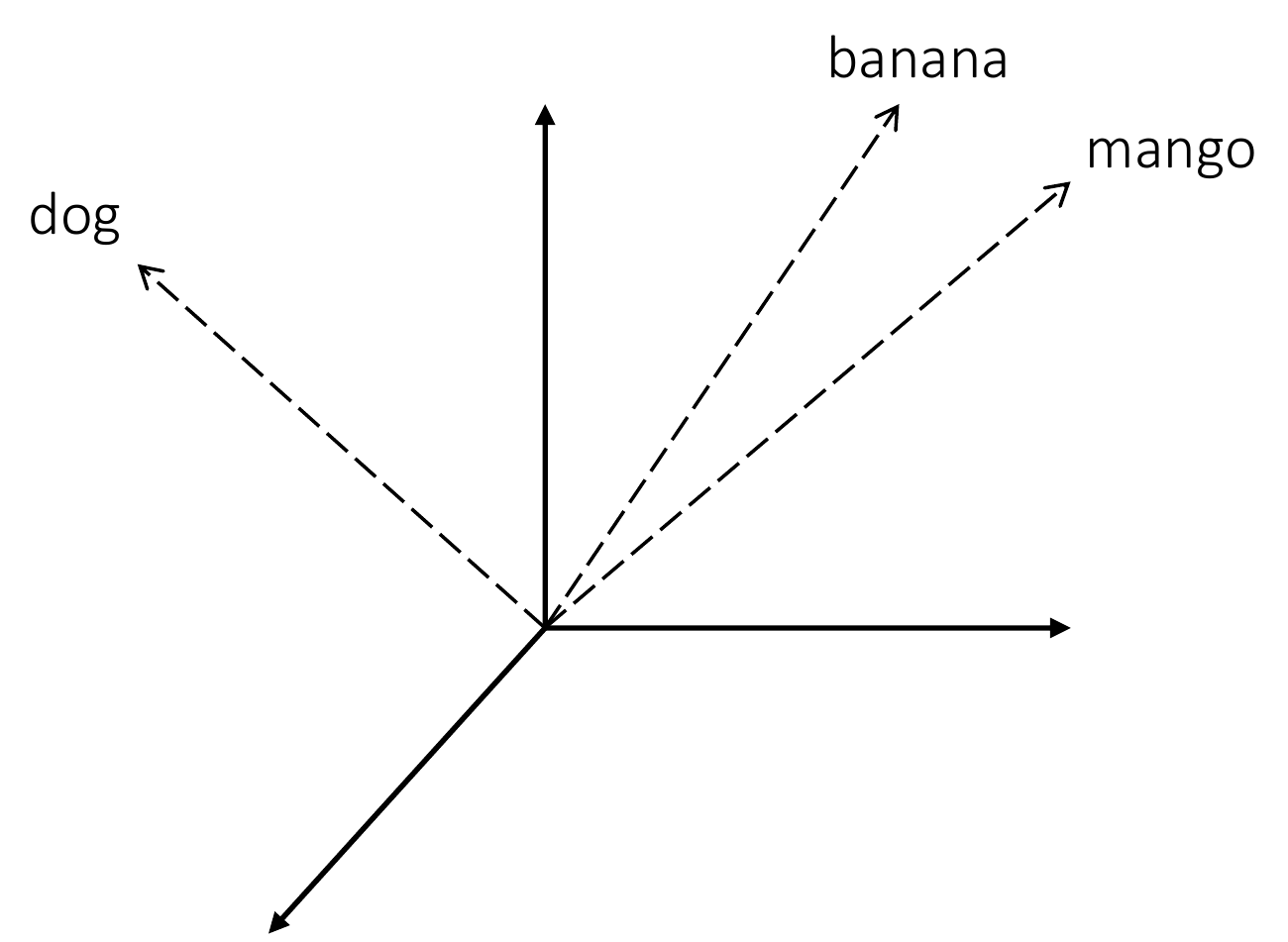}
\caption{A vector space representation of terms puts ``banana'' closer to ``mango'' because they share more common attributes than ``banana'' and ``dog''.}
\label{fig:embeddings}
\end{figure*}

With respect to \emph{compositionality}, it is important to understand that \emph{distributed} representations of items are often derived from local or distributed representation of its parts. For example, a document can be represented by the sum of the one-hot vectors or embeddings corresponding to the terms in the document. The resultant vector, in both cases, corresponds to a distributed bag-of-word representation. Similarly, the character trigraph representation of terms in Figure \ref{fig:banana4} is simply an aggregation over the one-hot representations of the constituent trigraphs.

In the context of neural models, distributed representations generally refer to learnt embeddings. The idea of `local' and `distributed' representations has a specific significance in the context of neural network models. Each concept, entity, or term can be represented within a neural network by the activation of a single neuron (local representation) or by the combined pattern of activations of several neurons (distributed representation) \cite{hinton1984distributed}.

\subsection{Notions of similarity}
\label{sec:wordrep-sim}

Any vector representation inherently defines some notion of relatedness between terms. Is ``Seattle'' closer to ``Sydney'' or to ``Seahawks''? The answer depends on the type of relationship we are interested in. If we want terms of similar \emph{type} to be closer, then ``Sydney'' is more similar to ``Seattle'' because they are both cities. However, if we are interested to find terms that co-occur in the same document or passage, then ``Seahawks''---Seattle's football team---should be closer. The former represents a \emph{Typical}, or type-based notion of similarity while the latter exhibits a more \emph{Topical} sense of relatedness.

\begin{table}
\renewcommand{\arraystretch}{1.1}\addtolength{\tabcolsep}{2.5pt}
\begin{center}
\caption{A toy corpus of short documents that we consider for the discussion on different notions of similarity between terms under different distributed representations. The choice of the feature space that is used for generating the distributed representation determines which terms are closer in the vector space, as shown in Figure \ref{fig:notionsofsimilarity}.}
\label{tbl:wordrep-tinycorpus}
\resizebox{0.9\textwidth}{!}{
\begin{tabular}{rlccrl}
\toprule
\multicolumn{6}{c}{Sample documents} \\
\midrule
doc 01 & Seattle map & & & doc 09 & Denver map \\
doc 02 & Seattle weather & & & doc 10 & Denver weather \\
doc 03 & Seahawks jerseys & & & doc 11 & Broncos jerseys \\
doc 04 & Seahawks highlights & & & doc 12 & Broncos highlights \\
doc 05 & Seattle Seahawks Wilson & & & doc 13 & Denver Broncos Lynch \\
doc 06 & Seattle Seahawks Sherman & & & doc 14 & Denver Broncos Sanchez \\
doc 07 & Seattle Seahawks Browner & & & doc 15 & Denver Broncos Miller \\
doc 08 & Seattle Seahawks Ifedi & & & doc 16 & Denver Broncos Marshall \\
\bottomrule
\end{tabular}
}
\end{center}
\end{table}

\begin{figure*}
\center
\begin{subfigure}{0.99\textwidth}
    \includegraphics[width=\textwidth]{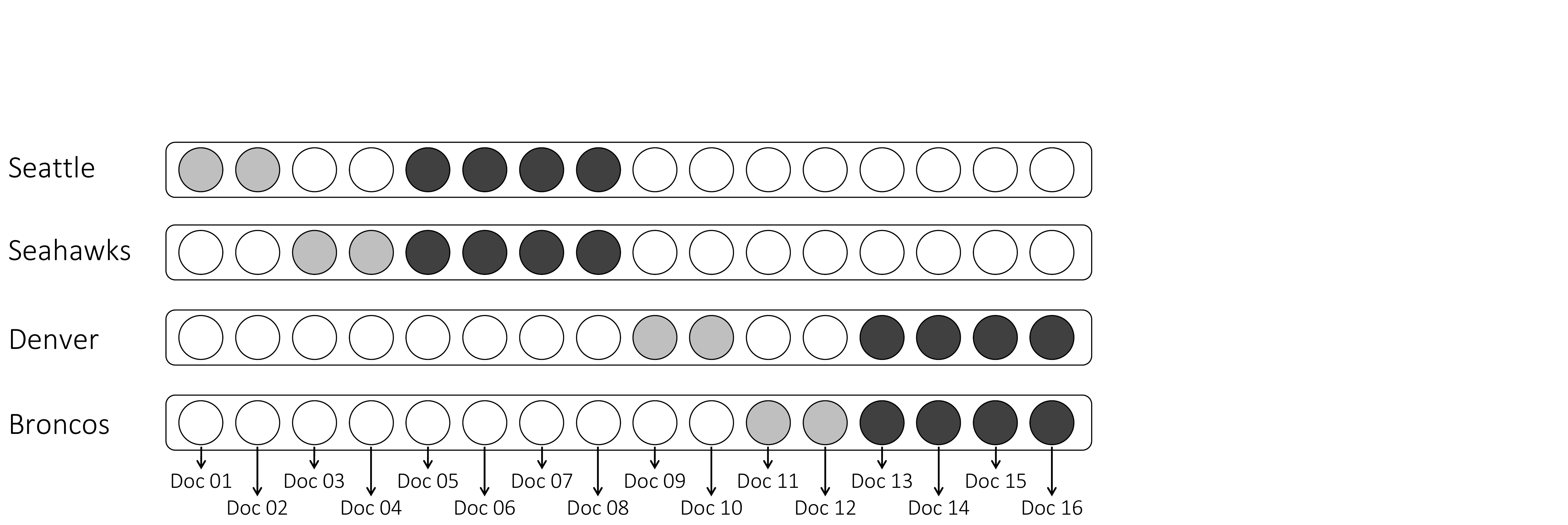}
    \caption{``In-documents'' features}
    \label{fig:notionsofsimilarity1}
\end{subfigure}
\begin{subfigure}{0.99\textwidth}
    \includegraphics[width=\textwidth]{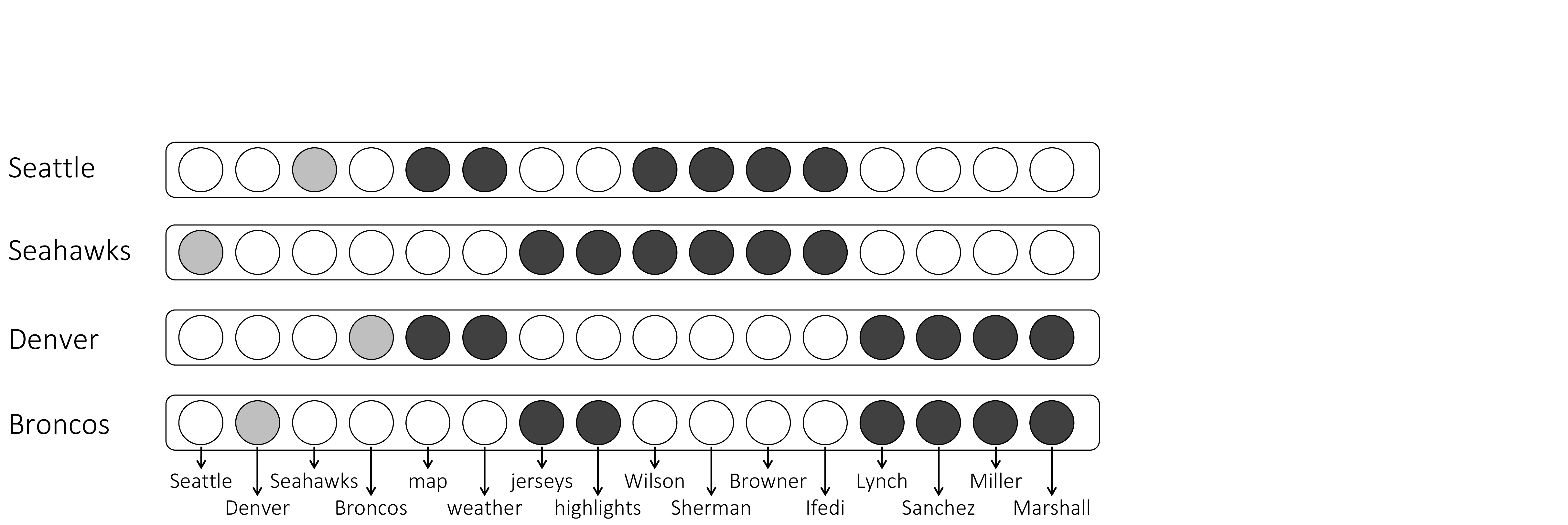}
    \caption{``Neighbouring terms'' features}
    \label{fig:notionsofsimilarity2}
\end{subfigure}
\begin{subfigure}{0.99\textwidth}
    \includegraphics[width=\textwidth]{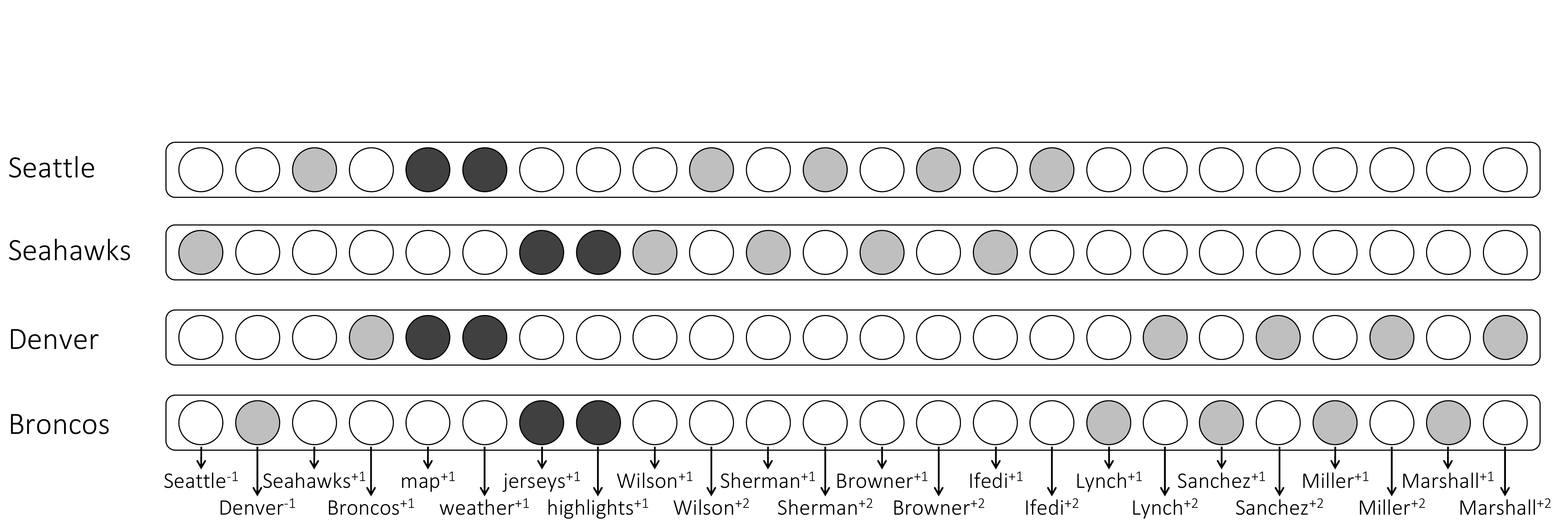}
    \caption{``Neighbouring terms w/ distances'' features}
    \label{fig:notionsofsimilarity3}
\end{subfigure}
\caption{The figure shows different distributed representations for the four terms---''Seattle'', ``Seahawks'', ``Denver'', and ``Broncos''---based on the toy corpus in Table \ref{tbl:wordrep-tinycorpus}. Shaded circles indicate non-zero values in the vectors---the darker shade highlights the vector dimensions where more than one vector has a non-zero value. When the representation is based on the documents that the terms occur in then ``Seattle'' is more similar to ``Seahawks'' than to ``Denver''. The representation scheme in (a) is, therefore, more aligned with a Topical notion of similarity. In contrast, in (c) each term is represented by a vector of neighbouring terms---where the distances between the terms are taken into consideration---which puts ``Seattle'' closer to ``Denver'' demonstrating a Typical, or type-based, similarity. When the inter-term distances are ignored, as in (b), a mix of Typical and Topical similarities is observed. Finally, it is worth noting that neighbouring-terms based vector representations leads to similarities between terms that do not necessarily occur in the same document, and hence the term-term relationships are less sparse than when only in-document features are considered.}
\label{fig:notionsofsimilarity}
\end{figure*}

If we want to compare ``Seattle'' with ``Sydeny'' and ``Seahawks based on their respective vector representations, then the underlying feature space needs to align with the notion of similarity that we are interested in. It is, therefore, important for the readers to build an intuition about the choice of features and the notion of similarity they encompass. This can be demonstrated by using a toy corpus, such as the one in Table \ref{tbl:wordrep-tinycorpus}. Figure \ref{fig:notionsofsimilarity1} shows that the ``in documents'' features naturally lend to a Topical sense of similarity between the terms, while the ``neighbouring terms with distances'' features in Figure \ref{fig:notionsofsimilarity3} gives rise to a more Typical notion of relatedness. Using ``neighbouring terms'' without the inter-term distances as features, however, produces a mixture of Topical and Typical relationships. This is because when the term distances are considered in feature definition then the document ``Seattle Seahawks Wilson'' produces the bag-of-features $\{Seahawks^{+1}, Wilson^{+2}\}$ for ``Seattle'' which is non-overlapping with the bag-of-features $\{Seattle^{-1}, Wilson^{+1}\}$ for ``Seahawks''. However, when the feature definition ignores the term-distances then there is a partial overlap between the bag-of-features $\{Seahawks, Wilson\}$ and $\{Seattle, Wilson\}$ corresponding to ``Seattle'' and ``Seahawks''. The overlap increases significantly when we use a larger window-size for identifying neighbouring terms pushing the notion of similarity closer to a Topical definition. This effect of the windows size on the Topicality of the representation space was reported by \citet{levy2014dependencybased} in the context of learnt embeddings.

Readers should take note that the set of all inter-term relationships goes far beyond the two notions of Typical and Topical that we discuss in this section. For example, vector representations could cluster terms closer based on linguistic styles---e.g., terms that appear in thriller novels versus in children's rhymes, or in British versus American English. However, the notions of Typical and Topical similarities popularly come up in discussions in the context of many IR and NLP tasks---sometimes under different names such as \emph{Paradigmatic} and \emph{Syntagmatic} relations\footnote{Interestingly, the notion of Paradigmatic (Typical) and Syntagmatic (Topical) relationships show up almost universally---not just in text. In vision, for example, the different images of ``noses'' bear a Typical similarity to each other, while they share a Topical relationship with images of ``eyes'' or ``ears''. Curiously, \citet{barthes1977elements} even extended this analogy to garments---where paradigmatic relationships exist between items of the same type (e.g., between hats and between boots) and the proper Syntagmatic juxtaposition of items from these different Paradigms---from hats to boots--- forms a fashionable ensemble .}---and the idea itself goes back at least as far as Saussure \cite{harris2001saussure, de2011course, chandler1994semiotics, sahlgren2006word}.

\subsection{Explicit vector representations}
\label{sec:wordrep-explicit}
Explicit vector representations can be broadly categorized based on their choice of distributional features (e.g., in documents, neighbouring terms with or without distances, etc.) and different weighting schemes (e.g., TF-IDF, positive pointwise mutual information, etc.) applied over the raw counts. We direct the readers to \cite{turney2010frequency, baroni2010distributional} which are good surveys of many existing explicit vector representation schemes.

\citet{levy2014linguistic} demonstrated that explicit vector representations are amenable to the term analogy task using simple vector operations. A term analogy task involves answering questions of the form ``\emph{man} is to \emph{woman} as \emph{king} is to \_\_\_\_?''---the correct answer to which in this case happens to be ``queen''. In NLP, term analogies are typically performed by simple vector operations of the following form followed by a nearest-neighbour search,

\begin{align}
\vec{v}_{king} - \vec{v}_{man} + \vec{v}_{woman} \approx \vec{v}_{queen}
\end{align}

\begin{figure*}
\center
\includegraphics[width=0.99\textwidth]{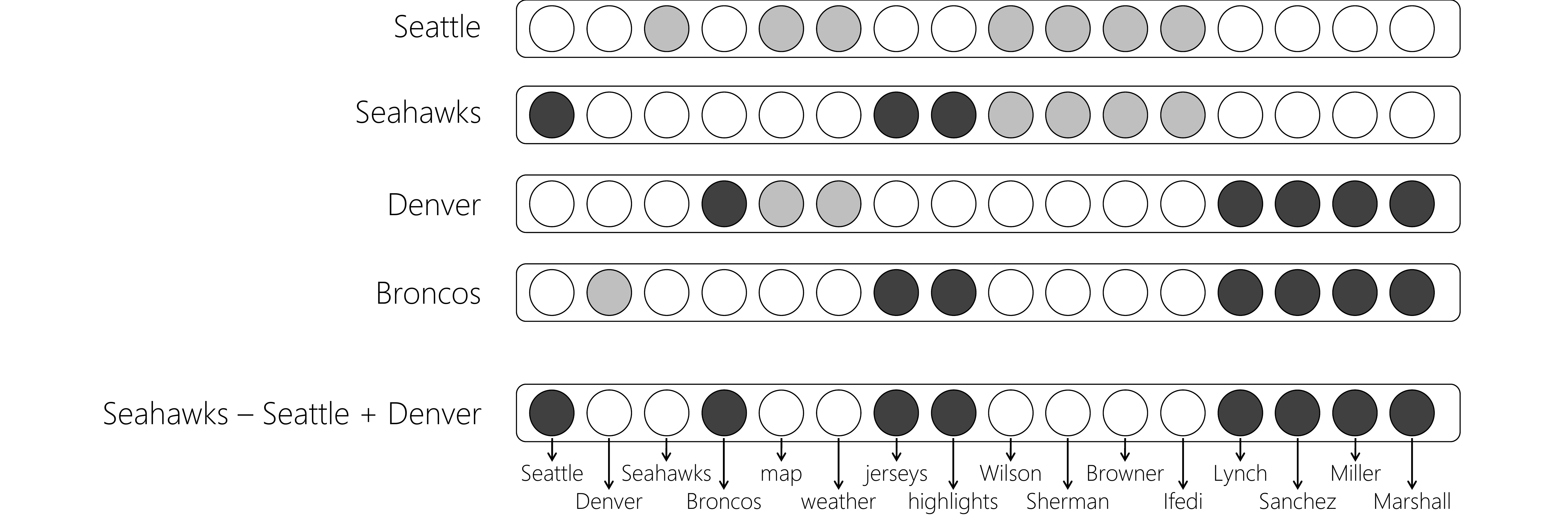}
\caption{A visual demonstration of term analogies via simple vector algebra. The shaded circles denote non-zero values. Darker shade is used to highlight the non-zero values along the vector dimensions for which the output of $\vec{v}_{Seahawks} - \vec{v}_{Seattle} + \vec{v}_{Denver}$ is positive. The output vector is closest to $\vec{v}_{Broncos}$ as shown in this toy example.}
\label{fig:analogy}
\end{figure*}

It may be surprising to some readers that the vector obtained by the simple algebraic operations $\vec{v}_{king} - \vec{v}_{man} + \vec{v}_{woman}$ produces a vector close to the vector $\vec{v}_{queen}$. We present a visual intuition of why this works in practice in Figure \ref{fig:analogy}, but we refer the readers to \cite{levy2014linguistic, arora2015rand} for a more rigorous mathematical explanation.

\subsection{Embeddings}
\label{sec:wordrep-embeddings}

While explicit vector representations based on distributional features can capture interesting notions of term-term similarity they have one big drawback---the resultant vector spaces are highly sparse and high-dimensional. The number of dimensions is generally in the same order as the number of documents or the vocabulary size, which is unwieldy for most practical tasks. An alternative is to learn lower dimensional representations of terms from the data that retains similar attributes as the higher dimensional vectors.

An \emph{embedding} is a representation of items in a new space such that the properties of, and the relationships between, the items are preserved. \citet{goodfellow2016deep} articulate that the goal of an embedding is to generate a \emph{simpler} representation---where simplification may mean a reduction in the number of dimensions, an increase in the sparseness of the representation, disentangling the principle components of the vector space, or a combination of these goals. In the context of term embeddings, the explicit feature vectors---like those we discussed in Section \ref{sec:wordrep-explicit}---constitutes the original representation. An embedding trained from these features assimilate the properties of the terms and the inter-term relationships observable in the original feature space.

The most popular approaches for learning embeddings include either factorizing the term-feature matrix (e.g. LSA \cite{deerwester1990indexing}) or using gradient descent based methods that try to predict the features given the term (e.g., \cite{bengio2003neural, mikolov2013efficient}). \citet{baroni2014don} empirically demonstrate that these feature-predicting models that learn lower dimensional representations, in fact, also perform better than explicit counting based models on different tasks---possibly due to better generalization across terms---although some counter evidence the claim of better performances from embedding models have also been reported in the literature \cite{levy2015improving}.

The sparse feature spaces of Section \ref{sec:wordrep-explicit} are easier to visualize and leads to more intuitive explanations---while their corresponding embeddings are more practically useful. Therefore, it makes sense to \emph{think sparse, but act dense} in many scenarios. In the rest of this section, we will describe some of the popular neural and non-neural embedding models.

\paragraph*{Latent Semantic Analysis (LSA)}
LSA \cite{deerwester1990indexing} involves performing \emph{singular value decomposition} (SVD) \cite{golub1970singular} on a term-document (or term-passage) matrix $X$ to obtain its low-rank approximation \cite{markovsky2011low}. SVD on $X$ involves finding a solution to $\begin{matrix}X=U\Sigma V^{T}\end{matrix}$, where $U$ and $V$ are orthogonal matrices and $\Sigma$ is a diagonal matrix.\footnote{The matrix visualization is adapted from \url{https://en.wikipedia.org/wiki/Latent_semantic_analysis}.}

\begin{align}
\begin{smallmatrix}&X&&&U&&\Sigma &&V^\intercal\\&(\vec{\textbf{d}}_j)&&&&&&&({\vec {\textbf{d}}}_j)\\&\downarrow &&&&&&&\downarrow \\({\vec{\textbf{t}}}_i^{\mkern4mu\intercal})\rightarrow &{\begin{bmatrix}x_{1,1}&\dots &x_{1,|D|}\\\\\vdots &\ddots &\vdots \\\\x_{|T|,1}&\dots &x_{|T|,|D|}\\\end{bmatrix}}&=&({\vec {\textbf{t}}}_i^{\mkern4mu\intercal})\rightarrow &{\begin{bmatrix}{\begin{bmatrix}\,\\\,\\{\vec{\textbf{u}}}_1\\\,\\\,\end{bmatrix}}\dots {\begin{bmatrix}\,\\\,\\{\vec{\textbf {u}}}_l\\\,\\\,\end{bmatrix}}\end{bmatrix}}&\cdot &{\begin{bmatrix}\sigma_1&\dots &0\\\vdots &\ddots &\vdots \\0&\dots &\sigma_l\\\end{bmatrix}}&\cdot &{\begin{bmatrix}{\begin{bmatrix}&&{\vec{\textbf{v}}}_1&&\end{bmatrix}}\\\vdots \\{\begin{bmatrix}&&{\vec{\textbf {v}}}_l&&\end{bmatrix}}\end{bmatrix}}\end{smallmatrix}
\end{align}

where, $\sigma _{1},\dots ,\sigma _{l}$, $\vec{\textbf{u}}_{1},\dots ,\vec{\textbf{u}}_{l}$, and $\vec{\textbf{v}}_{1},\dots ,\vec{\textbf{v}}_{l}$ are the singular values, the left singular vectors, and the right singular vectors, respectively. The $k$ largest singular values, and corresponding singular vectors from $U$ and $V$, is the rank $k$ approximation of $X$ ($X_k=U_k\Sigma_k V_k^T$). The embedding for the $i_{th}$ term is given by $\Sigma_k\vec{\textbf{t}}_i$.

While LSA operate on a term-document matrix, matrix factorization based approaches can also be applied to term-term matrices \cite{rohde2006improved, bullinaria2012extracting, lebret2013word}.

\bigskip
\emph{Neural term embedding} models are typically trained by setting up a prediction task. Instead of factorizing the term-feature matrix---as in LSA---neural models are trained to predict the term from its features. Both the term and the features have \emph{one-hot} representations in the input and the output layers, respectively, and the model learns dense low-dimensional representations in the process of minimizing the prediction error. These approaches are based on the \emph{information bottleneck method} \cite{tishby2000information}---discussed in more details in Section \ref{sec:dnn-arch}---with the low-dimensional representations acting as the bottleneck. The training data may contain many instances of the same term-feature pair proportional to their frequency in the corpus (e.g., word2vec \cite{mikolov2013efficient}), or their counts can be pre-aggregated (e.g., GloVe \cite{pennington2014glove}).

\paragraph*{Word2vec}

\begin{figure}
\center
\begin{subfigure}{0.70\textwidth}
    \includegraphics[width=\textwidth]{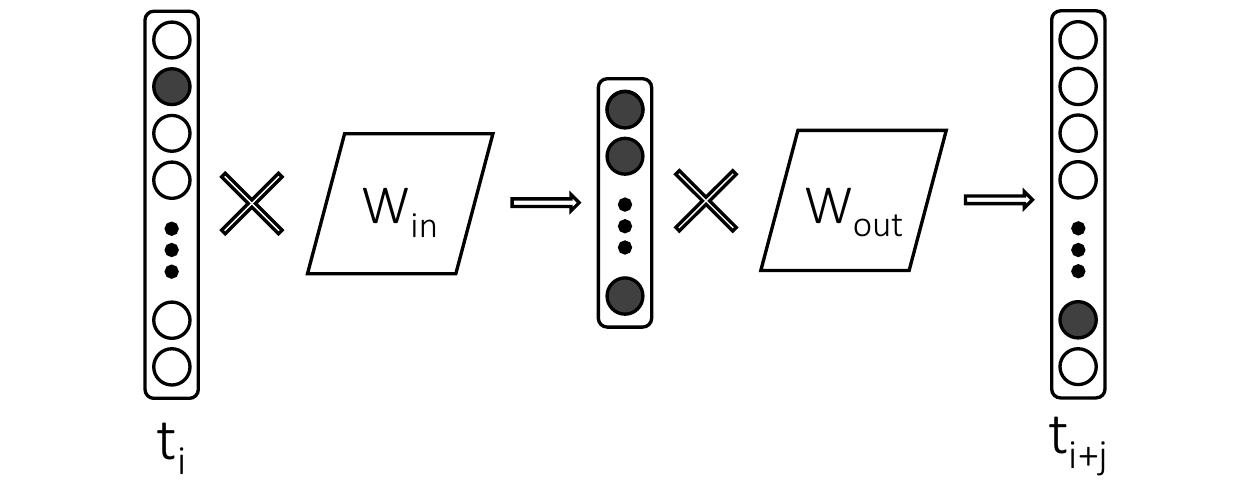}
    \caption{Skip-gram}
    \label{fig:word2vec1}
\end{subfigure}
\hfill
\begin{subfigure}{0.70\textwidth}
    \includegraphics[width=\textwidth]{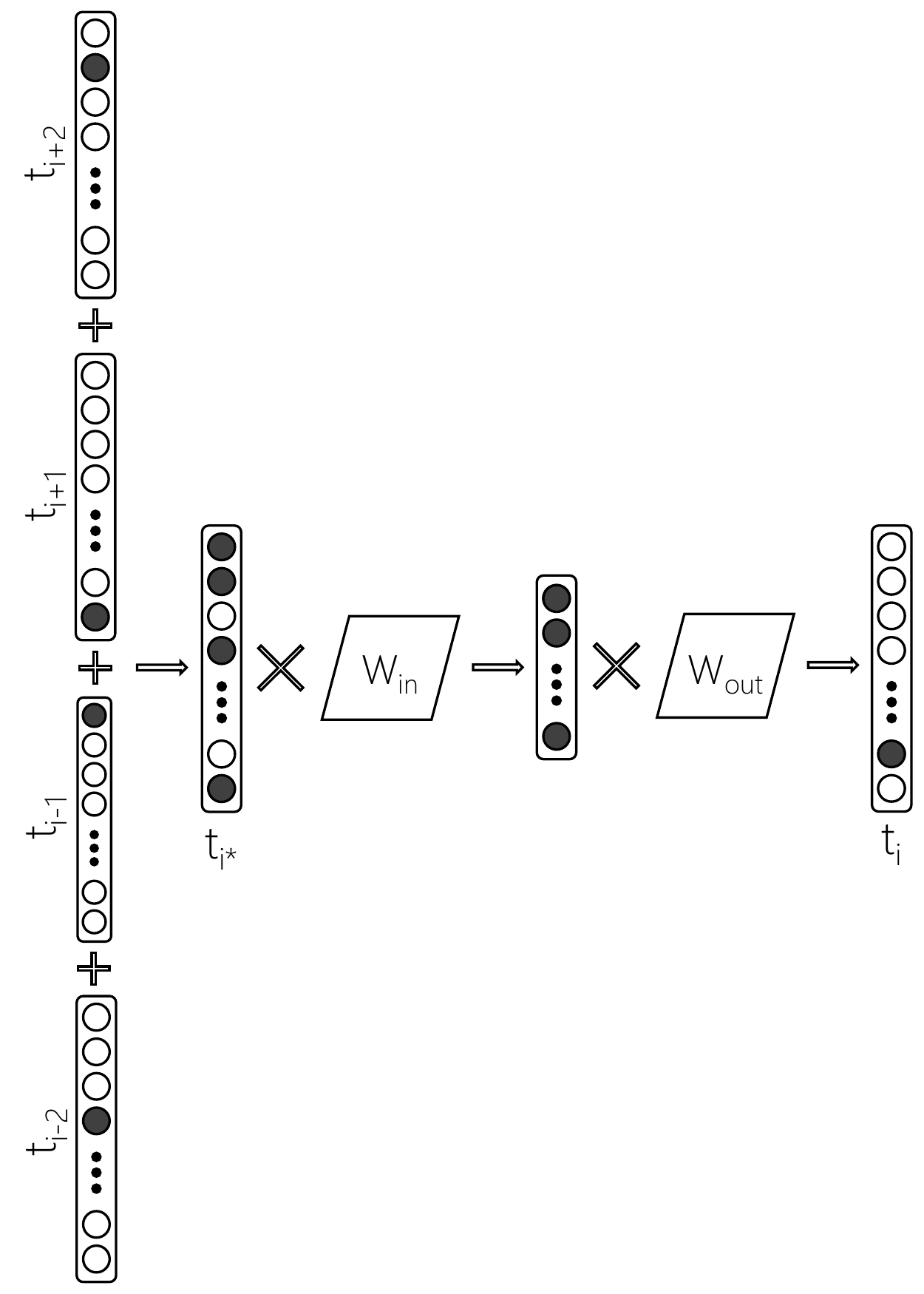}
    \caption{Continuous bag-of-words (CBOW)}
    \label{fig:word2vec2}
\end{subfigure}
\caption{The (a) skip-gram and the (b) continuous bag-of-words (CBOW) architectures of word2vec. The architecture is a neural network with a single hidden layer whose size is much smaller than that of the input and the output layers. Both models use one-hot representations of terms in the input and the output. The learnable parameters of the model comprise of the two weight matrices $W_{in}$ and $W_{out}$ that corresponds to the embeddings the model learns for the input and the output terms, respectively. The skip-gram model trains by minimizing the error in predicting a term given one of its neighbours. The CBOW model, in contrast, predicts a term from a bag of its neighbouring terms.}
\label{fig:word2vec}
\end{figure}

For word2vec \cite{mikolov2013efficient, mikolov2013distributed, mikolov2013linguistic, goldberg2014word2vec, rong2014word2vec}, the features for a term are made up of its neighbours within a fixed size window over the text from the training corpus. The \emph{skip-gram} architecture (see Figure \ref{fig:word2vec1}) is a simple one hidden layer neural network. Both the input and the output of the model is in the form of one-hot vectors and the loss function is as follows,

\begin{align}
\label{eqn:wordrep-skipgram}
\mathcal{L}_{skip-gram} &= -\frac{1}{|S|} \sum_{i=1}^{|S|}{\sum_{-c \leq j \leq +c, j \neq 0}{log(p(t_{i+j}|t_i))}} \\
\text{where,}\qquad p(t_{i+j}|t_i) &= \frac{\exp{((W_{out}\vec{v}_{t_{i+j}})^\intercal (W_{in}\vec{v}_{t_{i}}))}}{\sum_{k=1}^{|T|}{\exp{((W_{out}\vec{v}_{t_{k}}})^\intercal (W_{in}\vec{v}_{t_{i}}))}}
\end{align}

S is the set of all windows over the training text and c is the number of neighbours we need to predict on either side of the term $t_i$. The denominator for the softmax function for computing $p(t_{i+j}|t_i)$ sums over all the words in the vocabulary. This is prohibitively costly and in practice either hierarchical-softmax \cite{morin2005hierarchical} or negative sampling is employed. Also, note that the model has two different weight matrices $W_{in}$ and $W_{out}$ that are learnable parameters of the models. $W_{in}$ gives us the IN embeddings corresponding to all the input terms and $W_{out}$ corresponding to the OUT embeddings for the output terms. Generally, only $W_{in}$ is used and $W_{out}$ is discarded after training, but we will discuss an IR application that makes use of both the IN and the OUT embeddings later in Section \ref{sec:embir-allpairs}.

The \emph{continuous bag-of-words} (CBOW) architecture (see Figure \ref{fig:word2vec2}) is similar to the skip-gram model, except that the task is to predict the middle term given the sum of the one-hot vectors of the neighbouring terms in the window. Given a middle term $t_i$ and the set of its neigbours $\{t_{i-c}, \dots, t_{i-1}, t_{i+1}, \dots,t_{i+c} \}$, the CBOW model creates a single training sample with the sum of the one-hot vectors of all the neighbouring terms as input and the one-hot vector $\vec{v}_{t_i}$, corresponding to the middle term, as the expected output.

\begin{align}
\label{eqn:wordrep-cbow}
\mathcal{L}_{CBOW} &= -\frac{1}{|S|} \sum_{i=1}^{|S|}{log(p(t_i|\sum_{-c \leq j \leq +c, j \neq 0}{t_{i+j}}))}
\end{align}

Contrast this with the skip-gram model that creates $2 \times c$ samples by individually pairing each of the neighbouring terms with the middle term. During training, given the same number of windows of text, the skip-gram model, therefore, trains orders of magnitude slower than the CBOW model \cite{mikolov2013efficient} because it creates $2 \times c$ the number of training samples.

Word2vec gained particular popularity for its ability to perform word analogies using simple vector algebra, similar to what we have already discussed in Section \ref{sec:wordrep-explicit}. For domains where the interpretability of the embeddings may be important, \citet{sun2016sparse} introduced an additional constraint in the loss function to encourage more sparseness in the learnt representation.

\begin{align}
\label{eqn:wordrep-sparsecbow}
\mathcal{L}_{sparse-CBOW} &= \mathcal{L}_{sparse-CBOW} - \lambda \sum_{t \in T}{\norm{\vec{v}_t}_1}
\end{align}

\paragraph*{GloVe}
The skip-gram model trains on individual term-neighbour pairs. If we aggregate all the training samples into a matrix $X$, such that $x_{ij}$ is the frequency of the pair $\langle t_i, t_j\rangle$ in the training data, then the loss function changes to,

\begin{align}
\label{eqn:wordrep-w2v2glove}
\mathcal{L}_{skip-gram} &= -\sum_{i=1}^{|T|}{\sum_{j=1}^{|T|}{x_{ij}log(p(t_i|t_j))}} \\
                                    &= -\sum_{i=1}^{|T|}{x_i\sum_{j=1}^{|T|}{\frac{x_{ij}}{x_i}log(p(t_i|t_j))}} \\
                                    &= -\sum_{i=1}^{|T|}{x_i\sum_{j=1}^{|T|}{\bar{p}(t_i|t_j) log(p(t_i|t_j))}} \\
                                    &= \sum_{i=1}^{|T|}{x_iH(\bar{p}(t_i|t_j) log(p(t_i|t_j)))}
\end{align}

$H(\dots)$ is the cross-entropy error between the actual co-occurrence probability $\bar{p}(t_i|t_j)$ and the one predicted by the model $p(t_i|t_j)$. This is similar to the loss function for GloVe \cite{pennington2014glove} if we replace the cross-entropy error with a squared-error and apply a saturation function $f(\dots)$ over the actual co-occurrence frequencies.

\begin{align}
\label{eqn:wordrep-glove}
\mathcal{L}_{GloVe} &= -\sum_{i=1}^{|T|}{\sum_{j=1}^{|T|}{f(x_{ij})(log(x_{ij}-\vec{v}_{w_i}^{\;\intercal} \vec{v}_{w_j}))^2}}
\end{align}

GloVe is trained using AdaGrad \cite{duchi2011adaptive}. Similar to word2vec, GloVe also generates two different (IN and OUT) embeddings, but unlike word2vec it generally uses the sum of the IN and the OUT vectors as the embedding for each term in the vocabulary.

\paragraph*{Paragraph2vec}

\begin{figure*}
\center
\includegraphics[width=0.96\textwidth]{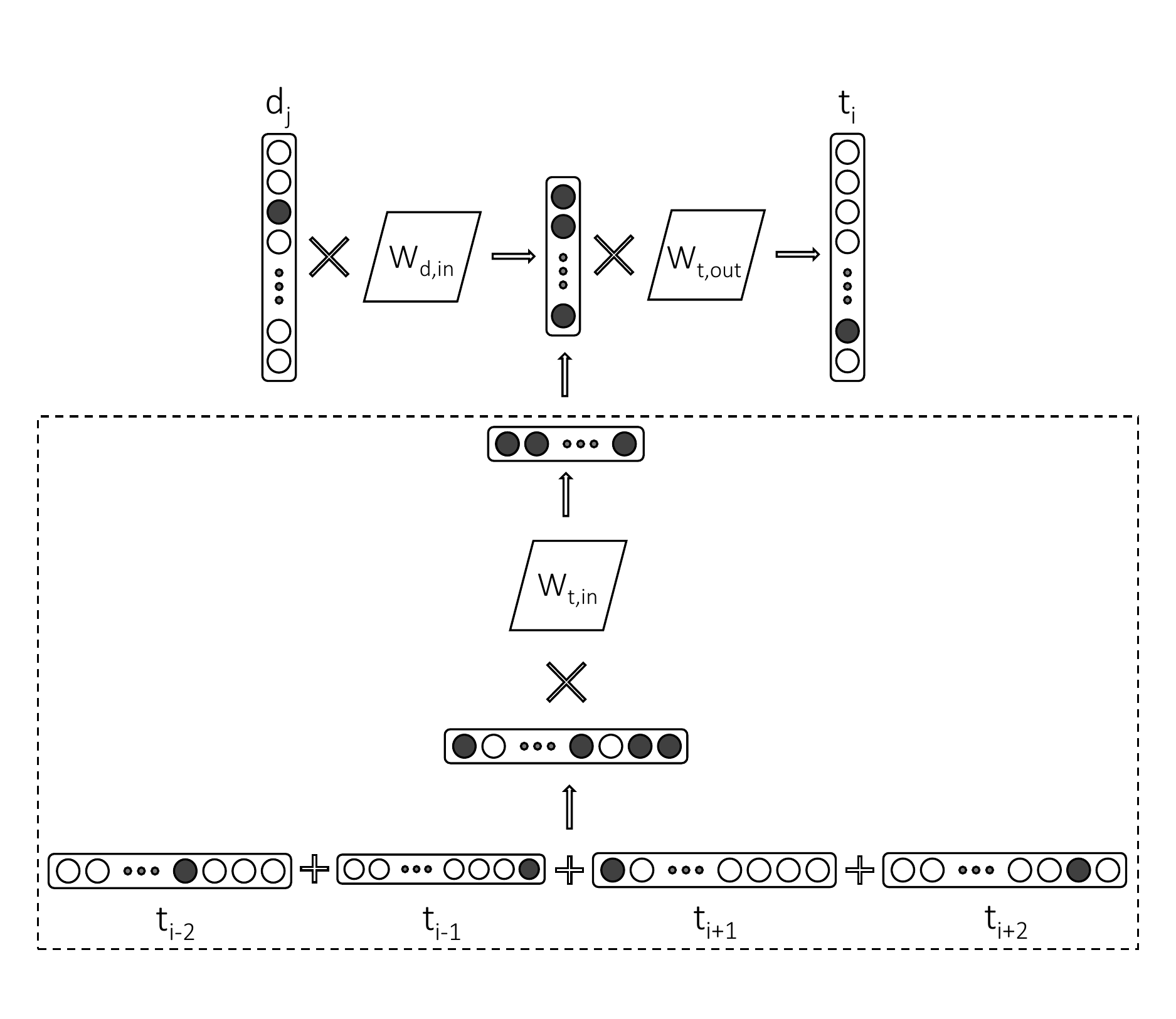}
\caption{The paragraph2vec architecture as proposed by \citet{le2014distributed} trains by predicting a term given a document (or passage) ID containing the term. By trying to minimize the prediction error, the model learns an embedding for the term as well as for the document. In some variants of the architecture, optionally the neighbouring terms are also provided as input---as shown in the dotted box.}
\label{fig:paragraph2vec}
\end{figure*}

Following the popularity of word2vec \cite{mikolov2013efficient, mikolov2013distributed}, similar neural architectures \cite{le2014distributed, grbovic2015context, grbovic2015search, sun2015learning, ai2016improving, ai2016analysis} have been proposed that trains on term-document co-occurrences. The training typically involves predicting a term given the ID of a document or a passage that contains the term. In some variants, as shown in Figure \ref{fig:paragraph2vec}, neighbouring terms are also provided as input.

The key motivation for training on term-document pairs is to learn an embedding that is more aligned with a Topical notion of term-term similarity---which is often more appropriate for IR tasks. The term-document relationship, however, tends to be more sparse \cite{yan2013learning}---including neighbouring term features may compensate for some of that sparsity.

In the context of IR tasks, \citet{ai2016improving, ai2016analysis} proposed a number of IR-motivated changes to the original Paragraph2vec \cite{le2014distributed} model training---including, document frequency based negative sampling and document length based regularization.


\section{Term embeddings for IR}
\label{sec:embir}

\begin{figure}[t]
\center
\begin{subfigure}{0.49\textwidth}
    \noindent
    \small
    \parbox{0.95\columnwidth}{%
    \input{about1.tex}
    \vspace{1em}}
    \caption{About Albuquerque}
    \label{fig:embir-aboutness1}
\end{subfigure}
\hfill
\begin{subfigure}{0.49\textwidth}
    \noindent
    \small
    \parbox{0.95\columnwidth}{%
    \input{about2.tex}
    \vspace{1em}}
    \caption{Not about Albuquerque}
    \label{fig:embir-aboutness2}
\end{subfigure}
\caption{Two passages both containing exactly a single occurrence of the query term ``Albuquerque''. However, the passage in (a) contains other terms such as ``population'' and ``area'' that are relevant to a description of the city. In contrast, the terms in passage (b) suggest that it is unlikely to be about the city, and only mentions the city potentially in a different context.}
\label{fig:embir-aboutness}
\end{figure}

Traditional IR models use local representations of terms for query-document matching. The most straight-forward use case for term embeddings in IR is to enable \emph{inexact} matching in the embedding space. In Section \ref{sec:ir-desiderata}, we argued the importance of inspecting non-query terms in the document for garnering evidence of relevance. For example, even from a shallow manual inspection, it is easy to conclude that the passage in Figure \ref{fig:embir-aboutness1} is \emph{about} Albuquerque because it contains ``metropolitan'', ``population'', and ``area'' among other informative terms. On the other hand, the passage in Figure \ref{fig:embir-aboutness2} contains ``simulator'', ``interpreter'', and ``Altair'' which seems to suggest that the passage is instead more likely related to computers and technology. In traditional term counting based IR approaches these signals are often ignored.

Most existing shallow neural methods for IR focus on inexact matching using term embeddings. These approaches can be broadly categorized as those that compare the query with the document directly in the embedding space; and those that use embeddings to generate suitable query expansion candidates from a global vocabulary and then perform retrieval based on the expanded query. We discuss both these classes of approaches in the remainder of this section.

\subsection{Query-document matching}
\label{sec:embir-allpairs}

A popular strategy for using term embeddings in IR involves deriving a dense vector representation for the query and the document from the embeddings of the individual terms in the corresponding texts. The term embeddings can be aggregated in different ways, although using the \emph{average word (or term) embeddings} (AWE) is quite popular \cite{kiros2014multiplicative, mitra2016desm, nalisnick2016improving, le2014distributed, vulic2015monolingual, kenter2016siamese, sun2016semantic}. Non-linear combinations of term vectors---such as using Fisher Kernel Framework \cite{clinchant2013aggregating}---have also been explored, as well as other families of aggregate functions of which AWE has been shown to be a special case \cite{zamani2016estimating}.

The query and the document embeddings themselves can be compared using a variety of similarity metrics, such as cosine similarity or dot-product. For example,

\begin{align}
\label{eqn:embir-centroid}
sim(q, d) = cos(\vec{v}_q, \vec{v}_d) &= \frac{\vec{v}_q^{\;\intercal} \vec{v}_d}{\norm{\vec{v}_q} \norm{\vec{v}_d}} \\
\text{where,}\quad\vec{v}_q &= \frac{1}{|q|} \sum_{t_q \in q}  \frac{\vec{v}_{t_q}} {\norm{\vec{v}_{t_q}}} \\
\vec{v}_d &= \frac{1}{|d|} \sum_{t_d \in d}  \frac{\vec{v}_{t_d}} {\norm{\vec{v}_{t_d}}}
\end{align}

An important consideration here is the choice of the term embeddings that is appropriate for the retrieval scenario. While, LSA \cite{deerwester1990indexing}, word2vec \cite{mikolov2013distributed}, and GloVe \cite{pennington2014glove} are popularly used---it is important to understand how the notion of inter-term similarity modelled by a specific vector space may influence its performance on a retrieval task. In the example in Figure \ref{fig:embir-aboutness}, we want to rank documents that contains related terms, such as ``population'' or ``area'' higher---these terms are Topically similar to the query term ``Albuquerque''. Intuitively, a document about ``Tucson''---which is Typically similar to ``Albuquerque''---is unlikely to satisfy the user intent. The discussion in Section \ref{sec:wordrep-sim} on how input features influence the notion of similarity in the learnt vector space is relevant here.

Models, such as LSA \cite{deerwester1990indexing} and Paragraph2vec \cite{le2014distributed}, that consider term-document pairs generally capture Topical similarities in the learnt vector space. On the other hand, word2vec \cite{mikolov2013distributed} and GloVe \cite{pennington2014glove} embeddings may incorporate a mixture of Topical and Typical notions of relatedness. These neural models behave more Typical when trained with short window sizes or on short text, such as on keyword queries \cite{levy2014dependencybased} (refer to Section \ref{sec:wordrep-sim} for more details).

In Section \ref{sec:wordrep-embeddings}, we made a note that the word2vec model learns two different embeddings---IN and OUT---corresponding to the input and the output terms. \citet{mitra2016desm} point out that when using word2vec embeddings for IR it is more appropriate to represent the query terms using the IN embeddings and the document terms using the OUT embeddings of the trained model. In this \emph{Dual Embedding Space Model} (DESM)\footnote{The dual term embeddings trained on Bing queries is available for download at \url{https://www.microsoft.com/en-us/download/details.aspx?id=52597}} \cite{mitra2016desm, nalisnick2016improving} the word2vec embeddings are trained on search queries, which empirically performs better than training on document body text. Training on short queries, however, makes the inter-term similarity more pronouncedly Typical (where, ``Yale'' is closer to ``Harvard'' and ``NYU'') when both terms are represented using their IN vectors---better retrieval performance is achieved instead by using the IN-OUT similarity (where, ``Yale'' is closer to ``faculty'' and ``alumni'') that mirrors more the Topical notions of relatedness.

\begin{align}
\label{eqn:embir-desm}
DESM_{in-out}(q, d) &= \frac{1}{|q|} \sum_{t_q \in q}  \frac{\vec{v}_{t_q,in}^{\;\intercal} \vec{v}_{d,out}}{\norm{\vec{v}_{t_q,in}}\norm{\vec{v}_{d,out}}} \\
\vec{v}_{d,out} &= \frac{1}{|d|} \sum_{t_d \in d}  \frac{\vec{v}_{t_d,out}} {\norm{\vec{v}_{t_d,out}}}
\end{align}

An alternative to representing queries and documents as an aggregate of their term embeddings is to incorporate the term representations into existing IR models, such as the ones we discussed in Section \ref{sec:ir-models}. \citet{zuccon2015integrating} proposed the \emph{Neural Translation Language Model} (NTLM) that uses the similarity between term embeddings as a measure for term-term translation probability $p(t_q|t_d)$ in Equation \ref{eqn:ir-tm}.

\begin{align}
p(t_q|t_d) &= \frac{cos(\vec{v}_{t_q}, \vec{v}_{t_d})}{\sum_{t \in T}{cos(\vec{v}_{t}, \vec{v}_{t_d})}}
\label{eqn:embir-ntlm}
\end{align}

On similar lines, \citet{ganguly2015word} proposed the \emph{Generalized Language Model} (GLM) which extends the Language Model based approach in Equation \ref{eqn:ir-lm} to,

\begin{align}
\begin{split}
\label{eqn:embir-glm}
p(d|q) = \prod_{t_q \in q} \bigg(&\lambda \frac{tf(t_q,d)}{|d|} + \alpha \frac{\sum_{t_d \in d}{(sim(\vec{v}_{t_q}, \vec{v}_{t_d}) \cdot tf(t_d,d))}}{\sum_{t_{d_1} \in d}{\sum_{t_{d_2} \in d}{sim(\vec{v}_{t_{d_1}}, \vec{v}_{t_{d_2}})}} \cdot {|d|}^2} \\ &+ \beta \frac{\sum_{\bar{t} \in N_t}{(sim(\vec{v}_{t_q}, \vec{v}_{\bar{t}}) \cdot \sum_{\bar{d} \in D}{tf(\bar{t},\bar{d})})}}{\sum_{t_{d_1} \in N_t}{\sum_{t_{d_2} \in N_t}{sim(\vec{v}_{t_{d_1}}, \vec{v}_{t_{d_2}})}} \cdot \sum_{\bar{d} \in D}{|\bar{d}|} \cdot |N_t|} + (1 - \alpha - \beta - \lambda) \frac{\sum_{\bar{d} \in D}{tf(t_q,\bar{d})}}{\sum_{\bar{d} \in D}{|\bar{d}|}} \bigg)
\end{split}
\end{align}

where, $N_t$ is the set of nearest-neighbours of term $t$. \citet{ai2016analysis} incorporate paragraph vectors \cite{le2014distributed} into the query-likelihood model \cite{ponte1998language}.

Another approach, based on the Earth Mover's Distance (EMD) \cite{rubner1998metric}, involves estimating similarity between pairs of documents by computing the minimum distance in the embedding space that each term in the first document needs to travel to reach the terms in the second document. This measure, commonly referred to as the \emph{Word Mover's Distance} (WMD), was originally proposed by Wan et al. \cite{wan2005earth, wan2007novel}, but used WordNet and topic categories instead of distributed representations for defining distance between terms. Term embeddings were later incorporated into the model by Kusner et al. \cite{kusnerword, huang2016supervised}. Finally, \citet{guo2016semantic} incorporated similar notion of distance into the \emph{Non-linear Word Transportation} (NWT) model that estimates relevance between a a query and a document. The NWT model involves solving the following constrained optimization problem,

\begin{align}
\label{eqn:embir-nwt}
\text{max} \quad &\sum_{t_q \in q}{\log\bigg(\sum_{t_d \in u(d)}{f(t_q, t_d)\cdot{max\big(cos(\vec{v}_{t_q}, \vec{v}_{t_d}), 0\big)}^{idf(t_q)+b}}\bigg)} \\
\text{subject to}\quad &f(t_q, t_d) \ge 0, \quad \forall t_q \in q, t_d \in d \\
\text{and}\quad & \sum_{t_q \in q}{f(t_q, t_d)} = \frac{tf(t_d)+\mu\frac{\sum_{\bar{d} \in D}{tf(t_q,\bar{d})}}{\sum_{\bar{d} \in D}{|\bar{d}|}}}{|d| + \mu}, \quad \forall t_d \in d \\
\text{where,}\quad &idf(t) = \frac{|D| - df(t) + 0.5}{df(t) + 0.5}
\end{align}

$u(d)$ is the set of all unique terms in document $d$, and $b$ is a constant.

Another term-alignment based distance metric was proposed by \citet{kenter15short} for computing short-text similarity. The design of the \emph{saliency-weighted semantic network} (SWSN) is motivated by the BM25 \cite{robertson2009probabilistic} formulation.

\begin{align}
\label{eqn:embir-ss}
swsn(s_l,s_s) &= \sum_{t_l \in s_l}{idf(t_l)\cdot {\frac {sem(t_l,s_s) \cdot (k_1+1)}{sem(t_l,s_s)+k_1 \cdot \left(1-b+b \cdot {\frac{|s_s|}{avgsl}}\right)}}} \\
\text{where,}\quad sem(t, s) &= \max_{\bar{t} \in s}{cos(\vec{v}_t, \vec{v}_{\bar{t}})}
\end{align}

Here $s_s$ is the shorter of the two sentences to be compared, and $s_l$ the longer sentence.

\begin{figure}
\center
\begin{subfigure}{0.99\textwidth}
    \noindent
    \parbox{0.95\columnwidth}{%
    \vspace{1.5em}
    \input{cambridge1.tex}}
    \caption{Passage about the city of Cambridge}
    \label{fig:embir-cambridge1}
\end{subfigure}
\begin{subfigure}{0.99\textwidth}
    \noindent
    \parbox{0.95\columnwidth}{%
    \vspace{1.5em}
    \input{cambridge2.tex}}
    \caption{Passage about the city of Oxford}
    \label{fig:embir-cambridge2}
\end{subfigure}
\begin{subfigure}{0.99\textwidth}
    \noindent
    \parbox{0.95\columnwidth}{%
    \vspace{1.5em}
    \input{cambridge3.tex}}
    \caption{Passage about giraffes}
    \label{fig:embir-cambridge3}
\end{subfigure}
\begin{subfigure}{0.99\textwidth}
    \noindent
    \parbox{0.95\columnwidth}{%
    \vspace{1.5em}
    \input{cambridge4.tex}}
    \caption{Passage about giraffes, but 'giraffe' is replaced by 'Cambridge'}
    \label{fig:embir-cambridge4}
\end{subfigure}
\vspace{1.5em}
\caption{A visualization of IN-OUT similarities between terms in different passages with the query term ``Cambridge''. The visualization---adapted from \url{https://github.com/bmitra-msft/Demos/blob/master/notebooks/DESM.ipynb}---reveal that, besides the term ``Cambridge'', many other terms in the passages about both Cambridge and Oxford have high similarity to the query term. The passage (d) is adapted from the passage (c) on giraffes by replacing all the occurrences of the term ``giraffe'' with ``cambridge''. However, none of the other terms in (d) are found to be relevant to the query term. An embedding based approach may be able to determine that passage (d) is non-relevant to the query ``Cambridge'', but fail to realize that passage (b) is also non-relevant. A term counting-based model, on the other hand, can easily identify that passage (b) is non-relevant, but may rank passage (d) incorrectly high.}
\label{fig:embir-cambridge}
\end{figure}

\paragraph*{Telescoping evaluation}
Figure \ref{fig:embir-cambridge} highlights the distinct strengths and weaknesses of matching using local and distributed representations of terms for retrieval. For the query ``Cambridge'', a local representation (or exact matching) based model can easily distinguish between the passage on Cambridge (Figure \ref{fig:embir-cambridge1}) and the one on Oxford (Figure \ref{fig:embir-cambridge2}). However, the model is easily duped by an non-relevant passage that has been artificially injected with the term ``Cambridge'' (Figure \ref{fig:embir-cambridge4}). The distributed representation based matching, on the other hand, can spot that the other terms in the passage provide clear indication that the passage is not \emph{about} a city, but fails to realize that the the passage about Oxford (Figure \ref{fig:embir-cambridge2}) is inappropriate for the same query.

Embedding based models often perform poorly when the retrieval is performed over the full document collection \cite{mitra2016desm}. However, as seen in the example of Figure \ref{fig:embir-cambridge}, the errors made by embedding based models and exact matching models are typically different---and the combination of the two performs better than exact matching models alone \cite{mitra2016desm, ganguly2015word, ai2016analysis}. Another popular technique is to use the embedding based model to re-rank only a subset of the documents retrieved by a different---generally an exact matching based---IR model. The chaining of different IR models where each successive model re-ranks a smaller number of candidate documents is called \emph{Telescoping} \cite{matveeva2006high}. Telescoping evaluations are popular in the neural IR literature \cite{mitra2016desm, huang2013learning, shen2014latent, guo2016deep, mitra2016learning} and the results are representative of performances of these models on re-ranking tasks. However, as \citet{mitra2016desm} demonstrate, good performances on re-ranking tasks may not be indicative how the model would perform if the retrieval involves larger document collections.

\subsection{Query expansion}
\label{sec:embir-qexp}

\begin{figure}
\center
\begin{subfigure}{0.49\textwidth}
    \includegraphics[width=\textwidth]{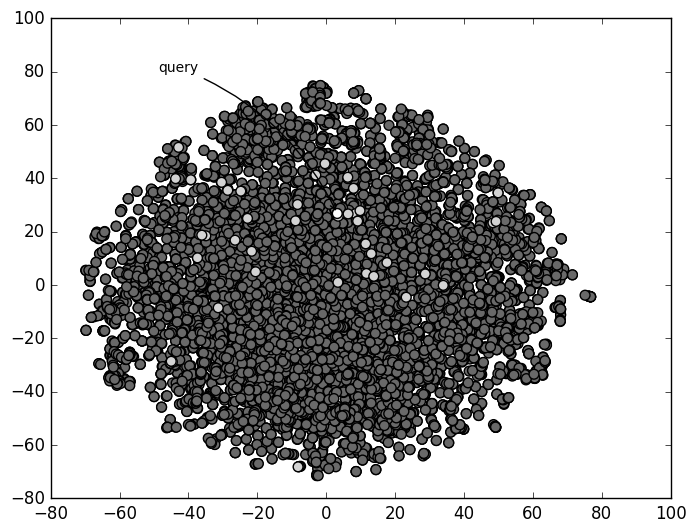}
    \caption{Global embedding}
    \label{fig:globalemb}
\end{subfigure}
\hfill
\begin{subfigure}{0.49\textwidth}
    \includegraphics[width=\textwidth]{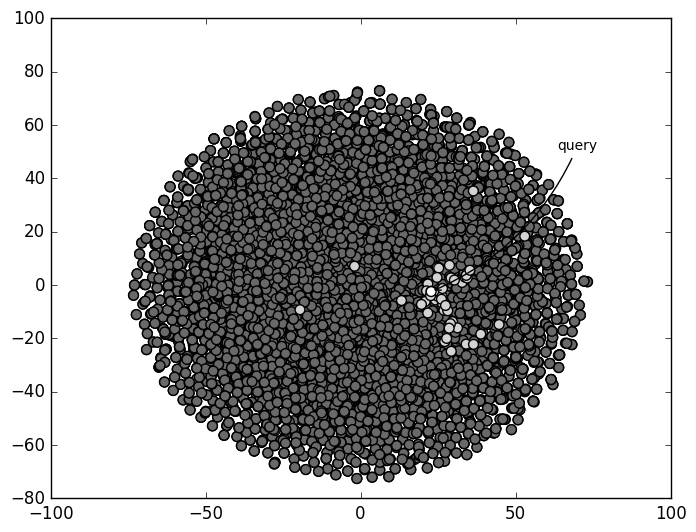}
    \caption{Local embedding}
    \label{fig:localemb}
\end{subfigure}
\caption{A two-dimensional visualization of term embeddings when the vector space is trained on a (a) global corpus and a (b) query-specific corpus, respectively. The grey circles represent individual terms in the vocabulary. The white circle represents the query ``‘ocean remote sensing'' by averaging the embeddings of the individual terms in the query, and the light grey circles correspond to good expansion terms for this query. When the representations are query-specific then the meaning of the terms are better disambiguated, and more likely to result in the selection of good expansion terms.}
\label{fig:globallocalemb}
\end{figure}

Instead of comparing the query and the document directly in the embedding space, an alternative approach is to use term embeddings to find good expansion candidates from a global vocabulary, and then retrieving documents using the expanded query. Different functions \cite{diaz2016query, roy2016using, zamani2016embedding} have been proposed for estimating the relevance of candidate terms to the query---all of them involves comparing the candidate term individually to every query term using their vector representations, and then aggregating the scores. For example, \cite{diaz2016query, roy2016using} estimate the relevance of candidate term $t_c$ as,

\begin{align}
\label{eqn:embir-qexp}
score(t_c,q) &= \frac{1}{|q|} \sum_{t_q \in q}{cos(\vec{v}_{t_c},\vec{v}_{t_q})}
\end{align}

Term embedding based query expansion on its own performs worse than pseudo-relevance feedback \cite{roy2016using}. But like the models in the previous section, shows better performances when used in combination with PRF \cite{zamani2016embedding}.

\citet{diaz2016query} explored the idea of query-specific term embeddings and found that they are much more effective in identifying good expansion terms than a global representation (see Figure \ref{fig:globallocalemb}). The local model proposed by \citet{diaz2016query} incorporate relevance feedback in the process of learning the term embeddings---a set of documents is retrieved for the query and a query-specific term embedding model is trained. This \emph{local} embedding model is then employed for identifying expansion candidates for the query for a second round of document retrieval.

\bigskip

Term embeddings have also been explored for re-weighting query terms \cite{zheng2015learning} and finding relevant query re-writes \cite{grbovic2015context}, as well as in the context of other IR tasks such as cross-lingual retrieval \cite{vulic2015monolingual} and entity retrieval \cite{van2016learning}. In the next section, we move on to neural network models with deeper architectures and their applications to retrieval.


\section{Deep neural networks}
\label{sec:dnn}

\begin{figure*}
\center
\begin{subfigure}{0.95\textwidth}
    \includegraphics[width=\textwidth]{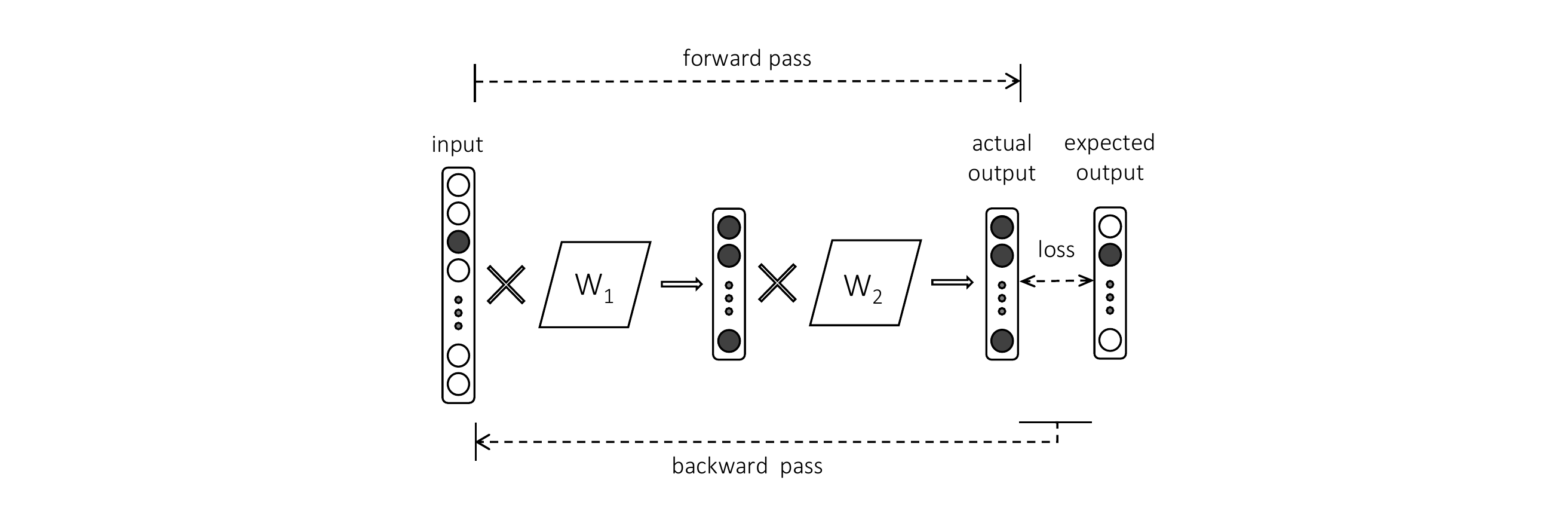}
    \caption{A neural network with a single hidden layer.}
    \label{fig:neuralnet1}
\end{subfigure}
\hfill
\begin{subfigure}{0.95\textwidth}
    \includegraphics[width=\textwidth]{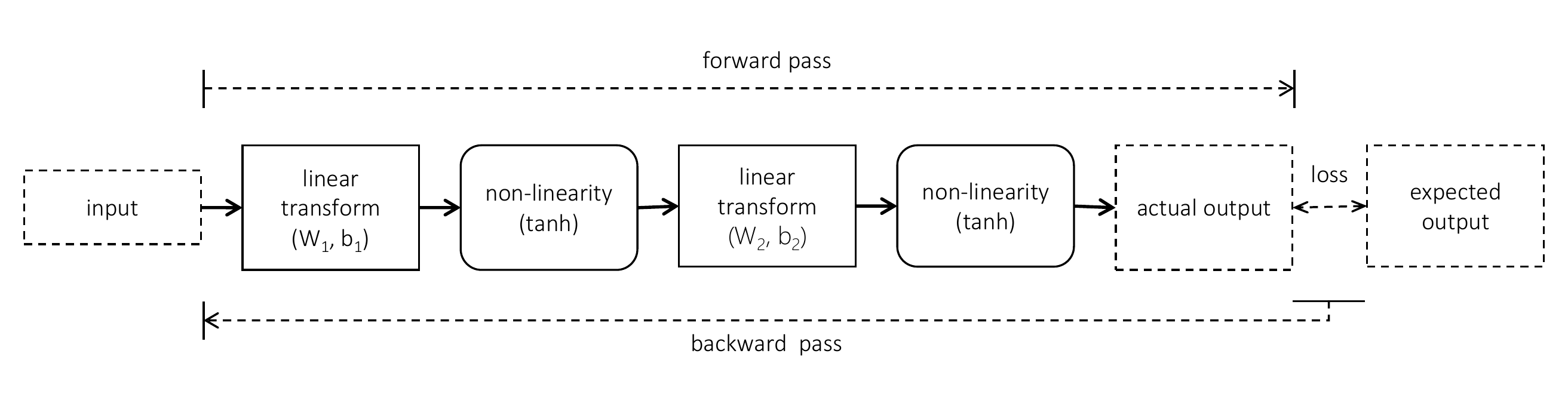}
    \caption{The same neural network viewed as a chain of computational steps.}
    \label{fig:neuralnet2}
\end{subfigure}
\caption{Two different visualizations of a feed-forward neural network with a single hidden layer. In (a), the addition of the bias vector and the non-linearity function is implicit. Figure (b) shows the same network but as a sequence of computational nodes. Most popular neural network toolkits implement a set of standard computational nodes that can be connected to build more sophisticated neural architectures.}
\label{fig:neuralnet}
\end{figure*}

Deep neural network models consist of chains of tensor operations. The tensor operation can range from parameterized linear transformations (e.g., multiplication with a weight matrix, addition of a bias vector) to elementwise application of non-linear functions, such as \emph{tanh} or \emph{rectified linear units} (ReLU) \cite{hahnloser2000digital, nair2010rectified, jarrett2009best}. Figure \ref{fig:neuralnet} shows a simple \emph{feed-forward} neural network with \emph{fully-connected} layers. For an input vector $\vec{x}$, the model produces the output $\vec{y}$ as follows,

\begin{align}
\label{eqn:neuralnet}
\vec{y} = tanh(W_2\cdot tanh(W_1\cdot \vec{x} + \vec{b}_1) + \vec{b}_2)
\end{align}

The model training involves tuning the parameters $W_1$, $\vec{b}_1$, $W_2$, and $\vec{b}_2$ to minimize the loss between the expected output and the actual output of the final layer. The parameters are usually trained discriminatively using backpropagation \cite{schmidhuber2015deep, bengio2009learning, hecht1988theory}. During forward-pass each layer generates an output conditioned on its input, and during backward pass each layer computes the error gradient with respect to its parameters and its inputs.

\begin{figure}
\center
\includegraphics[width=\linewidth]{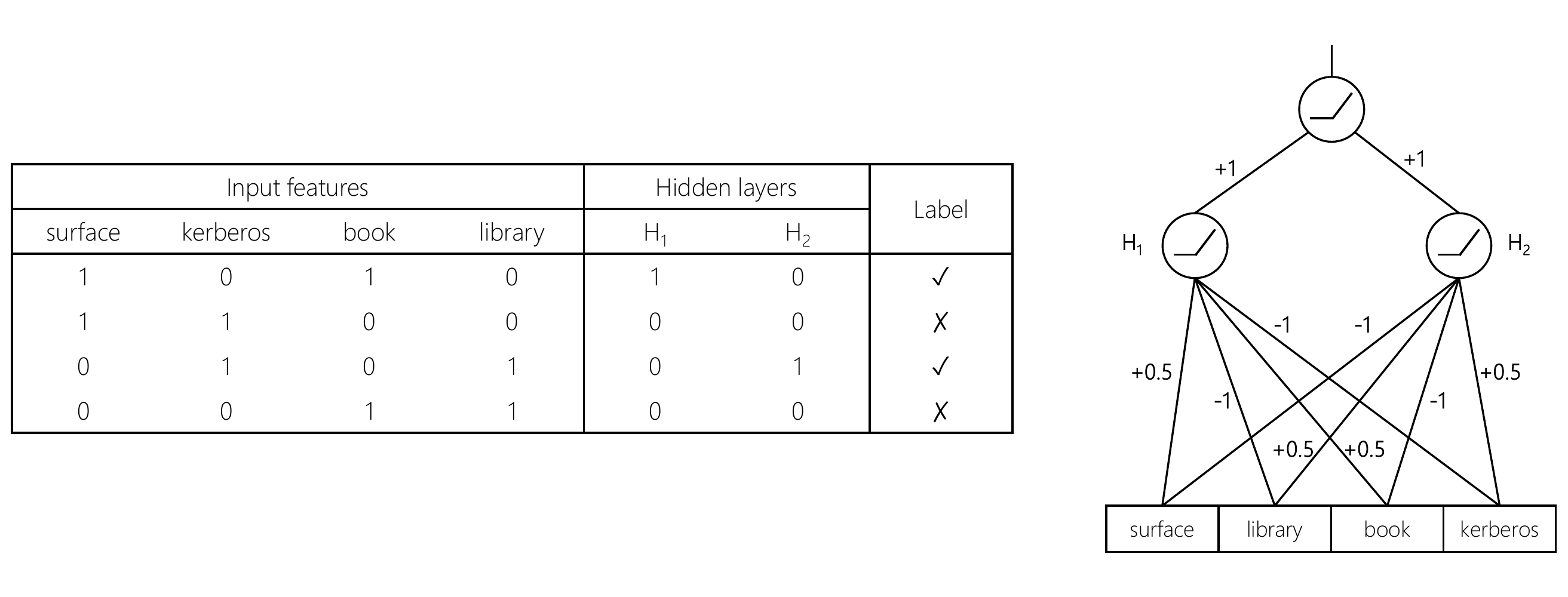}
\caption{Consider a toy binary classification task on a corpus of four short texts---``surface book'', ``kerberos library'', ``library book'', and ``kerberos surface''---where the model needs to predict if the text is related to computers. The first two texts---``Surface Book'' and ``kerberos library''---are positive under this classification, and the latter two negative. The input feature space consists of four binary features that indicate whether each of the four terms from the vocabulary is present in the text. The table shows that the specified classes are not linearly separable with respect to the input feature space. However, if we add couple of hidden nodes, as shown in the diagram, then the classes can be linearly separated with respect to the output of the hidden layer.}
\label{fig:hiddenlayer}
\end{figure}

The design of a DNN typically involves many choices of architectures and hyper-parameters. Neural networks with as few a single hidden layer---but with sufficient number of hidden nodes---can theoretically approximate any function \cite{hornik1989multilayer}. In practice, however, deeper architectures---sometimes with as many as 1000 layers \cite{he2016deep}---have been shown to perform significantly better than shallower networks. For readers who are less familiar with neural network models, we present a simple example in Figure \ref{fig:hiddenlayer} to illustrate how hidden layers enable these models to capture non-linear relationships. We direct readers to \cite{montufar2014number} for further discussions on how additional hidden layers help.

The rest of this section is dedicated to the discussion of input representations and popular architectures for deep neural models.

\subsection{Input text representations}
\label{sec:dnn-inputs}

\begin{figure*}
\center
\begin{subfigure}{0.9\textwidth}
    \includegraphics[width=\textwidth]{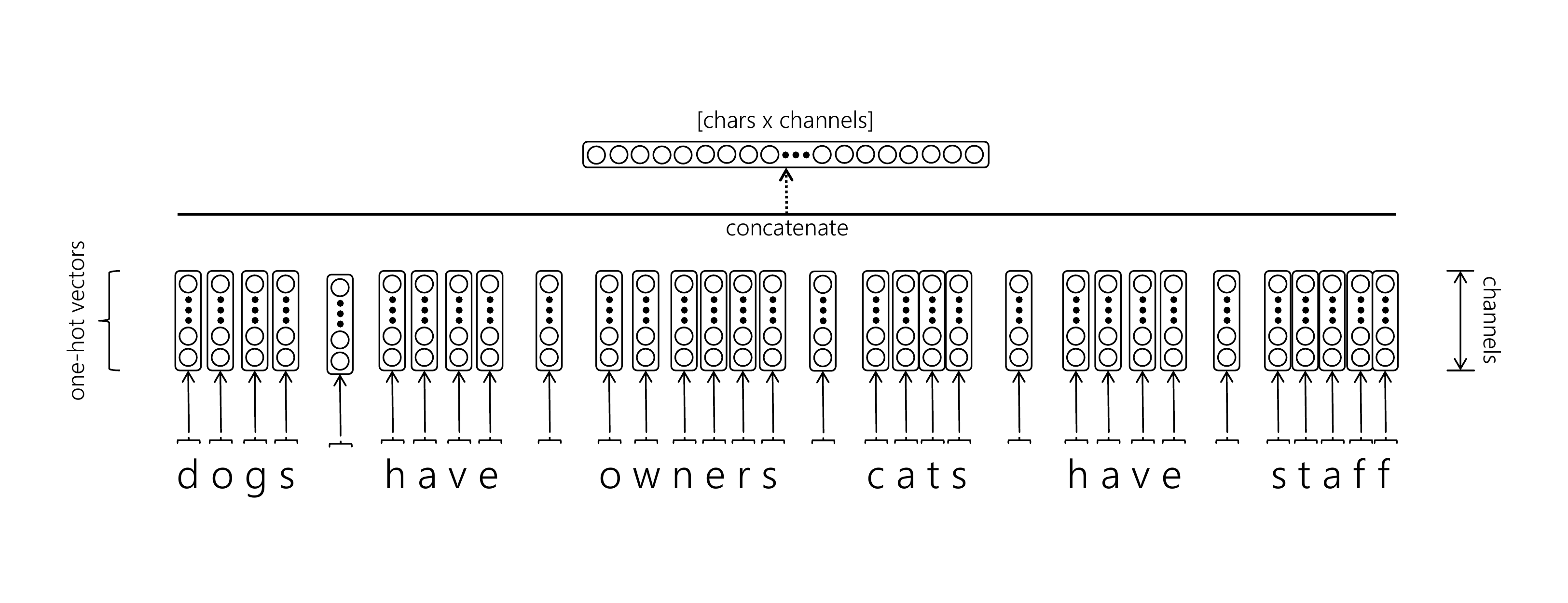}
    \caption{Character-level input}
    \label{fig:textinput1}
\end{subfigure}
\begin{subfigure}{0.9\textwidth}
    \includegraphics[width=\textwidth]{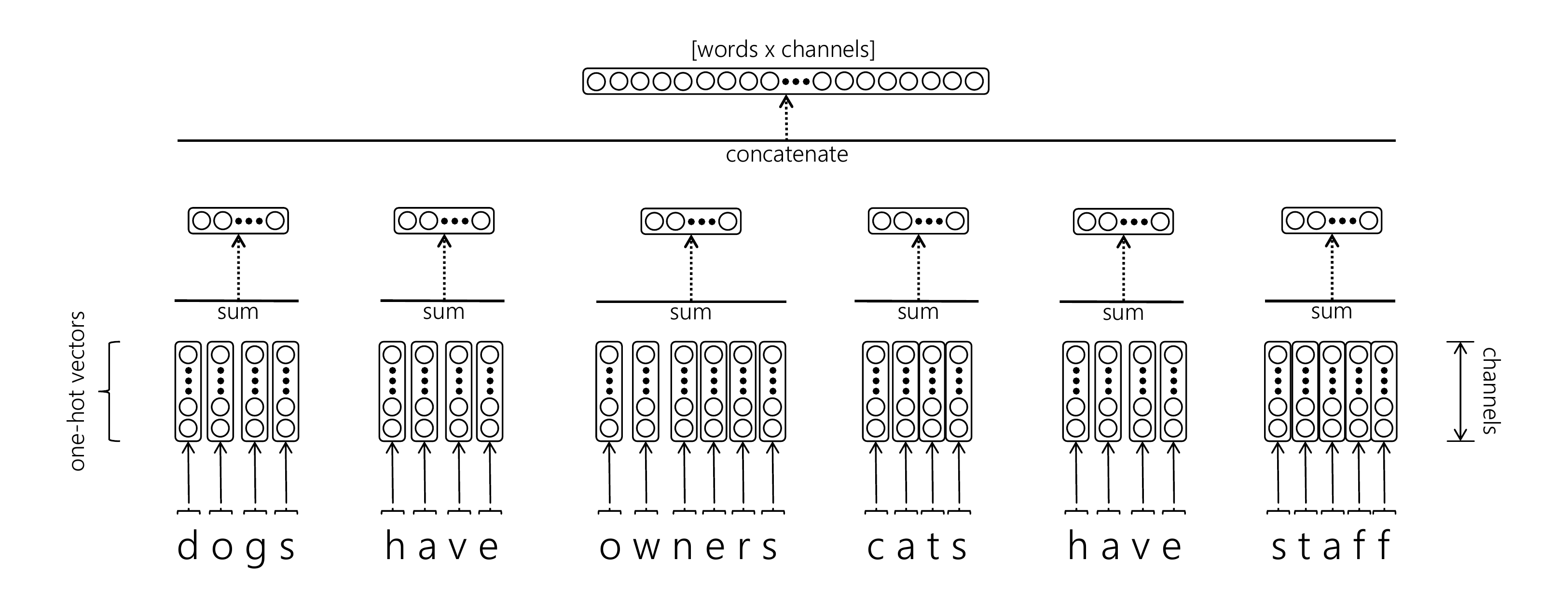}
    \caption{Term-level input w/ bag-of-characters per term}
    \label{fig:textinput2}
\end{subfigure}
\begin{subfigure}{0.9\textwidth}
    \includegraphics[width=\textwidth]{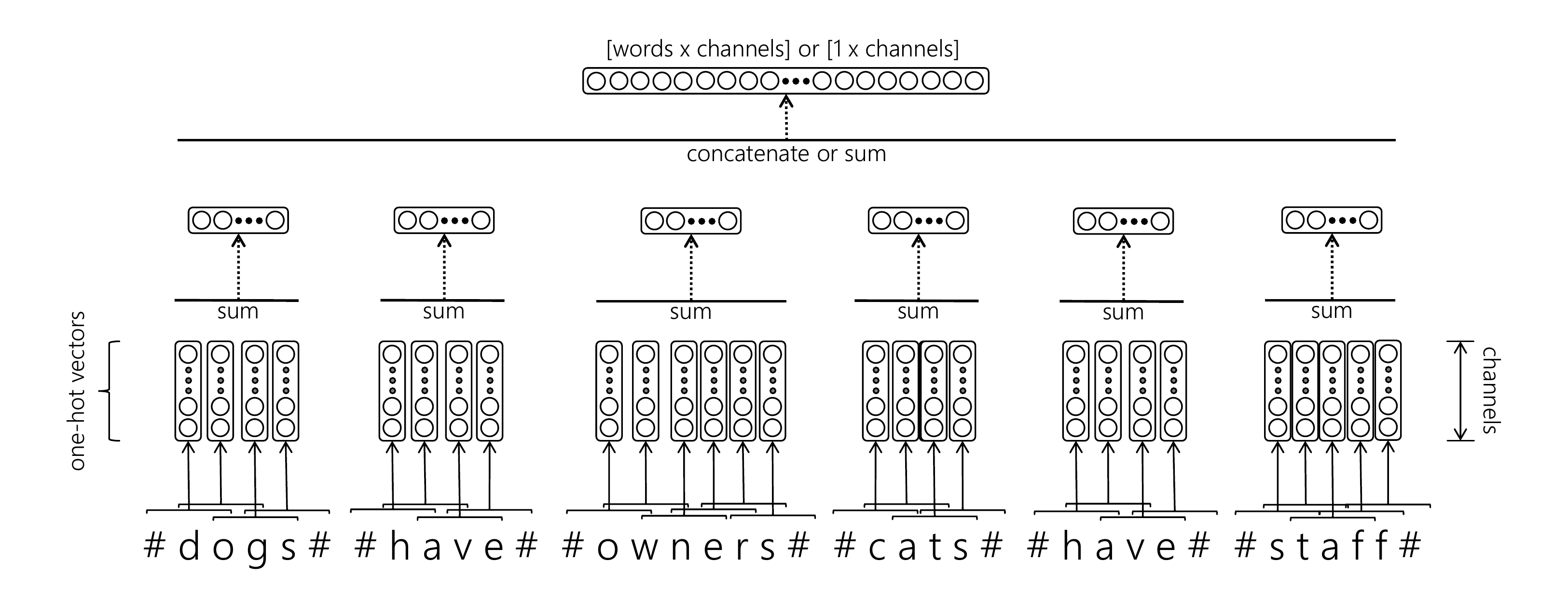}
    \caption{Term-level input w/ bag-of-trigraphs per term}
    \label{fig:textinput3}
\end{subfigure}
\begin{subfigure}{0.9\textwidth}
    \includegraphics[width=\textwidth]{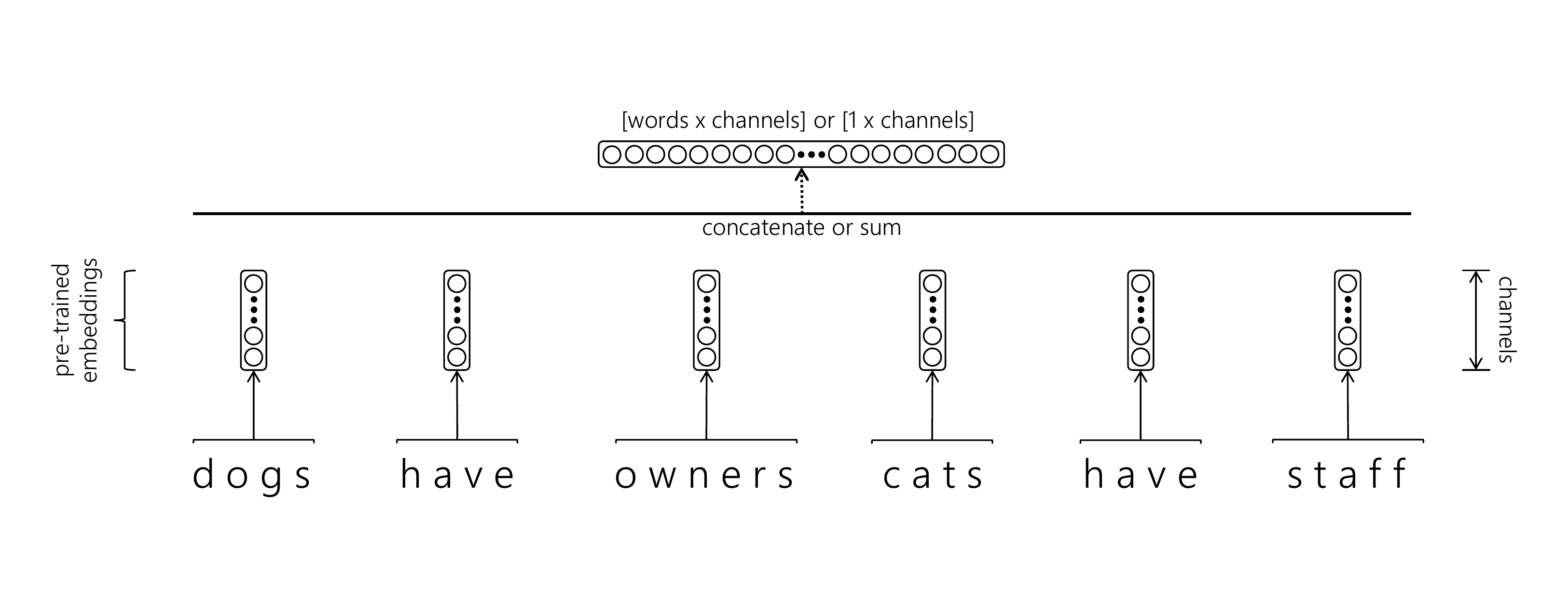}
    \caption{Term-level input w/ pre-trained term embeddings}
    \label{fig:textinput4}
\end{subfigure}
\caption{Examples of different representation strategies for text input to deep neural network models. The smallest granularity of representation can be a character or a term. The vector can be a sparse local representation, or a pre-trained embedding.}
\label{fig:textinput}
\end{figure*}

Neural models that learn representations of text take raw text as input. A key consideration is how the text should be represented at the input layer of the model. Figure \ref{fig:textinput} shows some of the popular input representations of text.

Some neural models \cite{jozefowicz2016exploring, graves2013generating, sutskever2011generating, kim2015character} operate at the character-level. In these models, each character is typically represented by a one-hot vector. The vector dimensions---referred to as \emph{channels}---in this case equals the number of allowed characters in the vocabulary. These models incorporate the least amount of prior knowledge about the language in the input representation---for example, these models are often required to learn about tokenization from scratch by treating space as just another character in the vocabulary. The representation of longer texts, such as sentences, can be derived by concatenating or summing the character-level vectors as shown in Figure \ref{fig:textinput1}.

The input text can also be pre-tokenized into terms---where each term is represented by either a sparse vector or using pre-trained term embeddings (Figure \ref{fig:textinput4}). Terms may have a one-hot (or local) representation where each term has an unique ID (Figure \ref{fig:textinput2}), or the term vector can be derived by aggregating one-hot vectors of its constituting characters (or character $n$-graphs) as shown in Figure \ref{fig:textinput3}. If pre-trained embeddings are used for term representation, then the embedding vectors can be further tuned during training, or kept fixed.

Similar to character-level models, the term vectors are further aggregated (by concatenation or sum) to obtain the representation of longer chunks of text, such as sentences. While one-hot representations of terms (Figure \ref{fig:textinput2}) are common in many NLP tasks, pre-trained embeddings (e.g., \cite{pang2016text, hu2014convolutional}) and character $n$-graph based representations (e.g., \cite{huang2013learning, mitra2016learning}) are more popularly employed in IR.

\subsection{Popular architectures}
\label{sec:dnn-arch}

In this section, we describe few neural operations and architectures popular in IR. For broader overview of different neural architectures and design patterns please refer to \cite{goodfellow2016deep, lecun2015deep, schmidhuber2015deep}.

\paragraph*{Shift-invariant neural operations}
Convolutional \cite{lecun2004learning, jarrett2009best, krizhevsky2012imagenet, lecun2010convolutional} and recurrent \cite{mikolov2010recurrent, graves2009novel, sak2014long, hochreiter1997long} architectures are commonplace in most deep learning applications. These neural operations are part of a broader family of shift-invariant architectures. The key intuition behind these architectures stem from the natural regularities observable in most inputs. In vision, for example, the task of detecting a face should be invariant to whether the image is shifted, rotated, or scaled. Similarly, the meaning of an English sentence should, in most cases, stay consistent independent of which part of the document it appears in. Therefore, intuitively a neural model for object recognition or text understanding should not learn an independent logic for the same action applied to different parts of the input space.

\begin{figure*}
\center
\begin{subfigure}{0.45\textwidth}
    \includegraphics[width=\textwidth]{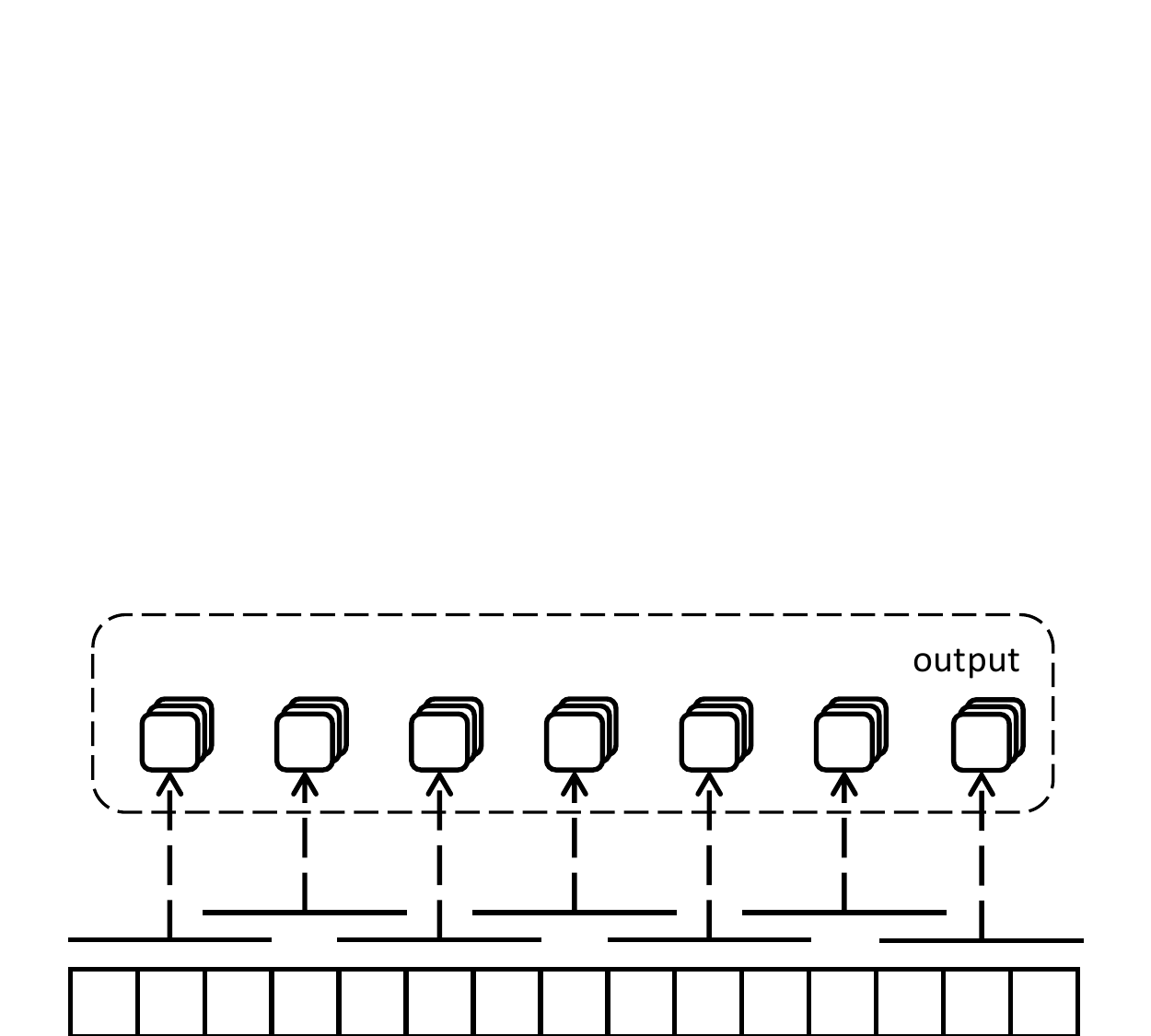}
    \caption{Convolution or pooling}
    \label{fig:neuralops1}
\end{subfigure}
\hfill
\begin{subfigure}{0.45\textwidth}
    \includegraphics[width=\textwidth]{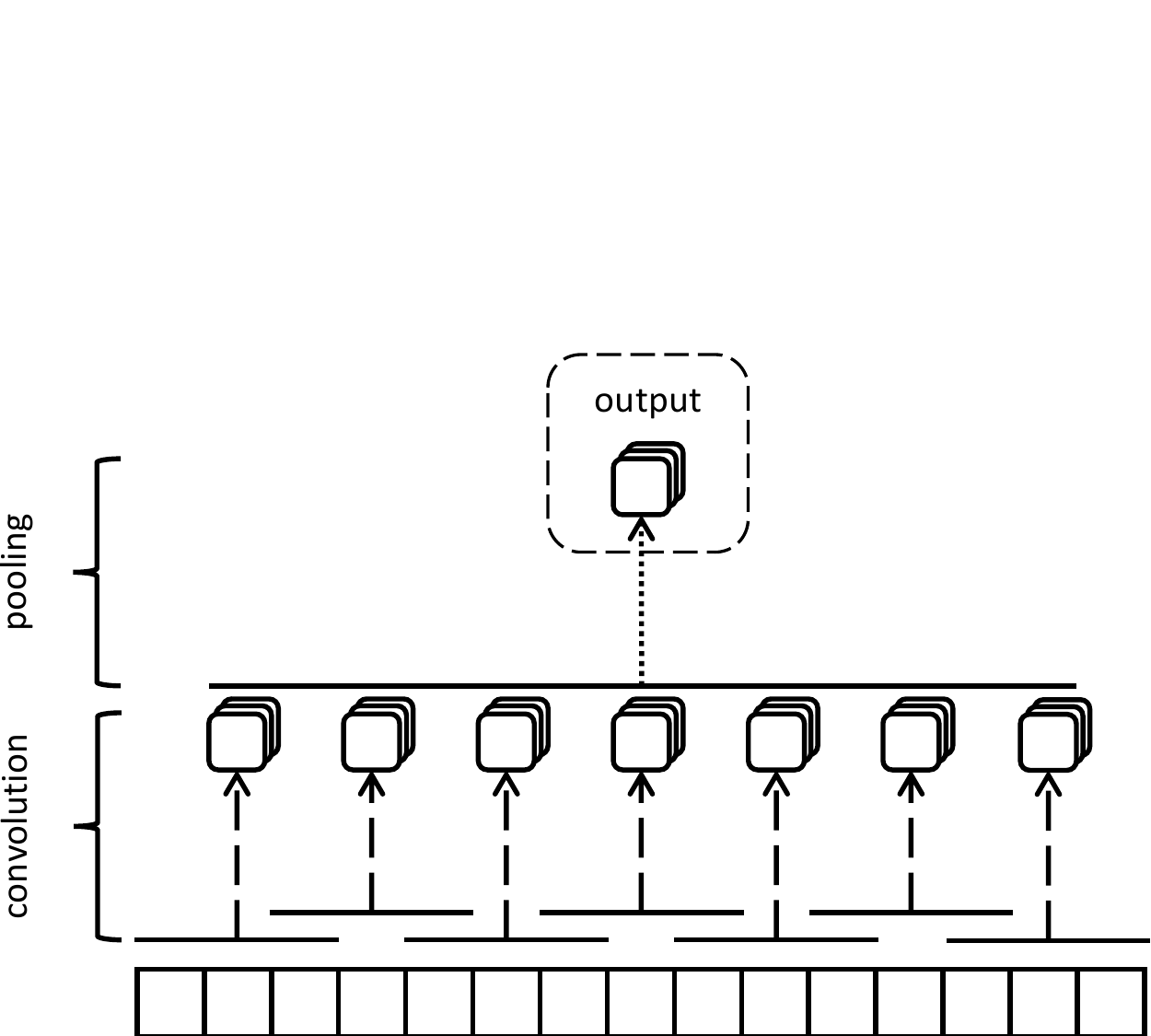}
    \caption{Convolution w/ global pooling}
    \label{fig:neuralops2}
\end{subfigure}
\begin{subfigure}{0.45\textwidth}
    \includegraphics[width=\textwidth]{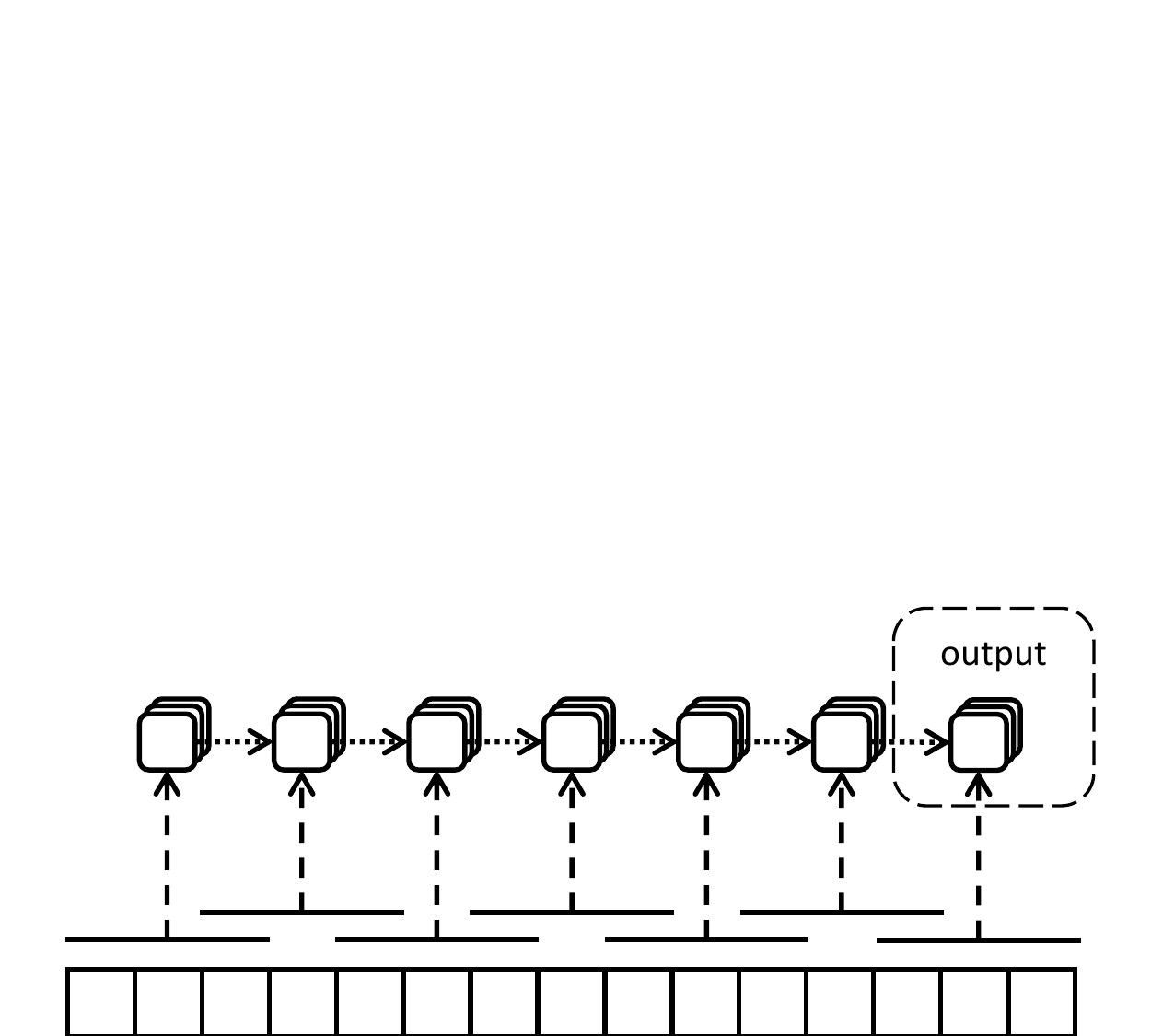}
    \caption{Recurrent}
    \label{fig:neuralops3}
\end{subfigure}
\hfill
\begin{subfigure}{0.45\textwidth}
    \includegraphics[width=\textwidth]{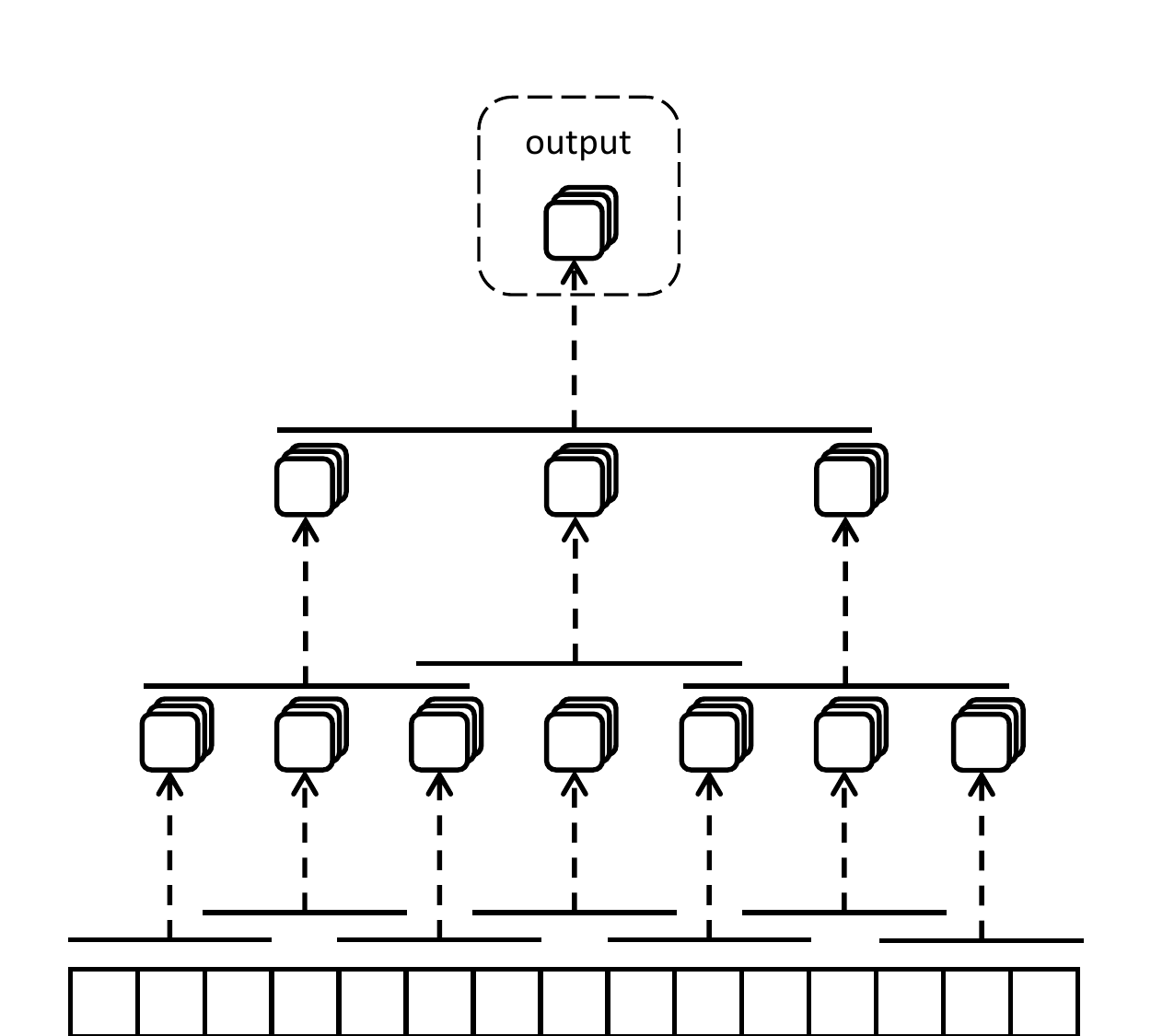}
    \caption{Recursive or tree}
    \label{fig:neuralops4}
\end{subfigure}
\caption{Popular shift-invariant neural architectures including convolutional neural networks (CNN), recurrent neural networks (RNN), pooling layers, and tree-structured neural networks.}
\label{fig:neuralops}
\end{figure*}

All shift-invariant neural operations fundamentally employ a window-based approach. A fixed size window is moved over the input space with fixed stride in each step. A (typically parameterized) function---referred to as a \emph{kernel}, or a \emph{filter}, or a \emph{cell}---is applied over each instance of the window. The parameters of the cell are shared across all the instances of the input window. The shared parameters not only implies less number of total parameters in the model,, but also more supervision per parameter per training sample due to the repeated application.

Figure \ref{fig:neuralops1} shows an example of a cell being applied on a sequence of terms---with a window size of three terms---in each step. A popular cell implementation involves multiplying with a weight matrix---in which case the architecture in Figure \ref{fig:neuralops1} is referred as \emph{convolutional}. An example of a cell without any parameters is \emph{pooling}---which consists of aggregating (e.g., by computing the max or the average) over all the terms in the window\footnote{If the input has multiple channels per term then the aggregation is performed per channel.}. Note, that the length of the input sequence can be variable in both cases and the length of the output of a convolutional (or pooling) layer is a function of the input length. Figure \ref{fig:neuralops2} shows an example of \emph{global pooling}---where the window spans over the whole input---being applied on top of a convolutional layer. The global pooling strategy is common for generating a fixed size output from a variable length input.\footnote{It is obvious, but may be still worth pointing out, that a \emph{global convolutional} layer is exactly the same as a fully-connected layer.}

In convolution or pooling, each window is applied independently. In contrast, in the \emph{recurrent} architecture of Figure \ref{fig:neuralops3} the cell not only considers the input window but also the output of the previous instance of the cell as its input. Many different cell architectures have been explored for recurrent neural networks (RNN)---although Elman network \cite{elman1990finding}, Long Short-Term Memory (LSTM) \cite{hochreiter1997long}, and Gated Recurrent Unit (GRU) \cite{chung2014empirical, cho2014properties} are popular. RNNs are popularly applied to sequences, but can also be useful for two (and higher) dimensional inputs \cite{wan2016match}.

One consideration when using convolutional or recurrent layers is how the window outputs are aggregated. Convolutional layers are typically followed by pooling or fully-connected layers that perform a global aggregation over all the window instances. While a fully-connected layer is aware of each window position, a global pooling layer is typically agnostic to it. However, unlike a fully-connected layer, a global max-pooling operation can be applied to a variable size input. Where a global aggregation strategy may be less appropriate (e.g., long sequences), recurrent networks with memory \cite{weston2014memory, sukhbaatar2015end, bordes2015large} and/or attention \cite{mnih2014recurrent, xu2015show, luong2015effective, hermann2015teaching, chorowski2015attention} may be useful.

Finally, Figure \ref{fig:neuralops3} shows \emph{tree-structured} (or \emph{recursive}) neural networks \cite{goller1996learning, socher2011semi, bowman2016fast, tai2015improved, socher2011parsing} where the same cell is applied at multple levels in a tree-like hierarchical fashion.

\paragraph*{Auto-encoders}

\begin{figure*}
\center
\begin{subfigure}{0.45\textwidth}
    \includegraphics[width=\textwidth]{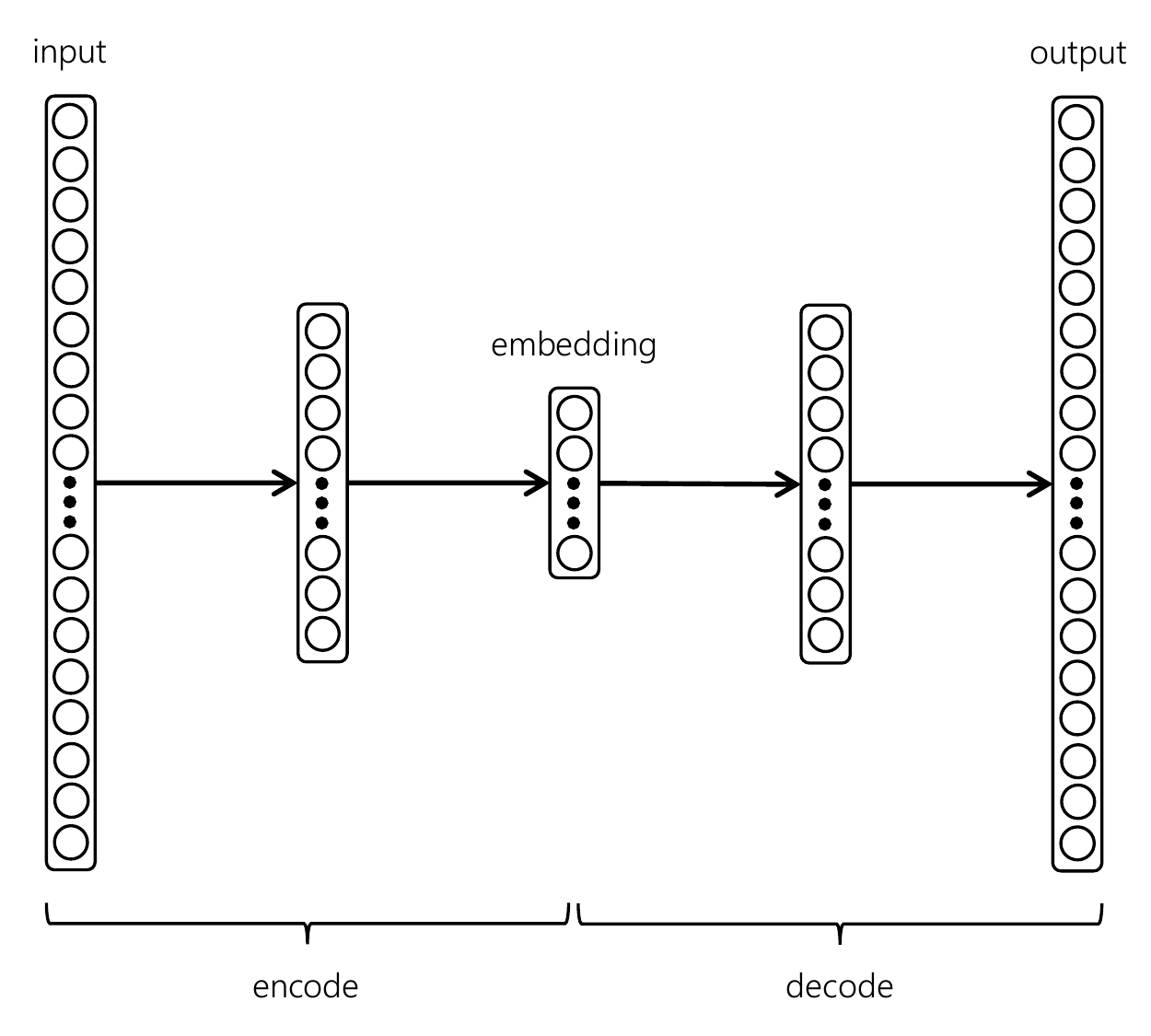}
    \caption{Auto-encoder}
    \label{fig:autoencoder}
\end{subfigure}
\hfill
\begin{subfigure}{0.45\textwidth}
    \includegraphics[width=\textwidth]{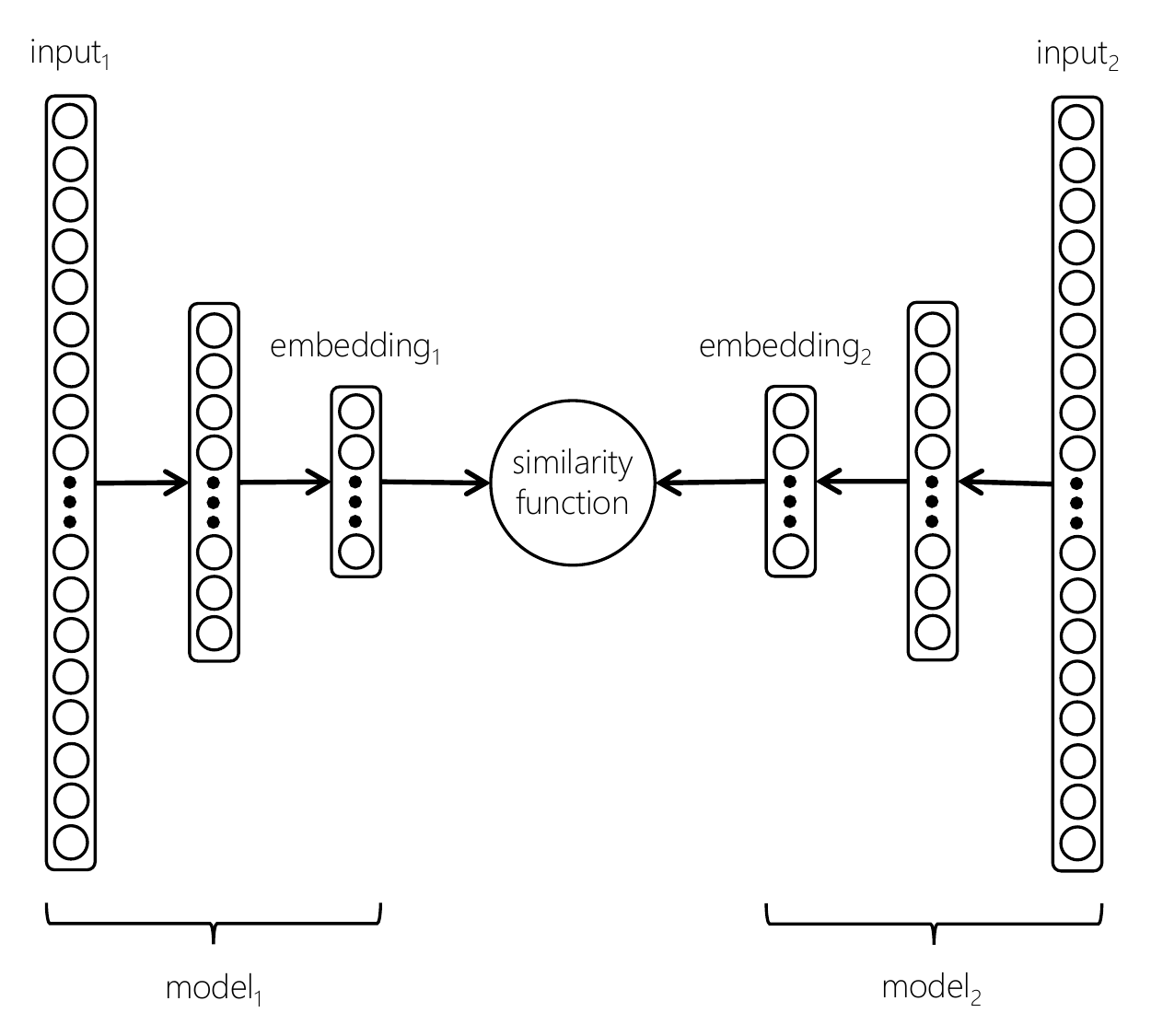}
    \caption{Siamese network}
    \label{fig:siamese}
\end{subfigure}
\caption{Both (a) the auto-encoder and (b) the Siamese network architectures are designed to learn compressed representations of inputs. In an auto-encoder the embeddings are learnt by minimizing the self-reconstruction error, whereas a Siamese network focuses on retaining the information that is necessary for determining the similarity between a pair of items (say, a query and a document).}
\label{fig:bottleneck}
\end{figure*}

The auto-encoder architecture \cite{bengio2007greedy, ranzato2006efficient, bengio2009learning} is based on the \emph{information bottleneck method} \cite{tishby2000information}. The goal is to learn a compressed representation $\vec{x} \in \mathbb{R}^{k}$ of items from their higher-dimensional vector representations $\vec{v} \in \mathbb{R}^{K}$, such that $k \ll K$. The model has an hour-glass shape as shown in Figure \ref{fig:autoencoder} and is trained by feeding in the high-dimensional vector inputs and trying to re-construct the same representation at the output layer. The lower-dimensional middle layer forces the encoder part of the model to extract the \emph{minimal sufficient statistics} of $\vec{v}$ into $\vec{x}$, such that the decoder part of the network can reconstruct the original input back from $\vec{x}$. The model is trained by minimizing the reconstruction error between the input $\vec{v}$ and the actual output of the decoder $\vec{v'}$. The squared-loss is popularly employed.

\begin{align}
\label{eqn:autoencoder}
\mathcal{L}_{auto-encoder}(\vec{v}, \vec{v'}) &= \norm{\vec{v} - \vec{v'}}^2
\end{align}

\paragraph*{Siamese networks}

Siamese networks were originally proposed for comparing fingerprints \cite{baldi1993neural} and signatures \cite{bromley1993signature}. \citet{yih2011learning} later adapted the same architecture for comparing short texts.

The siamese network, as seen in Figure \ref{fig:siamese}, resembles the auto-encoder architecture (if you squint hard enough!)---but unlike the latter is trained on pairs of inputs $\langle input_1, input_2\rangle$. The architecture consists of two models ($model_1$ and $model_2$) that project $input_1$ and $input_2$, respectively, to $\vec{v_1}$ and $\vec{v_2}$ in a common embedding space. A pre-defined metric (e.g., cosine similarity) is used to then compute the similarity between $\vec{v_1}$ and $\vec{v_2}$. The model parameters are optimized such that $\vec{v_1}$ and $\vec{v_2}$ are closer when the two inputs are expected to be similar, and further away otherwise.

One possible loss function is the logistic loss. If each training sample consist of a triple $\langle \vec{v_{q}}, \vec{v_{d1}}, \vec{v_{d2}} \rangle$, such that $sim(\vec{v_{q}}, \vec{v_{d1}})$ should be greater than $sim(\vec{v_{q}}, \vec{v_{d2}})$, then we minimize,

\begin{align}
\label{eqn:siamese}
\mathcal{L}_{siamese}(\vec{v_{q}}, \vec{v_{d1}}, \vec{v_{d2}}) &= log\Big(1 + e^{\gamma(sim(\vec{v_{q}}, \vec{v_{d2}}) - sim(\vec{v_{q}}, \vec{v_{d1}}))}\Big)
\end{align}

where, $\gamma$ is a constant that is often set to 10. Typically both the models---$model_1$ and $model_2$---share identical architectures, but can also choose to share the same parameters.

It is important to note that, unlike the auto-encoder, the minimal sufficient statistics retained by a Siamese network is dictated by which information it deems important for determining the similarity between the paired items.

\subsection{Neural toolkits}
\label{sec:dnn-toolkits}

In recent years, the advent of numerous flexible toolkits \cite{jia2014caffe, yu2014introduction, abadi2016tensorflow, collobert2011torch7, chen2015mxnet, al2016theano, tokui2015chainer, neubig2017dynet} has had a catalytic influence on the area of neural networks. Most of the toolkits define a set of common neural operations that---like Lego\footnote{\url{https://en.wikipedia.org/wiki/Lego}} blocks---can be composed to build complex network architectures.\footnote{\url{http://www.inference.vc/content/images/2016/01/9k-.jpg}} Each instance of these neural operations or \emph{computation nodes} can have associated learnable parameters that are updated during training, and these parameters can be shared between different parts of the network if necessary. Every computation node under this framework must implement the appropriate logic for,

\begin{itemize}
  \item computing the output of the node given the input (forward-pass)
  \item computing the gradient of the loss with respect to the inputs, given the gradient of the loss with respect to the output (backward-pass)
  \item computing the gradient of the loss with respect to its parameters, given the gradient of the loss with respect to the output (backward-pass)
\end{itemize}

A deep neural network, such as the one in Figure \ref{fig:neuralnet} or ones with much more complex architectures (e.g., \cite{he2016deep, szegedy2015going, larsson2016fractalnet}), can then be specified by chaining instances of these available computation nodes, and trained end-to-end on large datasets using backpropagation over GPUs or CPUs. In IR, various application interfaces \cite{van2017pyndri, mitra2017luandri} bind these neural toolkits with existing retrieval/indexing frameworks, such as Indri \cite{strohman2005indri}.

Refer to \cite{shi2016benchmarking} for a comparison of different neural toolkits based on their speed of training using standard performance benchmarks.


\section{Deep neural models for IR}
\label{sec:dnnir}

Traditionally, deep neural network models have much larger number of learnable parameters than their shallower counterparts. A DNN with a large set of parameters can easily overfit to smaller training datasets \cite{zhang2016understanding}. Therefore, during model design it is typical to strike a balance between the number of model parameters and the size of the data available for training. Data for ad-hoc retrieval mainly consists of,

\begin{itemize}
  \item Corpus of search queries
  \item Corpus of candidate documents
  \item Ground truth---in the form of either explicit human relevance judgments or implicit labels (e.g., from clicks)---for query-document pairs
\end{itemize}

While both large scale corpora of search queries \cite{Pass:2006, craswell2009proceedings} and documents \cite{callan2009clueweb09, craswell2003overview, bailey2003engineering} are publicly available for IR research, the amount of relevance judgments that can be associated with them are often limited outside of large industrial research labs---mostly due to user privacy concerns. We note that we are interested in datasets where the raw text of the query and the document is available. Therefore, this excludes large scale public labelled datasets for learning-to-rank (e.g., \cite{liu2007letor}) that don't contain the textual contents.

The proportion of labelled and unlabelled data that is available influences the \emph{level of supervision} that can be employed for training these deep models. Most of the models we covered in Section \ref{sec:embir} operate under the data regime where large corpus of documents or queries is available, but limited (or no) labelled data. Under such settings where no direct supervision or relevance judgments is provided, typically an \emph{unsupervised} approach is employed (e.g., \cite{salakhutdinov2009semantic}). The unlabelled document (or query) corpus is used to learn good text representations, and then these learnt representations are incorporated into an existing retrieval model or a query-document similarity metric. If small amounts of labelled data are available, then that can be leveraged to train a retrieval model with few parameters that in turn uses text representations that is pre-trained on larger unlabelled corpus. Examples of such \emph{semi-supervised} training includes models such as \cite{pang2016text, pang2016study, guo2016deep}. In contrast, \emph{fully-supervised} models such as \cite{huang2013learning, severyn2015learning, mitra2016learning, cohen2016end}, optimize directly for the target task by training on large number of labelled query-document pairs.

It is also useful to distinguish between deep neural models that focus on ranking long documents, from those that rank short texts (e.g., for the question-answering task, or for document ranking where the document representation is based on the title or on clicked queries). The challenges in short text ranking are somewhat distinct from those involved in the ad-hoc retrieval task \cite{cohen2016adaptability}. When computing similarity between pairs of short-texts, vocabulary mismatches are more likely than when the retrieved items contain long text descriptions \cite{metzler2007similarity}. Neural models that perform matching in an embedding space tends to be more robust towards the vocabulary mismatch problem compared to lexical term-based matching models. On the other hand, documents with long body texts may contain mixture of many topics and the query matches may be spread over the whole document. A neural document ranking model (NDRM) must effectively aggregate the relevant matches from different parts of a long document.

In the rest of this section, we discuss different types of NDRM architectures and approaches that have been explored in the literature.

\subsection{Document auto-encoders}
\label{sec:dnnir-autoencoder}

\citet{salakhutdinov2009semantic} proposed one of the earliest deep neural models for ad-hoc retrieval. The model is a deep auto-encoder trained on unlabelled document corpus. The model treats each document as a bag-of-terms and uses a one-hot vector for representing the terms themselves---considering only top two thousand most popular terms in the corpus after removing stopwords. \citet{salakhutdinov2009semantic} first pre-train the model layer-by-layer, and then train it further end-to-end for additional tuning. The model uses binary hidden units and therefore the learnt vector representations of documents are also binary.

The \emph{Semantic Hashing} model generates a condensed binary vector representation (or a hash) of documents. Given a search query, a corresponding hash is generated and the relevant candidate documents quickly retrieved that match the same hash vector. A standard IR model can then be employed to rank between the selected documents.

\begin{table}
\caption{Comparing the nearest neighbours for "seattle" and "taylor swift" in the CDSSM embedding spaces when the model is trained on query-document pairs vs. query prefix-suffix pairs. The former resembles a Topical notion of similarity between terms, while the latter is more Typical in the definition of inter-term similarities.}
\label{tbl:dnnir-nearestneighbors}
\begin{center}
\resizebox{.96\textwidth}{!}{
\begin{tabular}{c c c c c c c c}
\midrule
\multicolumn{3}{c}{\textbf{seattle}} & & & \multicolumn{3}{c}{\textbf{taylor swift}} \\
Query-Document & & Prefix-Suffix & & & Query-Document & & Prefix-Suffix \\ \cline{1-3} \cline{6-8}
weather seattle & & chicago & & & taylor swift.com & & lady gaga \\
seattle weather & & san antonio & & & taylor swift lyrics & & meghan trainor \\
seattle washington & & denver & & & how old is taylor swift & & megan trainor \\
ikea seattle & & salt lake city & & & taylor swift twitter & & nicki minaj \\
west seattle blog & & seattle wa & & & taylor swift new song & & anna kendrick \\
\bottomrule
\end{tabular}
}
\end{center}
\end{table}

Semantic hashing is an example of a document encoder based approach to IR. The vocabulary size of two thousand distinct terms may be too small for most practical IR tasks. A larger vocabulary or a different term representation strategy---such as the character trigraph based representation of Figure \ref{fig:textinput3}---may be considered. Another shortcoming of the auto-encoder architecture is that it minimizes the document reconstruction error which may not align exactly with the goal of the target IR task. A better alternative may be to train on query-document paired data where the choice of what constitutes as the minimal sufficient statistics of the document is influenced by what is important for determining relevance of the document to likely search queries. In line with this intuition, we next discuss the Siamese architecture based models.

\subsection{Siamese networks}
\label{sec:dnnir-siamese}

In recent years, several deep neural models based on the Siamese architecture have been explored especially for short text matching. The \emph{Deep Semantic Similarity Model} (DSSM) \cite{huang2013learning} is one such architecture that trains on query and document title pairs where both the pieces of texts are represented as bags-of-character-trigraphs. The DSSM architecture consists of two deep models---for the query and the document---with all fully-connected layers and cosine distance as the choice of similarity function in the middle. \citet{huang2013learning} proposed to train the model on clickthrough data where each training sample consists of a query $q$, a positive document $d^{+}$ (a document that was clicked by a user on the SERP for that query), and a set of negative documents $D^{-}$ randomly sampled with uniform probability from the full collection. The model is trained my minimizing the cross-entropy loss after taking a softmax over the model outputs for all the candidate documents,

\begin{align}
\label{eqn:dssm}
\mathcal{L}_{dssm}(q, d^{+}, D^{-}) &= -log\Big(\frac{e^{\gamma\cdot cos\big(\vec{q}, \vec{d^{+}}\big)}}{\sum_{d \in D}{e^{\gamma\cdot cos\big(\vec{q}, \vec{d}\big)}}}\Big) \\
\text{where,}\quad D &= \{d^{+}\}\cup D^{-}
\end{align}

While, DSSM \cite{huang2013learning} employs deep fully-connected architecture for the query and the document models, more sophisticated architectures involving convolutional layers \cite{shen2014learning, gao2014modeling, shen2014latent, hu2014convolutional}, recurrent layers \cite{palangi2015deep, palangi2014semantic}, and tree-structured networks \cite{tai2015improved} have also been explored. The similarity function can also be parameterized and implemented as additional layers of the neural network as in \cite{severyn2015learning}. Most of these models have been evaluated on the short text matching task, but \citet{mitra2016learning} recently reported meaningful performances on the long document ranking task from models like DSSM \cite{huang2013learning} and CDSSM \cite{shen2014latent}. \citet{mitra2016learning} also show that sampling the negative documents uniformly from the collection is less effective to using documents that are closer to the query intent but judged as non-relelvant by human annotators.

\begin{figure}
\center
\includegraphics[width=.7\linewidth]{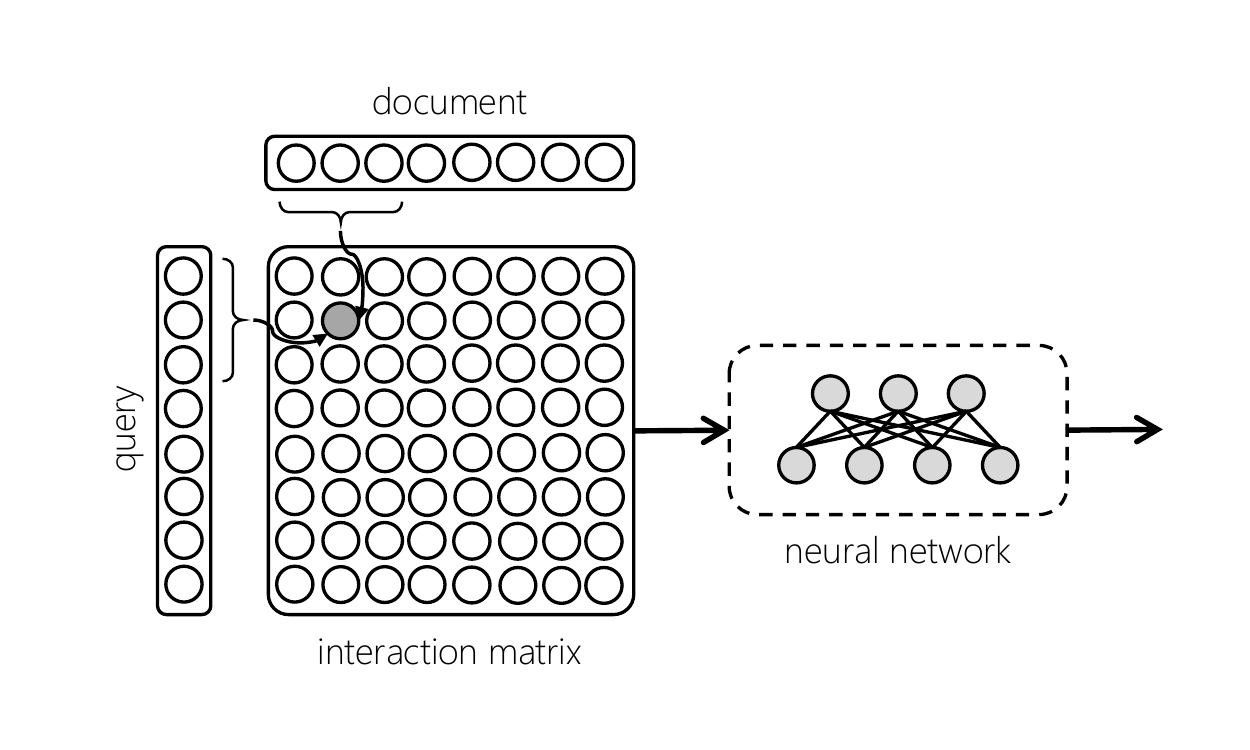}
\caption{Schematic view of an interaction matrix generated by comparing windows of text from the query and the document. A deep neural network---such as a CNN---operates over the interaction matrix to find patterns of matches that suggest relevance of the document to the query.}
\label{fig:interactionmatrix}
\end{figure}

\paragraph*{Notions of similarity}

It is important to emphasize that our earlier discussion in Section \ref{sec:wordrep-sim} on different notions of similarity between terms that can be learnt by shallow embedding models is also relevant in the context of these deeper architectures. In the case of Siamese networks, such as the convolutional-DSSM (CDSSM) \cite{shen2014latent}, the notion of similarity being modelled depends on the choice of the paired data that the model is trained on. When the CDSSM is trained on query and document title pairs \cite{shen2014latent} then the notion of similarity is more \emph{Topical} in nature. \citet{mitra2015suffixrank} trained the same CDSSM architecture on query prefix-suffix pairs which, in contrast, captures a more \emph{Typical} notion of similarity, as shown in Table \ref{tbl:dnnir-nearestneighbors}. In a related work, \citet{mitraexploring} demonstrated that the CDSSM model when trained on session-query pairs is amenable to vector-based text analogies.

\begin{align}
\vec{v}_{\text{things to do in london}} - \vec{v}_{\text{london}} + \vec{v}_{\text{new york}} &\approx \vec{v}_{\text{new york tourist attractions}} \\
\vec{v}_{\text{university of washington}} - \vec{v}_{\text{seattle}} + \vec{v}_{\text{denver}} &\approx \vec{v}_{\text{university of colorado}} \\
\vec{v}_{\text{new york}} + \vec{v}_{\text{newspaper}} &\approx \vec{v}_{\text{new york times}}
\end{align}

By modelling different notions of similarity these deep neural models tend to be more suitable for other IR tasks, such as query auto-completion \cite{mitra2015suffixrank} or session-based personalization \cite{mitraexploring}.

\subsection{Interaction-based networks}
\label{sec:dnnir-interaction}

Siamese networks represent both the query and the document using single embedding vectors. Alternatively, we can individually compare different parts of the query with different parts of the document, and then aggregate these partial evidence of relevance. Especially, when dealing with long documents---that may contain a mixture of many topics---such a strategy may be more effective than trying to represent the full document as a single low-dimensional vector. Typically, in these approaches a sliding window is moved over both the query and the document text and each instance of the window over the query is compared against each instance of the window over the document text (see Figure \ref{fig:interactionmatrix}). The terms within each window can be represented in different ways including, one-hot vectors, pre-trained embeddings, or embeddings that are updated during the model training. A neural model (typically convolutional) operates over the generated interaction matrix and aggregates the evidence across all the pairs of windows compared.

The interaction matrix based approach have been explored both for short text matching \cite{lu2013deep, hu2014convolutional, yin2015abcnn, pang2016text, yang2016anmm, wan2015deep}, as well as for ranking long documents \cite{mitra2016learning, pang2016study}.

\subsection{Lexical and semantic matching networks}
\label{sec:dnnir-lexical}

\begin{figure}[t]
\center
\begin{subfigure}{0.49\textwidth}
    \noindent
    \small
    \parbox{0.95\columnwidth}{%
    \input{uspresident1.tex}
    \vspace{1em}}
    \caption{Lexical model}
    \label{fig:dnnir-uspresident1}
\end{subfigure}
\hfill
\begin{subfigure}{0.49\textwidth}
    \noindent
    \small
    \parbox{0.95\columnwidth}{%
    \input{uspresident2.tex}
    \vspace{1em}}
    \caption{Semantic model}
    \label{fig:dnnir-uspresident2}
\end{subfigure}
\caption{Analysis of term importance for estimating the relevance of a passage to the query ``United States President'' by a lexical and a semantic deep neural network model. The lexical model only considers the matches of the query terms in the document, but gives more emphasis to earlier occurrences. The semantic model is able to extract evidence of relevance from related terms such as ``Obama'' and ``federal''.}
\label{fig:dnnir-duetviz}
\end{figure}

Much of the explorations in neural IR models have focused on learning good representations of text. However, these representation learning models tend to perform poorly when dealing with rare terms and search intents. In Section \ref{sec:ir-desiderata}, we highlighted the importance of modelling rare terms in IR. Based on similar motivaions, \citet{guo2016deep} and \citet{mitra2016learning} have recently emphasized the importance of modelling lexical matches using deep neural networks. \citet{mitra2016learning} argue that Web search is a ``tale of two queries''. For the query ``pekarovic land company'', it is easier to estimate relevance based on patterns of exact matches of the rare term ``pekarovic''. On the other hand, a neural model focused on matching in the embedding space is unlikely to have a good representation for this rare term. In contrast, for the query ``what channel are the seahawks on today'', the target document likely contains ``ESPN'' or ``Sky Sports''---not the term ``channel''. A representation learning neural model can associate occurrences of ``ESPN'' in the document as positive evidence towards the document being relevant to the query. Figure \ref{fig:dnnir-duetviz} highlights the difference between the terms that influence the estimation of relevance of the same query-passage pair by a lexical matching and a semantic matching model. A good neural IR model should incorporate both lexical and semantic matching signals \cite{mitra2016learning}. 

\begin{figure}
\center
\includegraphics[width=.9\linewidth]{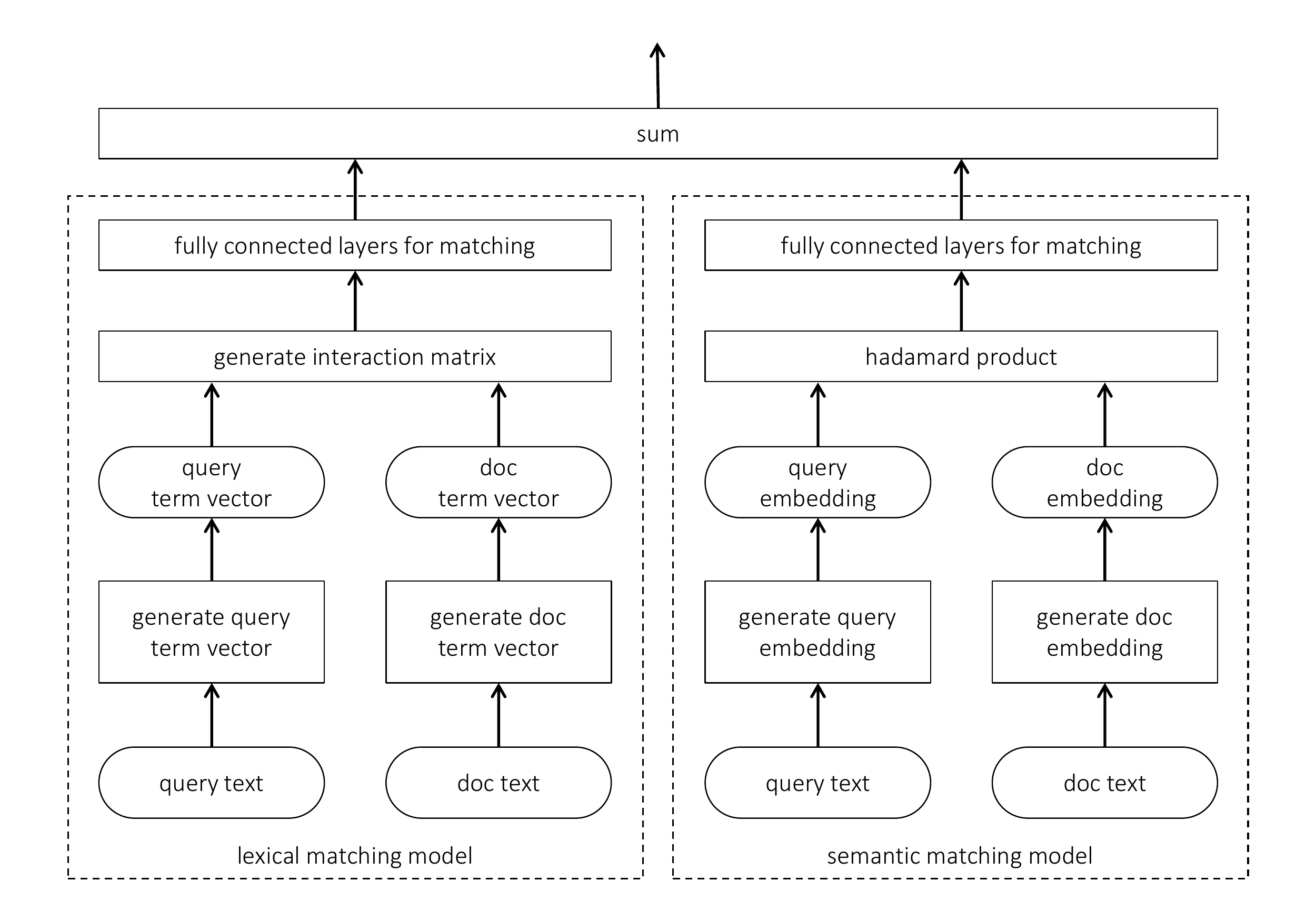}
\caption[Caption]{In the Duet architecture \cite{mitra2016learning}, the two sub-networks are jointly trained and the final output is a linear combination of the outputs of the lexical and the semantic matching sub-networks. The lexical matching sub-network (left) uses a convolutional model that operates over a binary interaction matrix.\footnotemark The semantic matching sub-network (right) learns representations of query and document text for effective matching in the embedding space. Cross-entropy loss is used to train the network similar to other models in Section \ref{sec:dnnir-siamese}.}
\label{fig:duet}
\end{figure}

\citet{guo2016deep} proposed to use histogram-based features in their DNN model to capture lexical notion of relevance. \citet{mitra2016learning} leverage large scale labelled data from Bing to train a \emph{Duet} architecture (Figure \ref{fig:duet}) that learns to identify good patterns of both lexical and semantic matches jointly. Neural models that focus on lexical matching typically have fewer parameters, and can be trained under small data regimes---unlike their counterparts that focus on learning representations of text.

\footnotetext{It is important to emphasize, that while \citet{mitra2016learning} and others have used interaction-based representation for modelling lexical matches, the two ideas are distinct. Some interaction-matrix based representations compare texts using their pre-trained embeddings \cite{hu2014convolutional, yin2015abcnn}. Similarly, lexical matching can be modelled without employing an interaction matrix based representation \cite{guo2016deep}.}

\begin{figure}
\center
\includegraphics[width=.5\linewidth]{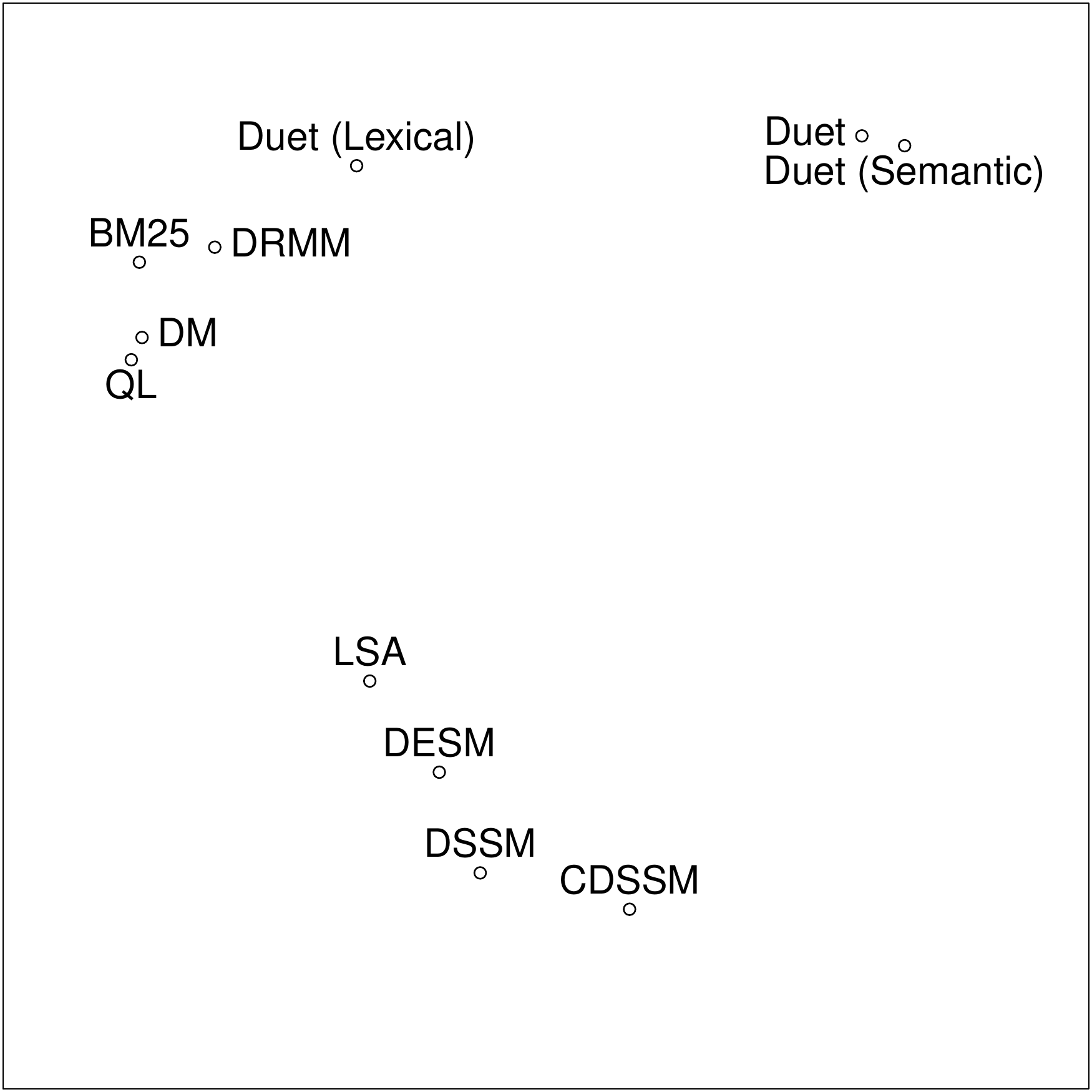}
\caption{A demonstration that IR models that focus on lexical matching tend to perform well on queries that are distinct from queries on which semantic matching models achieve good relevance. Each model is represented by a vector of NDCG scores achieved on a set of test queries. For visualization, t-SNE \cite{maaten2008visualizing} is used to plot the points in a two-dimensional space. Lexical matching models (BM25, QL, DM, DRMM, and Duet-Lexical) are seen to form a cluster---as well as the models that focus on representation learning.}
\label{fig:modelspace}
\end{figure}

Interestingly, a query level analysis seems to indicate that both traditional non-neural IR approaches and more recent neural methods tend to perform well on different segments of queries depending on whether they focus on lexical or semantic matching. Figure \ref{fig:modelspace} plots a few of these models based on their per-query NDCG values on a test set.

\section{Conclusion}
\label{sec:conclusion}

We present a tutorial on neural methods for information retrieval. For machine learning researchers who may be less familiar with IR tasks, we introduced the fundamentals  of traditional IR models and metrics. For IR researchers, we summarized key concepts related to representation learning with (shallow or deep) neural networks. Finally, we presented some of the recent neural methods for document ranking and question-answer matching.

We have focused on retrieval of long and short text. In the case of long text, the model must deal with variable length documents, where the relevant sections of a document may be surrounded by irrelevant text. For both long and short text, but particularly for short, IR models should also deal with the query-document vocabulary mismatch problem, by learning how patterns of query words and (different) document words can indicate relevance. Models should also consider lexical matches when the query contains rare terms---such as a person's name or a product model number---not seen during training, and to avoid retrieving semantically related but irrelevant results.

An ideal model for information retrieval would be able to infer the meaning of a query from context. Given a query about the Prime Minister of UK, for example, it may be obvious from context whether it refers to  John Major or Teresa May---perhaps due to the time period of the corpus, or it may need to be disambiguated based on other context such as the other query terms or the user's short or long-term history. The ideal IR model may need to encode this context, which means that the model is like a \emph{library} that effectively memorizes massive number of connections between entities and contexts. The number of learnable parameters of a ML model, however, is typically fixed, which may imply that there is a limited budget for how much real world knowledge the model can incorporate during training. An ideal model, therefore, may also need to learn to be like a \emph{librarian} with incomplete domain knowledge, but capable of reading documents related to the current query and reasoning about the meaning of the query as part of the retrieval process.

Many of the breakthroughs in deep learning have been motivated by the needs of specific application areas. Convolutional neural networks, for example, are particularly popular with the vision community, whereas recurrent architectures find more applications in speech recognition and NLP. It is likely that the specific needs and challenges of IR tasks may motivate novel neural architectures and methods. Future IR explorations may also be motivated by developments in related areas, such as NLP. For example, neural architectures that have been evaluated on non-IR tasks \cite{zhao2015self, kalchbrenner2014convolutional, denil2014modelling, kim2014convolutional, collobert2011natural} can be investigated in the retrieval context. Similarly, new methods for training deep models for NLP---e.g., using reinforcement learning \cite{ranzato2015sequence, yogatama2016learning} and generative adversarial networks (GANs) \cite{yu2016seqgan}---may carry over to the IR setup.

However, given the pace at which the area of deep learning is growing, in terms of the number of new architectures and training regimes, we should be wary of the combinatorial explosion of trying every model on every IR task. We should not disproportionately focus on maximizing quantitative improvements and in the process neglect theoretical understanding and qualitative insights. It would be a bad outcome for the field if these explorations do not grow our understanding of the fundamental principles of machine learning and information retrieval. Neural models should not be the hammer that we try on every IR task, or we may risk reducing every IR task to a nail.\footnote{\url{https://en.wikipedia.org/wiki/Law_of_the_instrument}} A better metaphor for the neural models may be a mirror that allows IR researchers to gain new insights into the underlying principles of IR. This may imply that we prefer neural models that, if not interpretable, then at least are amenable to analysis and interrogation. We may elicit more insights from simpler models while more sophisticated models may achieve state-of-the-art performances. As a community, we may need to focus on both to achieve results that are both impactful as well as insightful.

The focus of this article has been on ad-hoc retrieval and to a lesser extent on question-answering. However, neural approaches have shown interesting applications to other existing retrieval scenarios, including query auto-completion \cite{mitra2015suffixrank}, query recommendation \cite{sordoni2015hierarchical}, session modelling \cite{mitraexploring}, modelling diversity \cite{xia2016modeling}, modelling user click behaviours \cite{borisov2016neural}, knowledge-based IR \cite{nguyen2016toward}, and even optimizing for multiple IR tasks \cite{liu2015representation}. In addition, recent trends suggest that advancements in deep neural networks methods are also fuelling emerging IR scenarios such as conversational IR \cite{yan2016learning, zhou2016multi} and multi-modal retrieval \cite{ma2015multimodal}. Neural methods may have an even bigger impact on some of these other IR tasks.

IR also has a role in the context of the ambitions of the machine learning community. Retrieval is key to many one-shot learning approaches \cite{koch2015siamese, vinyals2016matching}. \citet{ghazvininejad2017knowledge} proposed to ``search'' external information sources in the process of solving complex tasks using neural networks. The idea of learning local representations proposed by \citet{diaz2016query} may be applicable to non-IR tasks. While we look at applying neural methods to IR, we should also look for opportunities to leverage IR techniques as part of---or in combination with---neural and other machine learning models.

Finally, we must also renew our focus on the fundamentals, including benchmarking and reproducibility. An important prerequisite to enable the ``neural IR train'' to steam forward is to build shared public resources---e.g., large scale datasets for training and evaluation, and repository of shared model implementations---and to ensure that appropriate bindings exist (e.g., \cite{van2017pyndri, mitra2017luandri}) between popular IR frameworks and popular toolkits from the neural network community. The emergence of new IR tasks also demands rethinking many of our existing metrics. The metrics that may be appropriate for evaluating document ranking systems may be inadequate when the system generates textual answers in response to information seeking questions. In the latter scenario, the metric should distinguish between whether the response differs from the ground truth in the information content or in phrasing of the answer \cite{mitraproposal, liu2016not, galley2015deltableu}. As multi-turn interactions with retrieval systems become more common, the definition of task success will also need to evolve accordingly. Neural IR should not only focus on novel techniques, but should also encompass all these other aspects.

\bibliographystyle{ACM-Reference-Format}
\bibliography{bibtex}

\end{document}